\documentstyle[eqsecnum,aps,smc,floats,twocolumn,psfig]{revtex}

\makeatletter

\def\thebibliography#1{\section*{\refname\@mkboth
 {\uppercase{\refname}}{\uppercase{\refname}}}\list
 {\@biblabel{\arabic{enumi}}}{\settowidth\labelwidth{\@biblabel{#1}}%
 \leftmargin\labelwidth
 \advance\leftmargin\labelsep
 \usecounter{enumi} 
 \def\theenumi{\arabic{enumi}}}%
 \def\newblock{\hskip .11em plus.33em minus.07em}%
 \sloppy\clubpenalty4000\widowpenalty4000
 \sfcode`\.=1000\relax}

\makeatother

\def\Aref#1{$^{\rm #1}$} 
\def\AAref#1#2{$^{\rm #1,#2}$} 
\def\IAref#1#2{$^{\Inst{#1},\rm #2}$} 
\def\IIref#1#2{$^{\Inst{#1},\Inst{#2}}$} 

\def\IIAref#1#2#3{$^{\Inst{#1},\Inst{#2},\rm #3}$}

\def\Iref#1{$^{\Inst{#1}}$}

\def\U1{\rm U(1)}
\def\SU3{\rm SU(3)}
\def\gpone{g_1^{\rm p}}
\def\gdone{g_1^{\rm d}}
\def\gnone{g_1^{\rm n}}
\def\gptwo{g_2^{\rm p}}

\def\gammap{\Gamma_1^{\rm p}}

\def\gamman{\Gamma_1^{\rm n}}

\def\dsigt {a_0}
\def\au    {a_{\mathrm{u}}}
\def\as    {a_{\mathrm{s}}}
\def\ad    {a_{\mathrm{d}}}
\def\ac    {a_{\mathrm{c}}}
\def\ab    {a_{\mathrm{b}}}

\def\gev{\,\mbox{GeV}}
\def\gevtwo{\,\mbox{GeV}^2}

\newcommand{\beq}  {\begin{equation}}
\newcommand{\eeq}  {\end{equation}}
\newcommand{\bmath}{\begin{eqnarray}}
\newcommand{\emath}{\end{eqnarray}}
\newcommand{\TS}   {\textstyle}
\newcommand{\DS}   {\displaystyle}
\newcommand{\rme}  {\rm e}
\newcommand{\rmme}  {\rm \mu e}
\begin{document}
\begin{titlepage}

\draft

\preprint{HEP/123-qed; CERN--PPE/96--x}
\title{SPIN STRUCTURE OF THE PROTON FROM POLARIZED 
       INCLUSIVE DEEP--INELASTIC MUON--PROTON SCATTERING\\}
\vskip 2cm
\author{D.~Adams\Iref{a17},
B.~Adeva\Iref{a19},
E.~Arik\Iref{aa1},
A.~Arvidson\IAref{a22}{a},
B.~Badelek\IIref{a22}{a24},
M.K.~Ballintijn\Iref{a14},
G.~Bardin\IAref{a18}{\dagger},
G.~Baum\Iref{a1},
P.~Berglund\Iref{a7},
L.~Betev\Iref{a12}
% ...
\\
\vskip 0.3cm
(Spin Muon Collaboration)}
\address{~\vskip 0.5cm
\Instfoot {a1} {University of Bielefeld, Physics Department,
                33501 Bielefeld, Germany\Aref{aaa} }
\Instfoot {aa1} {Bogazi\c{c}i University and Istanbul Technical University,
                 Istanbul, Turkey\Aref{bbb} }
\Instfoot {a2} {University of California, Department of Physics,
                Los Angeles, 90024~CA, USA\Aref{ccc}}
\Instfoot {a4} {CERN, 1211 Geneva 23, Switzerland}
\Instfoot {a5} {University of Freiburg, Physics Department,
                79104 Freiburg, Germany\Aref{aaa}}
\Instfoot {a6} {GKSS, 21494 Geesthacht, Germany\Aref{aaa}}
\Instfoot {a7} {Helsinki University of Technology, Low Temperature
                Laboratory and Institute of Particle Physics Technology,
                Espoo, Finland}
\Instfoot {a8} {University of Houston, Department of Physics,
                and Institute for Beam Particle Dynamics,
                Houston, 77204 TX, USA\AAref{ccc}{ddd}}
\Instfoot {a9} {JINR, Dubna, RU-141980 Dubna, Russia} 
\Instfoot {a10} {University of Mainz, Institute for Nuclear Physics,
                 55099 Mainz, Germany\Aref{aaa}}
\Instfoot {a11} {University of Mons, Faculty of Science,
                 7000 Mons, Belgium}
\Instfoot {a12} {University of Munich, Physics Department,
                 80799 Munich, Germany\Aref{aaa}}
\Instfoot {a13} {Nagoya University, CIRSE and Department of Physics, Furo-Cho,
                 Chikusa-Ku, 464 Nagoya, Japan\Aref{eee}}
\Instfoot {a14} {NIKHEF, Delft University of Technology, FOM and 
                 Free University,
                 1009 AJ Amsterdam, The Netherlands\Aref{f{}f{}f}}
\Instfoot {a15} {Northeastern University, Department of Physics,
                 Boston, 02115 MA, USA\Aref{ddd}}
\Instfoot {a16} {Northwestern University, Department of Physics,
                 Evanston, 60208 IL, USA\AAref{ccc}{ddd}}
\Instfoot {a17} {Rice University, Bonner Laboratory,
                 Houston, 77251-1892 TX, USA\Aref{ccc}}
\Instfoot {a18} {C.E.A.~Saclay, DAPNIA, 91191 Gif-sur-Yvette, 
                 France\Aref{ggg}}
\Instfoot {a19} {University of Santiago, Department of Particle Physics,
                 15706 Santiago de Compostela, Spain\Aref{hhh}}
\Instfoot {a20} {Tel Aviv University, School of Physics,
                 69978 Tel Aviv, Israel\Aref{iii}}
\Instfoot {a21} {INFN Trieste and
                 University of Trieste, Department of Physics,
                 34127 Trieste, Italy}
\Instfoot {a22} {Uppsala University, Department of Radiation Sciences,
                 75121 Uppsala, Sweden}
\Instfoot {a23} {University of Virginia, Department of Physics,
                 Charlottesville, 22901 VA, USA\Aref{ccc}}
\Instfoot {a24} {Soltan Institute for Nuclear Studies
                 and Warsaw University,
                 00681 Warsaw, Poland\Aref{jjj}}
\Instfoot {a26} {Yale University, Department of Physics,
                 New Haven, 06511 CT, USA\Aref{ccc}}
\Anotfoot {a} {Now at Gammadata, Uppsala, Sweden}
\Anotfoot {b} {Now at INFN Torino, I-10125 Torino, Italy}
\Anotfoot {c} {Now at CEBAF, Newport News, VA 23606, USA}
\Anotfoot {d} {Now at Ericsson Infocom AB, Sweden}
\Anotfoot {e} {Now at University of Montreal,  H3C 3J7,
               Montreal, PQ, Canada}
\Anotfoot {f} {Now at CERN, 1211 Geneva 23, Switzerland}
\Anotfoot {g} {Now at University of Virginia, Department of Physics,
                 Charlottesville, 22901 VA, USA\Aref{ccc}}
\Anotfoot {h} {Now at INFN Trieste, 34127 Trieste, Italy}
\Anotfoot {i} {Now at Fermi National Accelerator Laboratory,
               Batavia, 60510 Illinois, USA} 
\Anotfoot {j} {Now at University of Wisconsin, USA} 
\Anotfoot {k} {Now at SLAC, Stanford CA 94309, USA}
\Anotfoot {l} {Now at Yale University, Department of Physics,
                 New Haven, 06511 CT, USA\Aref{ccc}}
\Anotfoot {m} {Permanent address: Miyazaki University, Faculty of Engineering,
               889-21 Miyazaki-Shi, Japan}
\Anotfoot {n} {Permanent address: Paul Scherrer Institut, 5232 Villigen,
                   Switzerland}
\Anotfoot {o} {Permanent address: The Institute of Physical and 
               Chemical Research (RIKEN), wako 351-01, Japan}
\Anotfoot {p} {Permanent address: University of California,
                 Institute of Particle Physics,
                 Santa Cruz, 95064 CA, USA}
\Anotfoot {q} {Permanent address: KEK, Tsukuba-Shi, 305 Ibaraki-Ken, Japan}
\Anotfoot {s} {Now at Ericsson Telecommunication, 5120 AA Rijen, 
               The Netherlands}
\Anotfoot {t} {Now at University of Bielefeld, Physics Department,
                33501 Bielefeld, Germany\Aref{aaa} }
\Anotfoot {u} {Permanent address: University of Buenos Aires,
               Physics Department, 1428 Buenos Ai\-res, Argentina }
\Anotfoot {v} {Permanent address: The American University, 
                Washington D.C. 20016, USA}
\Anotfoot {w} {Now at DESY}
\Anotfoot {x} {Now at Dresden Technical University, 01062 Dresden, 
               Germany}
\Anotfoot {y} {Permanent address: Brookhaven National Laboratory,
               Upton, 11973 NY, USA}
\Anotfoot {z} {Present address: ESFR, F-38043 Grenoble, France.}
\Anotfoot {aa} {Now at Los Alamos National Laboratory, Los Alamos,
               NM 87545, USA}
\Anotfoot {bb} {Now at NIKHEF, 1009 AJ Amsterdam, The Netherlands}
\Anotfoot {cc} {Now at F.Hoffmann-La Roche Ltd., CH-4070 Basel, Switzerland}
%
%
%
%
\Anotfoot {aaa} {Supported by the Bundesministerium f\"ur Bildung,
               Wissenschaft, Forschung und
               Technologie}
\Anotfoot {bbb} {Partially supported by TUBITAK and the Centre for 
               Turkish-Balkan Physics Research and Application 
               (Bogazi\c{c}i University)}
\Anotfoot {ccc}  {Supported by the U.S. Department of Energy}
\Anotfoot {ddd}  {Supported by the U.S. National Science Foundation}
\Anotfoot {eee}  {Supported by  Monbusho Grant-in-Aid
                for Scientific Research (International Scientific Research
                Program and Specially Promoted Research)}
\Anotfoot {f{}f{}f} {Supported by the National Science Foundation (NWO)
               of the Netherlands}
\Anotfoot {ggg} {Supported by the Commissariat \`a l'Energie Atomique}
\Anotfoot {hhh} {Supported by Comision Interministerial de Ciencia
               y Tecnologia}
\Anotfoot {iii} {Supported by the Israel Science Foundation.}
\Anotfoot {jjj} {Supported by KBN SPUB/P3/209/94.}
\Anotfoot {\dagger} {Deceased.}
}
\date{February 10, 1997}
\maketitle

\begin{abstract}
We have measured the spin-dependent structure function $g_1^{\rm p}$ 
in inclusive deep-inelastic scattering of polarized muons off 
polarized protons, in the kinematic range $0.003 < x < 0.7$ and
$1\gevtwo < Q^2 < 60\gevtwo$. 
A next-to-leading order QCD analysis is used to evolve the measured
$\gpone(x,Q^2)$ to a fixed $Q^2_0$. 
The first moment of $\gpone$ at $Q^2_0 = 10\gevtwo$ is  
$\gammap = 0.136\pm 0.013 \,(\mbox{stat.}) \pm 0.009\,(\mbox{syst.})
\pm 0.005\,(\mbox{evol.})$.
This result is below the prediction of the Ellis--Jaffe sum rule by more than
two standard deviations.
The singlet axial charge $\dsigt$ is found to be $0.28 \pm 0.16$. 
In the Adler--Bardeen factorization scheme, 
$\Delta g \simeq 2$ is required to bring 
$\Delta \Sigma$ in agreement with the Quark-Parton Model.
A combined analysis of all available proton and deuteron
data confirms the Bjorken sum rule.
\end{abstract}

\end{titlepage}


\section{INTRODUCTION}

Deep-inelastic scattering  of leptons from nucleons has revealed much of
what is known about quarks and gluons. The scattering of high-energy  charged 
polarized leptons on polarized nucleons 
provides insight into the spin structure of the nucleon at the parton level. 
The spin-dependent nucleon  structure functions
determined from these measurements are fundamental properties of the 
nucleon as are the spin-independent structure functions, and they
provide crucial information for the
development and testing of perturbative and non-perturbative 
Quantum Chromodynamics~(QCD).  Examples are the QCD spin-dependent sum
rules and calculations by lattice gauge theory.

\par
The first experiments on polarized electron--proton scattering 
were carried out by
the E80 and E130 Collaborations at SLAC~\cite{e80}.
They measured significant spin-dependent asymmetries in deep-inelastic
electron--proton scattering cross sections, and their results 
were consistent with the Ellis--Jaffe and Bjorken sum rules
with some plausible models of proton spin structure.
Subsequently, a similar experiment with a polarized muon beam and polarized 
proton target was made by the European
Muon Collaboration (EMC) at CERN~\cite{As88}.
With a tenfold higher beam energy as compared to that at SLAC,
the EMC measurement covered a much larger kinematic range than the
electron scattering experiments and found the violation of the 
Ellis--Jaffe sum rule~\cite{EJ74}.  This implies, in the framework of 
the Quark-Parton Model~(QPM), that the total contribution of the  
quark spins to the proton spin is small.

\par
This result was a great surprise and posed a major problem for the QPM, 
particularly because of the success of the QPM in explaining the 
magnetic moments of  hadrons in terms of three valence quarks.
It stimulated a new series of polarized electron and
muon nucleon scattering experiments which by now have achieved
the following:
\begin{enumerate}
\item inclusive scattering measurements of the spin-dependent 
  struc\-ture function $g_1^{\rm p}$ of the proton with improved 
  accuracy over an enlarged kinematic range;
\item evaluation of the first moment of the proton spin structure 
  function, $\Gamma_1^{\rm p} = \int_0^1 \gpone (x) {\rm d} x$,
  with reduced statistical and systematic errors;
\item similar measurements with polarized deuteron and $^3$He 
  targets, in order to measure the neutron spin structure function  
  and test the fundamental  Bjorken sum rule for 
  $\gammap - \gamman$~\cite{Bj66};
\item measurements of the spin-dependent structure
  function $g_2$ for the proton and neutron;
\item semi-inclusive measurements of final states which allow 
  determination of the separate valence and sea quark contributions
  to the nucleon spin.
\end{enumerate}

\par
The recent measurements of polarized muon-nucleon scattering have been done
by the Spin Muon Collaboration (SMC) at CERN 
with polarized muon beams of 100~GeV and 190~GeV obtained from the CERN SPS 
450~GeV proton beam and with polarized proton and deuteron targets. 
Spin-dependent cross section asymmetries are measured over a wide 
kinematic range with relatively high $Q^2$ and extending to low $x$ values.
The determination of $g_1(x,Q^2)$ for the proton and deuteron has been the 
principal result of the SMC experiment, but $g_2$ and semi-inclusive 
measurements have also been made. 

\par
The recent measurements of polarized electron-nucleon scattering 
have been done principally at SLAC in experiment E142~\cite{e142} (beam energy
$E_e \sim 19, 23, 26~{\rm GeV}$, $^3$He target), E143~\cite{E143p,E143d} 
(beam energy 
$E_e\sim 9, 16, 29~{\rm GeV}$, H and D
targets) and E154 ($E_e \sim 48~{\rm GeV}$, $^3$He target).
SLAC E155 with $E_e \sim 50~{\rm GeV}$ and polarized proton and deuteron 
targets will take data soon.
The SLAC experiments provide inclusive measurements of $g_1$ and $g_2$
over a kinematic range of relatively low $Q^2$ and do not extend to very
low $x$ values. However, the electron scattering experiments involve
very high beam intensities and achieve excellent statistical accuracies.
Hence the electron and the muon experiments are complementary.
Recently the HERMES experiment at DESY has become operational
and has reported preliminary results with a polarized $^3$He 
target~\cite{HER_he3}.
This experiment uses a polarized electron beam of $27\gev$  in the 
electron ring at HERA and an internal polarized gas target.
Both inclusive and semi-inclusive data were obtained, and polarized 
H and D targets will be used in the future.

\par
In this paper, we present SMC results on the spin-dependent structure 
functions
$\gpone$ and $\gptwo$ of the proton, obtained from data taken in 1993 with a 
polarized butanol target. 
First results from these measurements were published 
in Refs.~\cite{SMC94p,SMC_g2}.
We use here the same data sample but present a more refined analysis; in
particular, we allow for a
$Q^2$-evolution of the $\gpone$ structure function 
as predicted by perturbative QCD. 
SMC has also published results on the deuteron structure function 
$\gdone$~\cite{SMCd93,SMCd95,SMCd96}
and on a measurement of semi-inclusive cross section 
asymmetries~\cite{SMC_semi}.
For a test of the Bjorken sum rule 
we refer to our measurement of $\gdone$.

\par
The paper is organized as follows.
In Section~\ref{theory}, we review the theoretical background.
The experimental set-up and the data-taking procedure
are described in Section~\ref{Exp_method}.
In Section~\ref{Ana_asy} we discuss the analysis of cross section 
asymmetries and in Section~\ref{g1_mom} we give the evaluation of the 
spin-dependent structure function $\gpone$ and its first moment.
The results for $\gptwo$ are discussed in Section~\ref{g2_mom}.
In Section~\ref{bj_sec} we combine proton and deuteron results to 
determine the structure function  $\gnone$ of the
neutron and to test the Bjorken sum rule. 
In Section~\ref{scon} we interpret our results in terms of the spin 
structure of the proton.
Finally, we present our conclusions in Section~\ref{conclusion}.

\section{THEORETICAL OVERVIEW}
\label{theory}
\subsection{The cross sections  for polarized lepton-nucleon scattering}

\par
The polarized deep-inelastic lepton--nucleon inclusive   
scattering cross section in the
one-photon exchange approximation can be written as the sum of a
spin-independent term $\bar{\sigma}$ and a 
spin-dependent term $\Delta \sigma $ and involves the lepton
helicity $h_{\ell}=\pm 1$:
\beq
  \sigma = \bar{\sigma} - \frac{1}{2} h_{\ell} \Delta\sigma.
\eeq
For longitudinally polarized leptons  
the spin $\bf{S_{\ell}}$ is along the lepton
momentum $\bf{k}$.
The spin-independent cross section for parity-conserving interactions 
can be expressed in terms of two unpolarized structure functions
$F_1$ and $F_2$.
These functions depend on the four momentum transfer squared $Q^2$ 
and the scaling variable $x = Q^2/2M\nu$, where $\nu$ is the energy
of the exchanged virtual photon, and $M$ is the nucleon mass.
The double differential cross section can be written 
as a function of $x$ and $Q^2$~\cite{tp_rw}:
\bmath
\nonumber
\frac{{\rm d^2}\bar{\sigma}} {{\rm d}x {\rm d}Q^2} =
\frac{4 {\pi}{\alpha}^2}{Q^{4} x} 
\Biggl[ xy^2(1-\frac{2m_{\ell}^2}{Q^2}) F_1(x,Q^2) + \\
(1 - y - \frac{{\gamma}^2 y^2}{4}) F_2(x,Q^2) \Biggr],
\emath
where $m_{\ell}$ is the lepton mass, 
$y=\nu/E$ in the laboratory system, and 
\beq
\gamma = {2Mx \over \sqrt{Q^2}} = {\sqrt{Q^2} \over \nu}. 
\label{eq:gamma}
\eeq

\par
\begin{figure}[t]
\begin{center}
\hspace{0.25cm}
\psfig{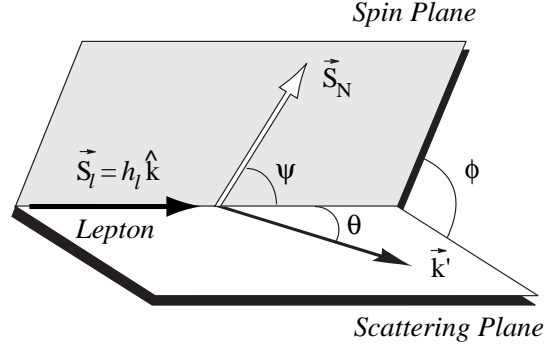}
\vspace{0.25cm}
\protect\caption[]{\label{plane_angles}Lepton and nucleon 
kinematic variables in polarized lepton scattering on a 
fixed polarized  nucleon  target.}
\end{center}
\end{figure}

The spin-dependent part of the cross section
can be written in terms of two structure functions $g_1$ and $g_2$ which
describe the interaction of lepton and hadron currents. 
When the lepton spin and the nucleon spin form an angle $\psi$, 
it can be expressed as~\cite{jaf_g2}
\beq
 \Delta \sigma=\cos\psi\,\Delta \sigma_{\parallel} 
  + \sin\psi\,\cos\phi\,\Delta \sigma_{\perp},
\label{delta_sigma}
\eeq 
where $\phi$ is the azimuthal angle between the scattering plane
and the spin plane (Fig.~\ref{plane_angles}).

\par
The cross sections $\Delta \sigma_{\parallel}$ and $\Delta \sigma_{\perp}$ 
refer to the two configurations where the
nucleon spin is (anti)parallel or orthogonal to the lepton spin;
$\Delta \sigma_{\parallel}$ is the difference 
between the cross sections for antiparallel and parallel spin orientations 
and $\Delta \sigma_{\perp} = -h_{\ell} \Delta \sigma_{\rm T}/\cos\phi$,
the difference between the cross sections at angles 
$\phi$ and $\phi + \pi$.
The corresponding differential cross sections are given by
\beq
  \frac{{\rm d}^2\Delta\sigma_{\parallel}}{{\rm d}x {\rm d}Q^2}
  = \frac{16\pi\alpha^2 y}{Q^4} \Biggl[(1 -{y\over 2} 
  -{\gamma^2 y^2\over 4})g_1 -{\gamma^2 y\over 2} g_2 \Biggr],
\label{delta_sigma_para}
\eeq
and
\beq
 \frac{{\rm d}^3\Delta\sigma_{\rm T}}{{\rm d}x {\rm d}Q^2{\rm d}\phi}
 = - \cos\phi\,\frac{8\alpha^2 y}{Q^4}\,\gamma\,
 \sqrt{1-y-{\gamma^2y^2 \over 4}}
 \Biggl({y \over 2} g_1 + g_2 \Biggr).
\label{delta_sigma_orto}
\eeq
For a high beam energy $E$, $\gamma$ is small since either $x$ is small or
$Q^2$ large. 
The structure function $g_1$ is therefore best measured
in the (anti)parallel configuration where it 
dominates the spin-dependent cross section;
$g_2$ is best obtained from a measurement in the orthogonal configuration, 
combined with a measurement of $g_1$.
In all formulae used in this article we consider only the 
single virtual-photon exchange.  The interference effects between virtual
Z$^0$ and photon exchange in deep-inelastic muon scattering have been 
measured~\cite{BCDMS} and found to be small and compatible with the
standard model expectations.  They can be neglected in the kinematic
range of current experiments.

\subsection{The cross section asymmetries}
\label{x_ass}

\par
The spin-dependent cross section terms, Eqs.~(\ref{delta_sigma_para})
and (\ref{delta_sigma_orto}), make only a small contribution to the total 
deep-inelastic scattering cross section and furthermore  their contribution 
is, in general, reduced by incomplete beam and target polarizations. 
Therefore they can best be determined from
measurements of cross section asymmetries in which the spin-independent 
contribution cancels.
The  relevant asymmetries are
\beq
      A_{\parallel} = {\Delta\sigma_{\parallel} \over 2 \bar{\sigma}},
   \hspace{1cm}
             A_{\perp}     = {\Delta\sigma_{\perp} \over 2 \bar{\sigma}},
\label{Asy_cross}
\eeq
which are related to the virtual photon-proton asymmetries 
$A_1$ and $A_2$ by 
\beq
             A_{\parallel} = D (A_1 + \eta A_2),  
   \hspace{1cm} 
      A_{\perp}     =  d (A_2 -\xi A_1),
\label{A_cross}
\eeq
where
\bmath
\label{A12}
          A_{1} &=& \frac{\sigma_{1/2} - \sigma_{3/2}}{\sigma_{1/2} +
               \sigma_{3/2}}
                  = \frac{g_1-\gamma^2\,g_2}{F_1}, \\
   A_{2} &=& \frac{2 \sigma^{\rm TL}} {\sigma_{1/2} + \sigma_{3/2}}
                  = \gamma \frac{g_1 + g_2}{F_1}.
\nonumber
\emath
In Eqs.~(\ref{A_cross}) and (\ref{A12}),
$D$ is the depolarization factor of the virtual photon defined below
and $d$, $\eta$ and $\xi$ are the kinematic factors:
\bmath  
             d    &=& \frac{\sqrt{1-y-\gamma^2 y^2/4}}{1 -y/2}\,D,
\\
             \eta &=& \frac{\gamma(1 - y - \gamma^2 y^2/4)}
                                {(1 - y/2)(1 + \gamma^2 y/2)},
\\
             \xi  &=& \frac{\gamma (1 - y/2)}{1 + \gamma^2 y/2}~.
\emath
The cross sections $\sigma_{1/2}$ and $\sigma_{3/2}$ 
refer to the absorption of a transversely polarized  virtual photon
by a polarized proton for total photon--proton angular momentum component
along the virtual photon axis of 1/2 and 3/2, respectively;
$\sigma^{\rm TL}$ is an interference cross section due to the helicity
spin-flip amplitude in forward Compton scattering~\cite{IOFFE}. 
The depolarization factor $D$ depends on $y$ and on the ratio
$R=\sigma_{\rm L}/\sigma_{\rm T}$ of longitudinal and transverse
photoabsorption cross sections:
\beq  
   D = \frac{y (2-y)(1+\gamma^2 y/2) }
            {y^2(1+\gamma^2) (1\!-\!2m_{\ell}^2/Q^2)\!+\!\!
        2(1\!-\!y\!-\!\gamma^2y^2/4)(1\!+\!R) }~.
\label{D_eq}
\eeq  

\par
From Eqs.~(\ref{A_cross}) and (\ref{A12}), we can express the virtual 
photon-proton  asymmetry $A_1$ in
terms of $g_1$ and $A_2$ and find the following relation for the
longitudinal asymmetry: 
\beq
        \frac{A_{\parallel}}{D} 
             = (1 + \gamma ^2)\frac{g_{1}}{F_{1}} + (\eta-\gamma ) A_2.
\label{aparall2}
\eeq
\mbox{The virtual-photon asymmetries} are bounded  by positivity relations 
$|A_1| \leq 1$ and $|A_2| \leq \sqrt{R}$~\cite{doncel}. 
When the term proportional to  $A_2$ is neglected in Eqs.~(\ref{A_cross}) 
and~(\ref{aparall2}), the longitudinal asymmetry 
is related to $A_1$ and $g_1$ by
\beq
   A_1 \simeq  \frac{A_{\parallel}}{D},  \hspace{1.5cm}
   \frac{g_{1}}{F_{1}} \simeq \frac{1}{1 + \gamma ^2}\, 
             \frac{A_{\parallel}}{D},
\label{aparall3}
\eeq
respectively,
where $F_1$ is usually expressed in terms of $F_2$ and $R$:     
\beq
   F_1 = \frac{1 + \gamma ^2}{ 2\,x\,(1+R)}{F_2}.
\label{f1f2}
\eeq
These relations are used in the present analysis for the evaluation of
$g_1$ in bins of $x$ and $Q^2$, starting from the
asymmetries measured in the parallel spin configuration and using
parametrizations of $F_2(x,Q^2)$ and $R(x,Q^2)$.

\par
The virtual photon-proton  asymmetry $A_2$  is evaluated from 
the measured transverse and longitudinal asymmetries $A_\parallel$ and
$A_\perp$:
\beq
   A_2 = \frac{1}{1+\eta \xi} \left( \frac{A_{\perp}}{d}
                 + \xi \frac{A_{\parallel}}{D} \right).
\label{q2a2}
\eeq
From Eqs.~(\ref{eq:gamma}) and~(\ref{A12}), $A_2$ has an explicit
$1/\sqrt{Q^2}$ dependence and is therefore expected to be small at high
energies.  
The structure function $g_2$ is obtained from the measured 
asymmetries using Eqs.~(\ref{A12}) and~(\ref{q2a2}).

\subsection{The spin-dependent structure function $g_1$}
\label{th_g1}

\par
The significance of the spin-dependent structure function $g_1$ can be
understood from the  virtual photon asymmetry $A_1$.  
As shown in Eq.~(\ref{A12}), 
$A_1\simeq g_1/F_1$, or $\sigma_{1/2}-\sigma_{3/2}\propto g_1$. 
In order to conserve angular momentum, 
a virtual photon with helicity $+1$ or $-1$ can only be absorbed by
a quark with a spin projection of $-\frac{1}{2}$ or $+\frac{1}{2}$,
respectively, if the quarks have no orbital angular momentum.
Hence, $g_1$ contains information on the quark spin
orientations with respect to the proton spin direction.

\par
In the simplest Quark-Parton Model (QPM), the quark densities
depend only on the momentum fraction~$x$ carried by the quark,
and $g_1$ is given by
\beq
   g_{1}(x) = {\TS\frac{1}{2}}\sum_{i=1}^{n_f} e_i^2 \Delta q_i(x),
\label{g1}
\eeq
where 
\beq
  \Delta q_{i}(x) = q_i^+(x) - q_i^-(x) + \bar q_i^{~+}(x) - \bar q_i^{~-}(x),
\eeq
$q_i^+(\bar q_i^{~+})$ and  $q_i^-(\bar q_i^{~-})$ are  
the distribution functions of quarks~(antiquarks) with spin
parallel and antiparallel to the nucleon spin,
respectively, $e_i$  is the electric charge of the quarks of flavor $i$;
and $n_f$ is the number of quark flavors involved.  

\par
In QCD, quarks interact by gluon exchange which gives rise to a weak
$Q^2$ dependence of the structure functions. 
The treatment of $g_1$ in perturbative QCD follows 
closely that of unpolarized parton distributions and structure 
functions~\cite{alta82}.
At a given scale $Q^2$, $g_1$ is related to the polarized quark and
gluon distributions by coefficient functions $C_q$ and $C_g$
through~\cite{alta82}
\bmath
   g_1(x,t) & = & {\TS\frac{1}{2}} \sum_{k=1}^{n_f} \frac{e_k^2 }{n_f}
        \int_x^1 \frac{{\rm d}y}{y} \Bigl [ 
        C_q^{\rm S}({\TS\frac{x}{y}},\alpha_s(t)) \Delta\Sigma(y,t)
      \nonumber \\
      & + & 2 n_f C_g({\TS\frac{x}{y}},\alpha_s(t)) \Delta g(y,t) 
      \nonumber \\
      & + & C_q^{\rm NS}({\TS\frac{x}{y}},\alpha_s(t)) 
     \Delta q^{\rm NS}(y,t) \Bigr].
\label{qcd:g1}
\emath
In this equation, $t = \ln(Q^2/\Lambda^2)$, $\alpha_s$ is the
strong coupling constant, and $\Lambda$ is the scale  parameter of QCD.
The superscripts S and NS, respectively, indicate flavor-singlet and 
non-singlet
parton distributions and coefficient functions;
$\Delta g(x,t)$ is the polarized gluon distribution and
$\Delta\Sigma$ and $\Delta q^{\rm NS}$ are the singlet and non-singlet
combinations of the polarized quark and antiquark distributions
\bmath
\label{qcd:qsi}
\Delta\Sigma(x,t) & = & \sum_{i=1}^{n_f} \Delta q_i(x,t),
\\
\Delta q^{\rm NS}(x,t) & = &
                {{\displaystyle \sum_{i=1}^{n_f}} 
                \left ( e_i^2 - 
                {\DS{1 \over n_f}}
                {\DS \sum_{k=1}^{n_f}} 
                e_k^2 \right ) 
                \over
                {\DS{1 \over n_f}}
                {\DS \sum_{k=1}^{n_f}}
                {e_k^2}}
                \Delta q_i(x,t).
\label{qcd:ns}
\emath
The $t$ dependence of the polarized quark and gluon distributions follows
the Gribov--Lipatov--Altarelli--Parisi (GLAP)
equations~\cite{alta77,lipatov}.  As for the unpolarized distributions,
the polarized singlet and gluon distributions are coupled by
\bmath
\label{qcd:apsi}
\nonumber
{{\rm d} \over {\rm d} t} \Delta\Sigma(x,t) 
     & = & \frac{\alpha_s(t)}{2\pi} 
     \int_x^1 \frac{{\rm d}y}{y} 
\Bigl [
  P_{qq}^{\rm S}(\protect{{\TS\frac{x}{y}}},\alpha_s(t)) \Delta\Sigma(y,t) \\
     & +&  2 n_f P_{qg}(\protect{{\TS\frac{x}{y}}},\alpha_s(t)) 
    \Delta g(y,t) 
\Bigr],
\\\nonumber
{{\rm d} \over {\rm d} t} \Delta g(x,t)  
     & = & \frac{\alpha_s(t)}{2\pi} \int_x^1 \frac{{\rm d}y}{y} 
\Bigl [
           P_{gq}(\protect{{\TS\frac{x}{y}}},\alpha_s(t)) \Delta\Sigma(y,t) \\
     & +&  P_{gg}(\protect{{\TS\frac{x}{y}}},\alpha_s(t)) \Delta g(y,t) 
\Bigr],
\label{qcd:apg}
\emath
whereas the non-singlet distribution evolves independently of the singlet
and gluon distributions:
\beq
   {{\rm d} \over {\rm d} t} \Delta q^{\rm NS}(x,t) = 
      \frac{\alpha_s(t)}{2\pi} \int_x^1 \frac{{\rm d}y }{y}\,
      P_{qq}^{\rm NS}({\TS\frac{x}{y}},\alpha_s(t)) \Delta q^{\rm NS}(y,t). 
\label{qcd:apns}
\eeq
Here, $P_{ij}$ are the QCD splitting functions 
for polarized parton distributions.

\par
Expressions~(\ref{qcd:g1}), (\ref{qcd:apsi}), (\ref{qcd:apg}) 
and~(\ref{qcd:apns}) are valid in all orders of perturbative QCD.
The quark and gluon distributions, coefficient functions, and splitting 
functions depend on the mass factorization scale and on the renormalization 
scale; we adopt here the simplest choice, setting both scales equal to~$Q^2$.
At leading order, the coefficient functions are
\bmath
   C_q^{0,\rm S}(\TS\frac{x}{y},\alpha_s) 
 & = & C_q^{0,\rm NS}(\TS\frac{x}{y},\alpha_s) =
      \delta(1-\TS\frac{x}{y}), \nonumber \\ 
   C_g^{0} (\TS\frac{x}{y},\alpha_s) & = & 0~.
\emath
Note that $g_1$ decouples from $\Delta g$ in this scheme.

\par
Beyond leading order, the coefficient functions and the splitting
functions are not uniquely defined; they depend on the renormalization
scheme.  The complete set of coefficient functions has been computed 
in the $\overline{\rm MS}$ renormalization scheme up to order
$\alpha_s^2$~\cite{zijl94}.
The ${\cal{O}}(\alpha_s^2)$ corrections to the polarized splitting
functions $P_{qq}$ and $P_{qg}$ have been computed in Ref.~\cite{zijl94}
and those to $P_{gq}$ and $P_{gg}$ in~\cite{mert95,vogelsang}. 
This formalism allows  a complete Next-to-Leading Order (NLO) QCD analysis
of the scaling violations of spin-dependent structure functions.

\par
In QCD, the ratio $g_1/F_1$ is $Q^2$-dependent because the splitting
functions, with the exception of $P_{qq}$, are different for polarized and
unpolarized parton distributions.
Both $P_{gq}$ and $P_{gg}$ are different in the two cases because of a soft
gluon singularity at $x=0$ which is only present in the unpolarized 
case.  However, in kinematic regions dominated by valence quarks, the $Q^2$
dependence of $g_1/F_1$ is expected to be small~\cite{ano_3}.

\subsection{The small-$x$ behavior of $g_1$}
\label{small_x_sec}
\par
The most important theoretical predictions for polarized deep-inelastic
scattering are the sum rules for the nucleon  structure functions $g_1$.
The evaluation of the first moment of $g_1$, 
\beq
\label{gamma_1}
\Gamma_1(Q^2) = \int_0^1 g_1(x,Q^2)\/{\rm d}x,
\eeq 
requires knowledge of $g_1$ over the entire $x$ region.
Since the experimentally accessible $x$ range is limited,
extrapolations to $x = 0$ and $x = 1$ are unavoidable.
The latter is not critical because it is constrained by the bound 
$|A_1| \le 1$ and gives only  a small contribution to the integral. 
However, the small-$x$ behavior of $g_1(x)$ is 
theoretically not well established and evaluation of $\Gamma_1$
depends critically on the assumption made for this extrapolation.

\par
From the Regge model it is expected that for 
$Q^2\ll 2M\nu$, i.e. $x\rightarrow 0$,
$g_1^{\rm p} + g_1^{\rm n}$ and $g_1^{\rm p} - g_1^{\rm n}$ behave 
like $x^{-\alpha}$~\cite{He73}, where $\alpha$ is the intercept of 
the lowest contributing Regge trajectories. These trajectories are those 
of the pseudovector mesons $f_1$ for the isosinglet combination, 
$g_1^{\rm p} + g_1^{\rm n}$ and of $a_1$ for the isotriplet combination, 
$g_1^{\rm p} - g_1^{\rm n}$, respectively. 
Their intercepts are negative  and assumed to be equal, and  in the range 
$-0.5 < \alpha < 0$. 
Such   behavior has been assumed in most analyses. 

\par
A flavor singlet  contribution to $g_1(x)$ that varies as 
$(2 \ln\frac{1}{x}-1)$~\cite{BaD94} was obtained from a model where 
an exchange of two nonperturbative gluons is assumed.
Even very divergent dependences like $g_1(x) \propto (x \ln^2 x )^{-1}$
were considered ~\cite{ClR94}.
Such dependences are not necessarily consistent  with the QCD evolution
equations.

\par
Expectations based on QCD calculations for the behavior  at small-$x$
of $ g_1(x,Q^2) $ are two-fold:
\begin{itemize}
  \item resummation of standard Altarelli-Parisi corrections 
        gives~\cite{AR75,bfr95b,ellis_smallx}
\beq
 g_1(x,Q^2) \sim \exp \left[ 
            A \sqrt{\ln (\alpha_x(Q^2_0)/\alpha_s(Q^2)) \ln(1/x) }
            \right]~,
\eeq
for the  non-singlet and singlet parts of $g_1$.
  \item resummation of  leading powers of $\ln 1/x$ gives
\begin{eqnarray}
g_1^{\rm NS}(x,Q^2)&\sim& x^{-w_{\rm NS}} 
 \hspace{.5cm} w_{\rm NS} \sim 0.4~,\\
g_1^{\rm S}(x,Q^2)&\sim& x^{-w_{\rm S}}
 \hspace{.7cm}w_{\rm S}\sim 3w_{\rm NS}~,
\end{eqnarray}
for  the non-singlet~\cite{Bartels1} and  singlet~\cite{Bartels2} parts, 
respectively.
\end{itemize}

\subsection{Sum-rule predictions}

\subsubsection{The first moment of $g_1$ and the Operator Product
               Expansion}
\label{qcd_cor}

\par
A powerful tool to study moments of structure functions is provided by the
Operator Product Expansion~(OPE), where the product of the 
leptonic and the
hadronic tensors describing polarized deep-inelastic lepton--nucleon
scattering reduces to the expansion of the product of two electromagnetic
currents. 
At leading twist, the only gauge-invariant contributions are due to the 
non-singlet and singlet axial currents~\cite{koda79,larin94}. 
If only the contributions from the three lightest quark flavors are
considered, the  axial current operator $A_k$ can be expressed in terms
of the $\SU3$ flavor matrices $\lambda_k\,(k=1, \ldots ,8)$ and
$\lambda_0=2I$ as~\cite{larin94}
\beq
   A_{\mu}^k = \overline{\psi} \frac{\lambda_k}{2} 
                   {\gamma_5} {\gamma_\mu} \psi,
\label{axial_ope}
\eeq
and the first moment of $g_1$ is given by
\bmath
\nonumber
        s_\mu\Gamma _1^{\rm p(n)}(Q^2)
        &=& \frac{C^{\rm S}_1(Q^2)}{9}
        \Bigl [ \left\langle ps\right| A_\mu ^0\left| ps\right\rangle 
       \Bigr ]\\
\nonumber
&+&
\frac{C^{\rm NS}_1(Q^2)}{6}
        \Bigl [ +(-) \left\langle ps\right| A_\mu ^3\left| ps\right\rangle\\
&+&
 \frac{1}{\sqrt{3}}
        \left\langle ps\right| A_\mu^8\left| ps\right\rangle \Bigr ],
\label{gamma_exp1}
\emath
where $C^{\rm NS}_1$ and $C^{\rm S}_1$ are the non-singlet and singlet
coefficient functions, respectively. 
The proton matrix elements for momentum $p$ and spin $s$, 
$\left\langle ps\right| A_\mu^i \left| ps\right\rangle$,
can be related to those of the neutron by assuming isospin symmetry. 
In terms of the axial charge matrix element (axial coupling) for flavor
$q_i$ and the covariant spin vector $s_\mu$, 
\bmath
   s_\mu a_i(Q^2)  = \left\langle ps\right| \bar q_i {\gamma_5}
      {\gamma_\mu} q_i \left| ps\right\rangle,
\emath
they can be written as
\bmath
\left\langle ps\right| A_\mu ^3\left| ps\right\rangle 
      & = & \frac{s_\mu}{2} a_3 
        =   \frac{s_\mu}{2} (\au - \ad) 
        =   \frac{s_\mu}{2} \left| \frac{g_A}{g_V} \right|,
\label{matrix_1} \\
\left\langle ps\right| A_\mu ^8\left| ps\right\rangle
      & = & \frac{s_\mu}{2\sqrt{3}} a_8 
        =   \frac{s_\mu}{2\sqrt{3}} 
            (\au + \ad - 2 \as), 
\label{matrix_2} \\
\left\langle ps \right| A_\mu ^0 \left| ps\right\rangle
      & = & s_\mu a_0 
        =   s_\mu (\au + \ad + \as), \\\nonumber
      & = & s_\mu\dsigt(Q^2),
\label{matrix_3}
\emath
where the $Q^2$ dependence of $a_u$, $a_d$ and $a_s$
is implied from now on and is   discussed in Section~\ref{U1}. 
The matrix element $a_3$ in Eq.~(\ref{matrix_1}) under isospin symmetry
is equal to the neutron $\beta$-decay constant $g_A/g_V$. 
If exact $\SU3$ symmetry is assumed for the axial-flavor octet current,
the axial couplings $a_3$ and $a_8$ in Eqs.~(\ref{matrix_1}) 
and~(\ref{matrix_2}) can be
expressed in terms of coupling constants $F$ and $D$, obtained from neutron 
and hyperon $\beta$-decays~\cite{EJ74}, as
\beq
\label{a3=F+D}
a_3 = F+D   \;\;\;\;\;\;\;\; a_8=3F-D .
\eeq  
The effects of a possible $\SU3$ symmetry breaking will be discussed
in Section~\ref{sec:su3}.

\par
The first moment of the polarized quark distribution for flavor $q_i$,
that is  $\Delta q_i =\int \Delta q_i(x){\rm d}x$, is the contribution 
of flavor $q_i$ to the spin of the nucleon.
In the QPM $a_i$ is interpreted as $\Delta q_i$ 
and $a_0$ as $\Delta \Sigma = \Delta u +\Delta d +\Delta s$.
In this framework, the moments $\au$, $\ad$, $\as$ $\ldots$ are bound by a
positivity limit given by the corresponding moments of $u, d, s, \ldots$
obtained from unpolarized structure functions.
In Section~\ref{U1} we will see that the $\U1$ anomaly
modifies this simple interpretation of the axial couplings.

\par
When $Q^2$ is above the charm threshold $(2m_c)^2$,
four flavors must be considered and an additional 
proton matrix element must be defined:
\beq
\left\langle ps \right| A_\mu ^{15} \left| ps\right\rangle
       =  \frac{s_\mu}{2\sqrt{6}}
          (\au + \ad + \as -3\ac)
       =  \frac{s_\mu}{2\sqrt{6}}a_{15},
\label{matrix_4}
\eeq
while   the singlet matrix element becomes 
$s_\mu (\au + \ad + \as + \ac)$.

\begin{table*}[th]
\protect\caption{Higher-order coefficients of the non-singlet
             and singlet coefficient functions $C_1^{\rm NS}$ and
             $C_1^{\rm S}$ in the $\overline{\rm MS}$ scheme.
             The coefficients $c_4^{\rm NS}$ and $c_3^{\rm S}$ are estimates;
             $c_3^{\rm S}$ is unknown for $n_f = 4$ flavors.
             The quantities $\dsigt^{\infty}$ and
             $\dsigt(Q^2)$ are discussed in
             Section~\protect\ref{EJ_sr}.
\label{t_coef}  }
\begin{center}
\begin{tabular}{c  rrrr  rrr rrr}    
\multicolumn{1}{c }{ $n_f$ } & 
\multicolumn{4}{c }{ non-singlet } &
\multicolumn{3}{c }{ singlet ({$\dsigt^{\infty}$})} & 
\multicolumn{3}{c}{ singlet ({$\dsigt(Q^2)$})} \\
& \multicolumn{1}{c }{$c_1^{\rm NS}$} 
& \multicolumn{1}{ c }{$c_2^{\rm NS}$}
& \multicolumn{1}{ c }{$c_3^{\rm NS}$}
& \multicolumn{1}{ c }{$c_4^{\rm NS}$}
& \multicolumn{1}{c }{$c_1^{\rm S} $}
& \multicolumn{1}{ c }{$c_2^{\rm S} $}
& \multicolumn{1}{ c }{$c_3^{\rm S} $}
& \multicolumn{1}{c }{$c_1^{\rm S} $} 
& \multicolumn{1}{ c }{$c_2^{\rm S} $}
& \multicolumn{1}{ c}{$c_3^{\rm S} $} \\\\
\hline
3 & 1.0 & 3.5833 & 20.2153 & 130 & 0.3333 & 0.5496 & 2 & 1 & 1.0959 & 3.7 \\
4 & 1.0 & 3.2500 & 13.8503 &  68 & 0.0400 & $-1.0815$ & --- & 1 & 
$-0.0666$ & --- \\
\end{tabular}
\end{center}
\end{table*}

\subsubsection{The Bjorken sum rule}

\par
The Bjorken sum rule~\cite{Bj66} is an immediate consequence of
Eqs.~(\ref{gamma_exp1}) and (\ref{matrix_1}). 
In the QPM where $C_1^{\rm NS}=1$,
\beq
          \gammap - \gamman = \frac{1}{6} \left| \frac{g_A}{g_V} \right| .
\label{bj_eq}
\eeq
In this form, the sum rule was first derived by Bjorken from current 
algebra and isospin symmetry,  and has since been recognized as a
cornerstone of the QPM.

\par
The Bjorken sum rule
is a rigorous prediction of QCD in the limit of infinite
momentum transfer. It is subject to QCD radiative corrections at
finite values of $Q^2$~\cite{koda79,koda80}.  
These QCD corrections have recently been computed up to
${\cal{O}}(\alpha_s^3)$~\cite{larin91} and the 
${\cal{O}}(\alpha_s^4)$ correction has been estimated~\cite{kata94}.
Since the Bjorken sum rule is a pure flavor non-singlet expression,
these corrections are given by the non-singlet coefficient function
$C_1^{\rm NS}$:
\beq
\gammap - \gamman = \frac{1}{6} \left | \frac{g_A}{g_V} \right | 
                    C_1^{\rm NS}.
\label{bj_sr}
\eeq
Beyond leading order, $C_1^{\rm NS}$ depends on the number of flavors and
on the renormalization scheme. 
Table~\ref{t_coef} shows the coefficients $c_i^{\rm NS}$ of the expansion
\bmath
C_1^{\rm NS} = 1 - c_1^{\rm NS} \left ( \frac{\alpha_s(Q^2)}{\pi} \right )
      -  c_2^{\rm NS} \left ( \frac{\alpha_s(Q^2)}{\pi} \right )^2
      \nonumber \\
      -  c_3^{\rm NS} \left ( \frac{\alpha_s(Q^2)}{\pi} \right )^3 
      -  {\cal O}(c_4^{\rm NS}) \left ( \frac{\alpha_s(Q^2)}{\pi} \right )^4,
\label{CNS1}
\emath
in the $\overline{\rm MS}$ scheme.

\subsubsection{The Ellis--Jaffe sum rules}
\label{EJ_sr}
\par
In the QPM the coefficient functions are equal to unity
and assuming exact $\SU3$ symmetry (Eq.~(\ref{a3=F+D}))
the expression~(\ref{gamma_exp1}) can be written: 
\beq
   \Gamma_1^{\rm p(n)} = +(-)\frac {1}{12}(F+D) + \frac {5}{36}(3F-D)
                       + \frac {1}{3}\as~.
\label{ellijaff}
\eeq
This relation was derived by Ellis and Jaffe~\cite{EJ74}.
With the additional assumption that $\as = 0$,
which in the QPM means $\Delta s=0$,
they obtained numerical predictions for $\gammap$ and $\gamman$.
The EMC measurement~\cite{As88} showed that $\gammap$ is smaller than 
their prediction which in the QPM implied that $\Delta \Sigma $, the 
contribution of quark spins to the proton spin, is small.
This result is at the
origin of the current interest in polarized  deep-inelastic scattering.

\par
The moments of $g_1$ and the Ellis--Jaffe predictions are also subject to
QCD radiative corrections.  
The coefficient function $C^{\rm NS}_1$ (Eq.~(\ref{CNS1})) used for the 
Bjorken sum rule also applies to the non-singlet part.
The additional coefficient function $C_1^{\rm S}$ for the singlet 
contribution in Eq.~(\ref{gamma_exp1}) has been computed up to 
${\cal{O}}(\alpha_s^2)$~\cite{larin94} 
and the ${\cal{O}}(\alpha_s^3)$ term has also been estimated
for $n_f = 3$ flavors~\cite{kata94a}:
\bmath
\label{CS1}
C_1^{\rm S} =  1 - c_1^{\rm S} \left ( \frac{\alpha_s(Q^2)}{\pi} \right )
                 - c_2^{\rm S} \left ( \frac{\alpha_s(Q^2)}{\pi} \right )^2
      \nonumber \\
       - {\cal{O}}(c_3^{\rm S})\left ( \frac{\alpha_s(Q^2)}{\pi} \right )^3 ,
\emath
and   the coefficients $c_i^{\rm S}$ are shown in Table~\ref{t_coef}.    
The QCD-corrected Ellis--Jaffe predictions for $a_s = 0$ become     
\bmath
\label{gamma_3}
\Gamma_1^{\rm p(n)} = C_1^{\rm NS} 
                  \left [ +(-) \frac{1}{12} \left | \frac{g_A}{g_V} \right | 
                  + \frac{1}{36} (3F-D) \right ]
      \nonumber \\
                  + \frac{1}{9} C_1^{\rm S} (3F-D).               
\emath
Since $\dsigt=a_8+3\as$, the assumption $\as=0$ is equivalent to
$\dsigt = a_8 = 3F-D$.  The quantity $3F-D$ is independent of~$Q^2$, so
the assumption $\dsigt = a_8$ should be made for 
$\dsigt^{\infty} = \dsigt(Q^2=\infty)$~\cite{larin94}~\footnote{In 
Ref.~\cite{larin94}, $\dsigt^{\infty}$ and $\dsigt(Q^2)$ 
are referred to as $\Sigma_{inv}$ and $\Sigma(Q^2)$, 
respectively.}.
The coefficients $c_i^{\rm S}$ 
in the third column of Table~\ref{t_coef} should be used to compute
the coefficient function $C_1^{\rm S}$ that appears in Eq.~(\ref{gamma_3}).

\subsubsection{Higher twist effects}
\par
As for unpolarized structure functions, spin-dependent structure 
functions measured
at small $Q^2$ are subject to higher twist~(HT) effects
due to nonperturbative contributions to the lepton--nucleon cross section.
In the analysis of moments and for not too low $Q^2$, such effects 
are expressed as a power series in $1/Q^2$:
\bmath
\nonumber
\Gamma_1 &=& \frac{1}{2} a^{(0)} + \frac{M^2}{9Q^2}(a^{(2)} 
+ 4d^{(2)}+4f^{(2)} )
                + {\cal{O}} \left ({M^4 \over Q^4} \right )\\
         &=& \frac{1}{2} a^{(0)} + {\rm HT}.
\emath
Here $a^{(0,2)}$, $d^{(2)}$ and $f^{(2)}$ are the reduced matrix elements of
the twist--2, twist--3 and twist--4 components, respectively, and~$M$ is the 
nucleon mass.
The values of $a^{(2)}$ and $d^{(2)}$ for proton and deuteron have 
recently been 
measured~\cite{e143_g2} from the second moment of $g_1$ and $g_2$, 
and found to be consistent with zero. 
Several authors have estimated the HT effects for 
$\Gamma_1$~\cite{BBK90,JU93,meyer} and for the Bjorken sum 
rule~\cite{ellis_alpha,mankiewicz}. 
In the literature, there is a consensus that such
effects are probably negligible in the
kinematic range  of the data used to evaluate $\Gamma_1$ in this paper.

\subsection{The physical interpretation of $a_0$ and the $\U1$ anomaly}
\label{U1}
\par
In the simplest approximation, the axial coupling $\dsigt(Q^2)$ is 
expected to be equal to 
$\Delta \Sigma$, the contribution of the quark spin to the nucleon spin.
However, in QCD the $\U1$ anomaly causes a gluon contribution to
$\dsigt(Q^2)$~\cite{ano_1,ano_2,ano_11} as well which makes
$\Delta \Sigma$ dependent on the factorization scheme, while
$\dsigt$ is not.  The total fraction of the 
nucleon spin carried by quarks is the sum of $\Delta \Sigma$ and $L_q$,
where $L_q$ is the contribution of quark orbital
angular momentum to the nucleon spin.
Recently, it was pointed out~\cite{ji_lz} that this sum is scheme-independent
because of an exact compensation between the anomalous 
contribution to $\Delta \Sigma$ and to $L_q$.

\par
The decomposition of $\dsigt$ into $\Delta \Sigma$ and a gluon contribution 
is scheme-dependent~\cite{cheng}.
In the Adler--Bardeen (AB)~\cite{AdlBar} factorization scheme~\cite{BaF95}
\beq
\dsigt(Q^2) 
        = \Delta \Sigma - n_{f} \frac{\alpha_{s} (Q^2)}{2 \pi} 
          \Delta g(Q^2),
\label{U1_A0}
\eeq
where the last term was originally identified as the anomalous
gluon contribution~\cite{ano_1,ano_2,ano_11}.
In this scheme $\Delta \Sigma$ is independent of $Q^2$; 
however it cannot be obtained from the measured 
$\dsigt$ without an input value for $\Delta g$.
In other schemes $\Delta \Sigma$ is equal to $\dsigt(Q^2)$ but then 
it depends on $Q^2$~\cite{cheng}.
The differences between these two schemes do  not vanish
when $Q^2 \rightarrow \infty$ 
because $\alpha_s(Q^2) \Delta g$ remains finite when
$Q^2 \rightarrow \infty$~\cite{ano_1}.

\subsection{The spin-dependent structure function $g_2$}
\label{g2_th}
\par
Phenomenologically, the structure function $g_2$ can be understood from the
spin-flip amplitude that gives rise to the interference asymmetry
$A_2 \propto g_1+g_2$ of Eq.~(\ref{A12}), owing to the absorption
of a longitudinally polarized photon by the nucleon. 
There are two mechanisms by which this can occur~\cite{ji_paris}. 
In the first, allowed in perturbative QCD, the photon is absorbed by a
quark, causing its helicity to flip, but 
since helicity is conserved for massless fermions, this process is
strongly suppressed for small quark masses.   
In the second, which is of a non-perturbative nature, the photon is
absorbed by coherent parton scattering where the final-state quark
conserves helicity by absorption of a helicity~$-1$ gluon.

\par
Wandzura and Wilczek have shown~\cite{gww} that $g_2$ can be decomposed as 
\beq
\label{g2_eq}
        g_2(x,Q^2) = g_{2}^{\rm WW}(x,Q^2) + \bar g_2(x,Q^2).
\eeq
The  term $g_2^{\rm WW}$ is a linear function of $g_1$,
\beq
\label{g2ww_eq}
        g_{2}^{\rm WW}(x,Q^2) 
        = - g_1(x,Q^2) +\int_{x}^{1} g_1(t,Q^2) \frac{{\rm d}t}{t}.
\eeq
The term $\bar g_2$ is due to a twist-3 contribution in the 
OPE~\cite{jaf_g2} and is a measure of quark--gluon correlations in the 
nucleon~\cite{rev_pol}.

\par
In the simplest QPM, $g_2$ vanishes because the masses and transverse 
momenta of quarks are neglected. 
The predictions of improved quark-parton models which take these aspects
into account depend critically on the assumptions made for the quark masses
and the nucleon wave function~\cite{rev_pol}.

\par
The Burkhardt--Cottingham sum rule predicts that the first moment of $g_2$ 
vanishes for both the proton and the  neutron~\cite{BUCO}:
\beq
   \Gamma_2 \equiv \int_{0}^{1} g_2(x) {\rm d} x = 0~.
\eeq
This sum rule is derived in Regge theory and relies on assumptions
that are not well established. 
Its validity has therefore been the subject of much debate
in the recent theoretical literature~\cite{jaf_g2,MvN93,ALNR94}.

\section{EXPERIMENTAL METHOD}
\label{Exp_method}

\subsection{Overview}     
\par
The experiment involves principally the measurement of 
cross section asymmetries for inclusive scattering of  
longitudinally polarized muons from polarized protons
in a solid butanol target (Fig.~\ref{f-spectrom}). 
The energy of the incoming positive muons, $190~\gev$, is measured 
with a magnetic spectrometer in the Beam Momentum Station~(BMS).
The scattered muons are detected in the Forward Spectrometer~(FS).
They are identified by coincident hits in arrays 
of hodoscopes located upstream and downstream of a hadron absorber;
their momenta are measured with a large-acceptance, high-resolution 
magnetic spectrometer.
The beam polarization is measured with a polarimeter located   
downstream of the~FS.
The high energy of the beam  provides a kinematic coverage down to 
$x\sim 0.003$ for $Q^2 > 1~{\rm GeV}^2$, and a high average $Q^2$.
A small data sample was collected with a beam energy of 100\,GeV and
transverse target polarization for the
measurement of the asymmetry $A^{\rm p}_2$.

\par
The counting-rate asymmetries measured in this experiment vary 
from~$0.001$ to~$0.05$ depending on the kinematic region. 
To assure that the asymmetries measured do not depend on 
the incident muon flux, the polarized target is subdivided 
into two cells which are polarized in opposite directions.
Frequent reversals of the target spin directions in both cells strongly
reduce systematic errors arising from time-dependent variations of 
the detector efficiencies.  
Such errors are further reduced by the  high redundancy of
detectors in the forward spectrometer.
The muon  beam polarization is not reversed in this experiment.

\par
The statistical errors of the counting-rate asymmetries are proportional to 
$(P_{\mu} P_{\rm t})^{-1}(N)^{-1/2}$, where $P_{\mu}$ and $P_{\rm t}$ are
the beam and target polarizations, respectively, and $N$ is the number of
events. Hence high values of $P_\mu$ and $P_{\rm t}$ as well as high~$N$ 
are important.

\subsection{The muon beam}
\par
The SMC experiment (NA47) is installed in the upgraded muon beam M2 of the 
CERN SPS~\cite{DGa94}.  A beryllium target is bombarded with 450\,GeV 
protons from the SPS and 
secondary pions and kaons are  momentum-selected and transported through a
600\,m long decay channel where for 200~GeV about 5 percent decay into muons
and neutrinos.
The remaining hadrons are stopped in a 9.9\,m  long beryllium absorber
for the 190~GeV muon beam. Downstream of the absorber, muons are 
momentum selected and transported into the experimental hall.

\par
\begin{figure}[t]
\begin{center}
\psfig{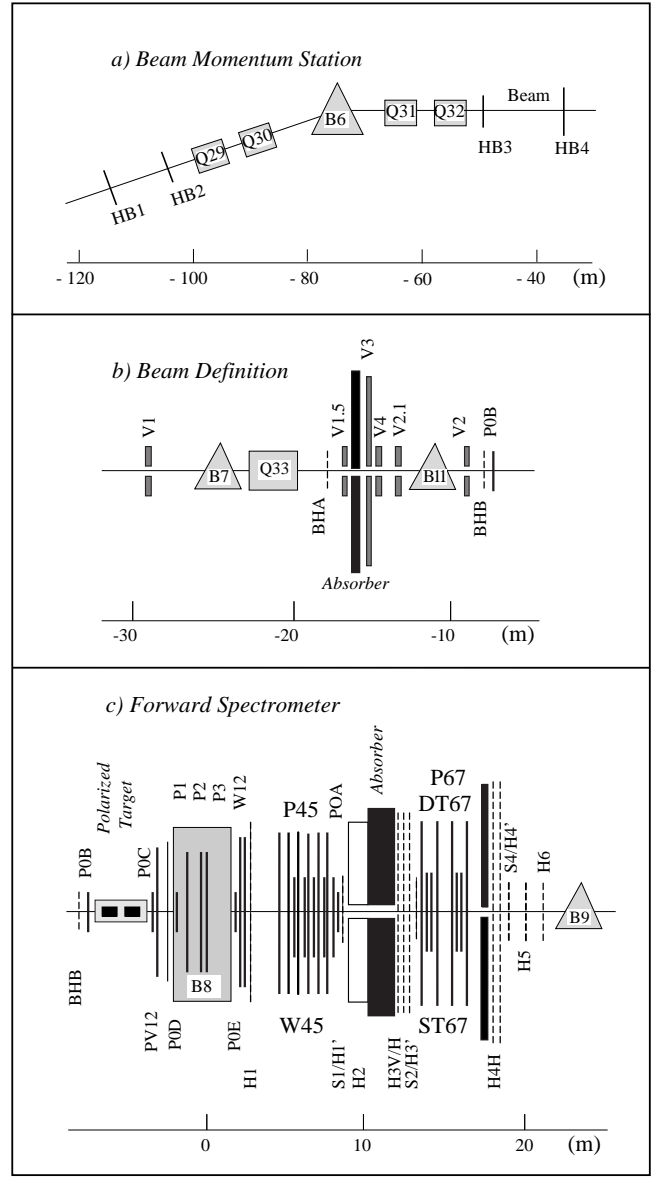}
\vspace{0.25cm}
\caption[]{Schematic layout of the muon beam and forward
 spectrometer. The individual detectors are discussed in the text 
 (see Table~\ref{t-detectors}).  In (b), 
 B11 is a compensating dipole that is used only when taking data 
 with transverse target polarization.  In (c), B8 is the forward 
 spectrometer magnet and referred to as the FSM in the text.
 A right-handed coordinate system is used with its origin at the 
 center of B8. The $x$-axis points along the 
 beam direction, and the $z$-axis points upwards (out of the page in
 (b) and (c)). 
\label{f-spectrom}}
\vspace{-0.25cm}
\end{center}
\end{figure}

The beam intensity was $4 \times 10^7$ muons per SPS pulse;
these pulses are 2.4\,s long with a repetition period of 14.4\,s. 
The beam spot on the target was approximately circular with a
r.m.s. radius of 1.6\,cm and a
r.m.s. momentum width of $\approx 2.5\%$.
The momentum of the incident muons is measured for each 
trigger in the BMS located upstream of the
experimental hall (Fig.~\ref{f-spectrom}).
The BMS employs a set of quadrupoles~(Q) and a 
dipole~(B6) in the beam line, with a nominal vertical
deflection of 33.7\,mrad.
Four planes of fast scintillator arrays~(HB) upstream and downstream of 
this magnet are used to measure the muon tracks.
The resolution of the momentum measurement is better than 0.5\%.

\par
\begin{figure}[t]
\begin{center}
\psfig{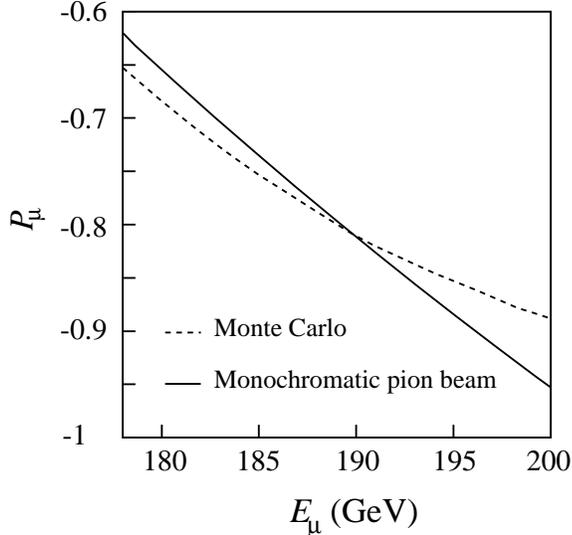}
\caption[]{
Muon polarization $P_{\mu}$ as a function of muon
beam energy $E_\mu$ \protect\cite{DGa94} for a
monochromatic pion beam of 205 GeV 
(solid line)(Eq.\,(\protect\ref{mono_p})),  
and mean $P_\mu$ vs. $E_\mu$  as calculated by beam transport
simulations \protect\cite{DGa94} (dashed line).
\label{mono}}
\end{center}
\end{figure}

The beam is naturally polarized because of  parity violation in the 
weak decays of the parent hadrons. 
For monochromatic muon and hadron beams, the polarization  is a function of
the ratio of muon and hadron energies~\cite{GLW57}:
\beq
\label{mono_p}
P_\mu = \pm \frac{m_{\pi,{\rm K}}^2
 +(1-\frac{2 E_{\pi,{\rm K}}}{E_\mu})m_\mu^2}
               {m_{\pi,{\rm K}}^2-m_\mu^2},
\eeq
where the $-$ and + signs refer  to positive and  negative 
muons, respectively (Fig.~\ref{mono}).
For a given pion energy, the muon intensity depends on the ratio
${E_{\pi,{\rm K}}}/{E_\mu}$;
this ratio was optimized  using Monte Carlo
simulations of the beam transport~\cite{MC1,MC2} to obtain the best 
combination of beam polarization and intensity.

\subsection{Measurement of the beam polarization }
\label{beam_sec}

\par
A polarimeter downstream of the muon spectrometer allows us
to determine the beam polarization by two different methods. 
The first involves measuring the energy spectrum of positrons 
from muon decay in flight, 
$\mu^+ \rightarrow \rme^+ \bar{\nu}_\mu \nu_{\rme}$, 
which depends on the parent-muon polarization~\cite{SMC94a}.
The second method involves measuring the spin-dependent
cross section asymmetry for elastic 
scattering of polarized muons on polarized electrons~\cite{Sch88}.
The two methods require different layouts for the polarimeter, and thus
cannot be run simultaneously.

\subsubsection{Polarized-muon decay}
\par
The energy spectrum of positrons from the decay
$\mu^+\rightarrow {\rme}^+\nu_{{\rme}}\bar{\nu}_{\mu}$~\cite{mic} can
be expressed in terms of the ratio of positron and muon energies 
$y_{{\rme}} = E_{{\rme}}/E_\mu$
and of the muon polarization $P_\mu$~\cite{CP74,CF78}:
\beq
   \frac{{\rm d}N}{{\rm d}y_{\rme}} = 
      N_0 \left[ \frac{5}{3} - 3 y_{\rme}^2 + \frac{4}{3} y_{\rme}^3 
      - P_\mu \left( \frac{1}{3}- 3 y_{\rme}^2 
      + \frac{8}{3} y_{\rme}^3 \right) \right],
\label{ThSpec}
\eeq
where $N_0$ is the number of muon decays.

\par
\begin{figure}[b]
\begin{center}
\psfig{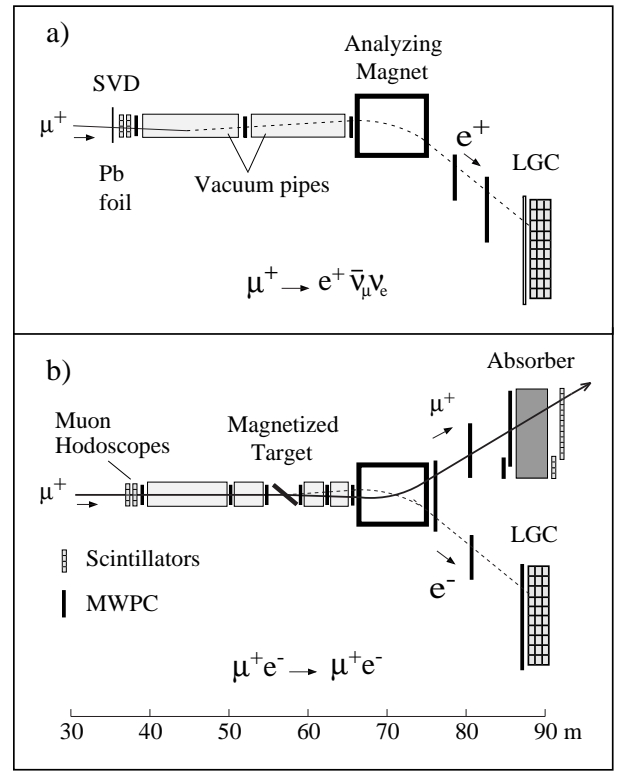}
\caption{
Schematic layout of the beam polarimeter for the 
 muon decay measurement~(a) and for the  muon--electron scattering 
 measurement~(b). The different components of the apparatus are discussed 
 in the text. The lead glass electromagnetic calorimeter and the shower 
 veto detector are labeled as LGC and SVD, respectively.
\label{fig-pol}}
\end{center}
\end{figure}

The polarimeter configuration for this measurement is shown in 
Fig.~\ref{fig-pol}\,(a). 
It consists of a 30\,m long evacuated decay volume, followed by a magnetic
spectrometer and an electromagnetic calorimeter to measure and identify 
the decay positrons.
The beginning of the decay path is defined by the  shower veto detector~(SVD)
which consists of a lead foil followed by two scintillator hodoscopes.
Along the decay path, tracks are measured with multiwire proportional
chambers~(MWPC). 
The decay positrons are momentum analyzed in a magnetic spectrometer 
consisting of a 6 meter long small-aperture dipole magnet followed 
by another set of MWPC. 
This spectrometer and the BMS, which measures the parent muon momentum,  
were intercalibrated in dedicated runs to 0.2\%. 
A lead glass calorimeter~(LGC) is used to identify the decay positrons.

\par
The trigger requires a hit in each SVD plane, in coincidence with a
signal from the LGC above a threshold of about~$15$\,GeV.  
Events with two or more hits in both planes within a 50\,ns time
window are rejected. This  suppresses background from incident positrons
originating upstream of the polarimeter and rejects events with more than 
one muon.

\par
In the off-line analysis, events whose energy $E_\mu$ was measured
in the BMS and experienced  a large energy loss in the SVD
are rejected.
A single track is required, both upstream and downstream of the magnet. 
To reject muon decays inside the magnetic field volume, the
upstream and downstream 
tracks are required to intersect in the center of the magnet. 
Decay positrons are identified by requiring that the momentum
measured by the polarimeter 
spectrometer matches the energy deposition in the LGC.

\par
The measured positron spectrum is corrected for the overall detector 
response.
The response function is obtained from a Monte Carlo simulation that
generates muons according to the measured beam phase space.
The simulation accounts for radiative effects at the vertex and external
bremsstrahlung, the geometry of the set-up, and chamber efficiencies.  
The Monte Carlo events were processed using the same procedure
applied to the real data. 
The response function is obtained by dividing the Monte Carlo
spectrum by the Michel spectrum of Eq.~(\ref{ThSpec}).

\par
\begin{figure}[t]
\begin{center}
\psfig{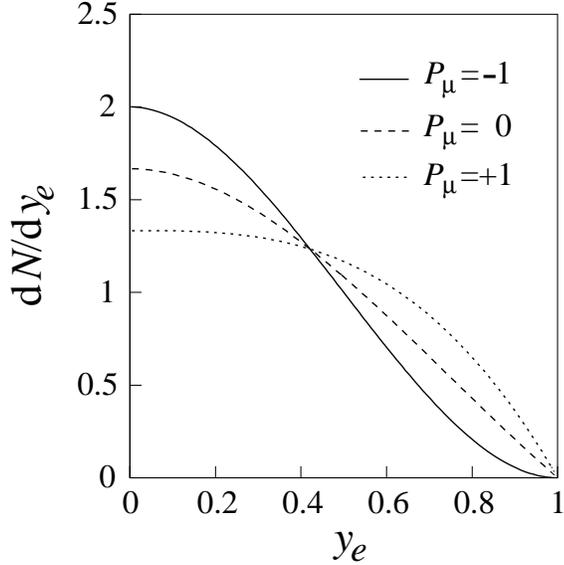}
\caption{The  Michel spectrum  predictions for  
$P_\mu=-1,\,0,$ and $+1$ are shown  by the solid,  
dashed and  dotted lines, respectively. \label{decay}}
\end{center}
\end{figure}

The polarization $P_\mu$ can be determined by fitting Eq.~(\ref{ThSpec}) 
to the measured decay spectrum corrected for the detector response.
Figure~\ref{decay} shows the sensitivity of the Michel spectrum to the
muon polarization. The systematic error in the $P_\mu$ determination  
is mainly due to  uncertainties in the response function,  
the main contributions to which are uncertainties in the MWPC 
efficiencies and in the background rejection.
Background due to external $\gamma$-conversion, 
$\mu^+ \rightarrow \mu^+ \gamma \rightarrow \mu^+ {\rme}^+ {\rme}^-$,
is measured using the charge-conjugate process with a $\mu^-$ beam
and was found to be negligible.
Other contributions to the systematic error arise from uncertainties in
$y_{\rme}$, in radiative effects at the vertex  
and in the alignment of the wire chambers.

\begin{figure*}[]
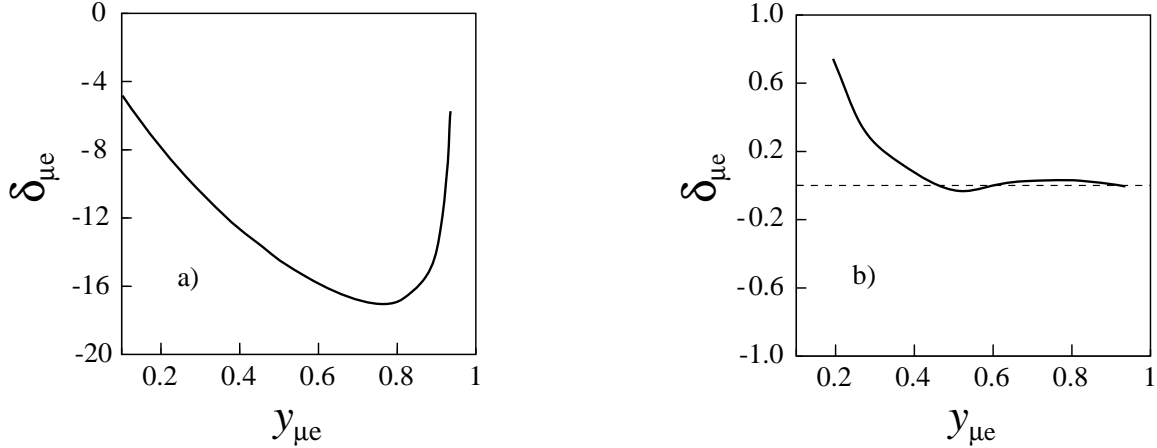

\begin{center}
\mbox{\psfig{file=fig6a.ai_smc,width=7.5cm}
\hspace{1.25cm}
\psfig{file=fig6b.ai_smc,width=7.5cm}}
\end{center}
\caption{The QED radiative corrections to the asymmetry $A_{\mu\mathrm{e}}$ 
(a) without experimental cuts. (b) The asymmetry if the
following experimental cuts are included in the calculation: 
(i) recoil electron energy greater than 35 GeV, 
(ii) energy difference between initial and final states less
than 40 GeV, (iii) angular cuts on both outgoing muon and electron. 
The corrections $\delta_{\rmme}$ are  given in percent.
\label{fig_rc_mue}}
\end{figure*}

\subsubsection{Polarized muon--electron scattering}
\par
In QED at first order, the differential cross section for elastic scattering
of longitudinally polarized muons off longitudinally polarized electrons
is~\cite{Bin57}
\beq
   \frac{{\rm d}\sigma}{{\rm d}y_{\rmme}} 
      = \frac{2{\pi} r^2_{\rme} m_{\rme}}{E_\mu}
        \left(\frac{1}{y_{\rmme}^2}-\frac{1}{y_{\rmme}Y} 
        + \frac{1}{2}\right) (1 + P_{\rme} P_\mu A_{\mu {\rme}}),
\label{eq:pmecs}
\eeq
where $m_{\rme}$ is the electron mass, $r_{\rme}$ the classical electron
radius, $y_{\rmme}= 1- E^\prime_\mu /E_\mu$, and 
$Y = (1 + m_\mu^2/2m_eE_\mu)^{-1}$ is the kinematic upper limit
of $y_{\rmme}$. 
The cross section asymmetry $A_{\mu {\rme}}$ for antiparallel ($\uparrow
\downarrow$) and parallel ($\uparrow \uparrow$) orientations of the
incoming muon and target electron spins is
\beq
   A_{\mu {\rme}} = \frac{{\rm d}\sigma^{\uparrow \downarrow} 
                    - {\rm d}\sigma^{\uparrow \uparrow}}
                    {{\rm d}\sigma^{\uparrow \downarrow} 
                    + {\rm d}\sigma^{\uparrow \uparrow}} 
                  = y_{\rmme}\,\frac{1-y_{\rmme}/Y+y_{\rmme}/2}
                                   {1-y_{\rmme}/Y+y_{\rmme}^2/2}.
\eeq
The measured  asymmetry $A_{\rm exp}$ is related to $ A_{\mu {\rme}}$ by
\beq
\label{Aexp_mue}
   A_{\rm exp}(y_{\rmme}) = P_{\rme} P_\mu A_{\mu {\rme}}(y_{\rmme}),     
\eeq
where $P_{\rme}$ and $P_\mu$ are the electron and muon polarizations,
respectively.
The measured asymmetries range from about $0.01$ at low $y_{\rmme}$
to $0.05$ at high $y_{\rmme}$.

\par
The experimental set-up for the $\mu$--$\rme$ scattering 
measurement is shown schematically in Fig.~\ref{fig-pol}\,(b).  
The lead foil is removed from the SVD and only   the hodoscopes of the 
SVD are used to tag the incident muon which is tracked in three MWPC
installed upstream of the magnetized target. 
Between the target and the spectrometer magnet, three additional chambers
measure the tracks of the scattered muon and of the knock-on electron.  
Downstream of the magnet, the muon and the electron are tracked in two
wire-chamber telescopes sharing a large MWPC. 
The electron is identified in the LGC and the muon is detected in a
scintillation-counter hodoscope located behind a 2\,m thick iron absorber.

\par
The polarized electron target is a 2.7\,mm thick foil made of a
ferromagnetic alloy consisting of 49\%~Fe, 49\%~Co and 2\%~V.  
It is installed in the gap of a soft-iron flat-magnet circuit with two
magnetizing solenoidal coils~\cite{BDM94}.  
The magnet circuit creates a saturated homogeneous field of 2.3\,T along the
plane of the target foil.  
In order to obtain a component of electron polarization parallel to the
beam, the target foil was positioned  at an angle of 25$^\circ$ 
to the beam axis.

\par 
To determine the target polarization, the magnetic flux in the foil under
reversal of the target-field orientation is measured with a pick-up coil
wound around the target. The magnetization of the target was
found to be constant along the foil to within 0.3\%.
The electron polarization is determined from the magneto-mechanical ratio
$g'$ of the foil material. 
A measurement of $g'$ for the  alloy used does not exist; 
a value of $g' = 1.916~\pm~0.002$ has been reported for an alloy of 50\%
Fe and  50\% Co~\cite{ScS69}. 
We assume that the addition of 2\% V does not affect $g'$ but we enlarge the
uncertainty to $\pm 0.02$. 
The resulting polarization along the beam axis is 
$|P_e| = 0.0756 \pm 0.0008$.
The loss of $\mu$--$\rme$  events because of  the internal motion of K-shell
electrons~\cite{Lev94} affects the asymmetry $A_{\rm exp}$ 
by less than --0.001 and was therefore neglected.

\par
To measure the cross section asymmetry, the target-field orientation was
changed between SPS pulses by reversing the current in the coil. 
The vertical component of the magnetizing field provides a bending power of
0.05\,Tm which gives rise to a false asymmetry.
This effect was compensated for by alternating the target angle every 
hour between $25^\circ$ and $-25^\circ$ and averaging the
asymmetries obtained with the two orientations.

\par
The trigger requires a coincidence between the two SVD hodoscope planes,
an energy deposition of 15\,GeV or more in the LGC, and a signal in the
muon hodoscope~(MH).  The scattering vertex is reconstructed from the 
track upstream and the two tracks downstream of the magnetized target. 
The three tracks were required to be in the same plane to within
20$^\circ$ and the reconstructed vertex to be within $\pm 50$\,cm of the
target position.  The two outgoing tracks were required to have an 
opening angle larger than 2\,mrad and to satisfy the two-body kinematics 
of elastic scattering to within 1\,mrad.  Since the electron radiates in 
the target, we use the scattered muon energy to calculate $y_{\rmme}$. 

\par
Background originates from bremsstrahlung ($\mu^+ \rightarrow \mu^+\gamma$) 
followed by conversion,  and pair production 
($\mu^+ \rightarrow \mu^+ {\rme}^+ {\rme}^-$).
It was determined experimentally by using a $\mu^-$ beam
with a similar set-up and triggering on $\mu^-{\rm{e}}^+$ coincidences. 
Most of the background was eliminated by requiring that the energy
conservation between the initial and final states be satisfied within
40\,GeV. This requirement rejects very few good events. 
The background correction to the beam polarization is $-0.012 \pm 0.004$.

\par
The  experimental asymmetry was obtained from data samples taken with
the two different target field orientations.  The data samples were 
normalized to the incident muon fluxes using a random trigger technique. 
A possible false asymmetry due to the target magnetic field was  studied
using both a Monte Carlo simulation of the apparatus and data taken with an
unpolarized polystyrene target under the same experimental conditions. 
In both cases the resulting asymmetry was found to be consistent with zero.
The radiative corrections $\delta_{\rmme}=(A_{\rmme}^{\rm QED}/A_{\rmme}-1)$
to the first order cross section of Eq.~(\ref{eq:pmecs}) are evaluated  using 
the program $\mu\rme la$~\cite{bardin}. The corrections are calculated up to  
${\cal{O}}(\alpha_{\rm QED}^3)$ with finite muon mass and found to be 
negligible once the experimental cuts are applied (Fig.~\ref{fig_rc_mue}). 

\par
\begin{figure}[b]
\begin{center}
\psfig{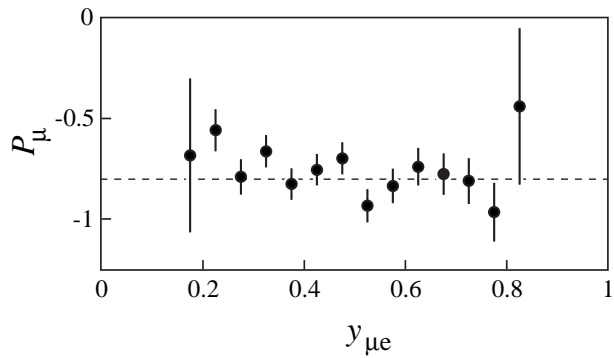}
\end{center}
\caption{Beam polarization vs. the ratio of electron and muon energies 
from polarized $\mu$--e scattering.
The dashed line represents the average value. \label{mue-data}}
\end{figure}

The polarization 
$P_{\mu} = A_{\rm exp}(y_{\rmme})/A_{\mu {\rme}}(y_{\rmme})\,P_{\rme}$ 
in bins of $y_{\rmme}$ is shown in Fig.~\ref{mue-data}. 
The main contributions to the systematic error are the 
uncertainty of the flux normalization, the false asymmetry, the
uncertainty of the target polarization, and the background subtraction.

\begin{figure*}[t]
\begin{center}
\hspace{-.5cm}
\psfig{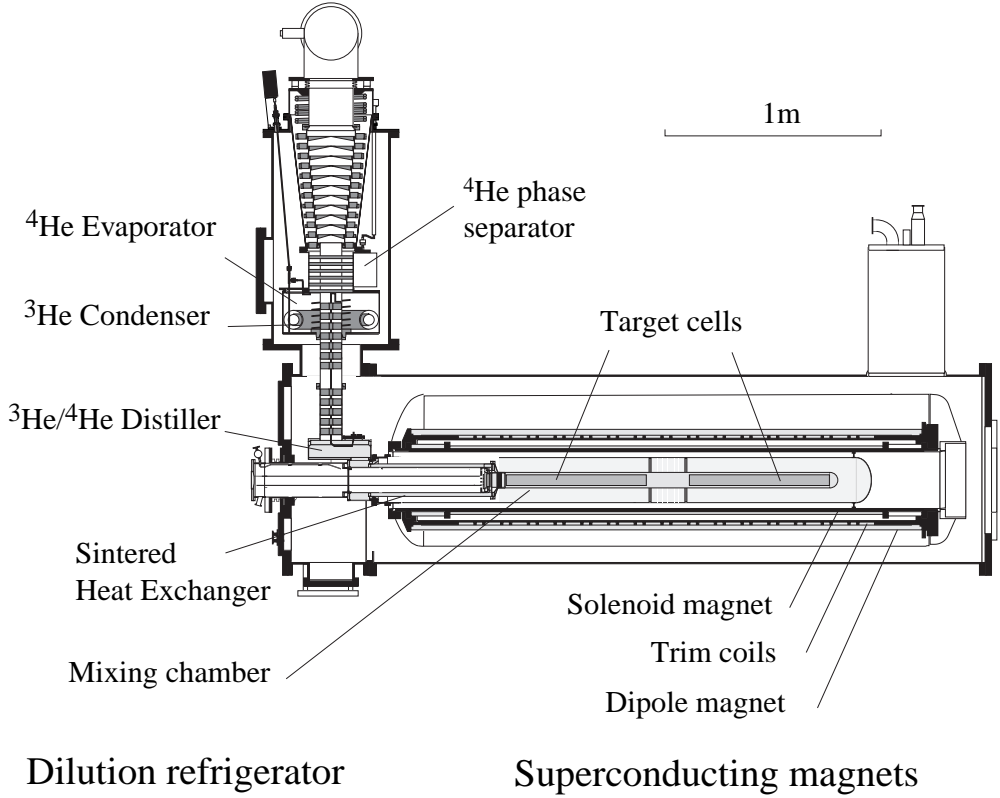}
\vspace{0.5cm}
\caption{Cross section of the  SMC polarized target.\label{cryos}}
\end{center}
\end{figure*}

\subsubsection{The beam polarization}
\par
The beam polarization obtained from the $\mu$--e scattering experiment
in 1993 is~\cite{jack,ethienne}:
\bmath
\label{pol_val}
  P_{\mu}=-0.779\pm 0.026\,(\mbox{stat.})\pm0.017 \,(\mbox {syst.}) 
\emath 
for $E_\mu=187.9~\rm GeV$. 
The polarization measured by the muon decay method in 1993,
$P_{\mu}= -0.803\pm0.029\,(\mbox{stat.})\pm0.020 \,(\mbox{syst.})$,
has been published earlier~\cite{SMC94p}.  
Both results are compatible. An alternative analysis with a larger data 
sample for the muon decay method is in progress and the systematic 
uncertainties of our previous analysis are being re-evaluated.
The result of the $\mu$--e scattering Eq.~(\ref{pol_val}) 
is used in this paper. For $E_\mu=100\,\rm GeV $
a value of $P_{\mu}=-0.82 \pm 0.06$ 
was used for the analysis of the $A_2$ measurement.  
This is  based on  the measurement reported in Ref.~\cite{SMC94a}.
Monte Carlo simulations of the muon beam~\cite{DGa94} are consistent with 
these measurements of $P_\mu$ for both beam energies.
We have evaluated the average polarization of our accepted event
sample taking into account the energy dependence of the muon polarization.
The polarization was calculated on an event-by-event basis using 
Eq.~(\ref{mono_p}) and assuming a monoenergetic pion beam (Fig.~\ref{decay}). 

\subsection{The polarized target}
\label{target_sec}
\par
The polarized proton target uses the 
method of dynamic nuclear polarization~(DNP)~\cite{DNP_MET}
and contains two oppositely polarized target cells exposed 
to the same  muon beam (Fig.~\ref{cryos})~\cite{As88}. 
The solid target material is butanol (CH$_3$(CH$_{2}$)$_3$OH) plus 5\% water
doped with paramagnetic EHBA-Cr(V) molecules.
A superconducting magnet system~\cite{DA91} and a $^3$He\,--$^4$He 
dilution refrigerator~(DR)~\cite{JK95-1} provide the strong magnetic field 
and the low temperature required for high polarization, and allow for  
frequent inversion of the field and thus of the polarization vectors.
Additional subsystems include a double microwave set-up needed for the DNP
and a 10-channel NMR system to measure the spin polarization~\cite{DK95-2}.
During data-taking, the nuclear spin axis is aligned either along or
perpendicular to the beam direction in order to measure $A_{\parallel}$ or
$A_{\perp}$, respectively.

\par
The two target cells were each 60\,cm long, cylindrical, polyester--epoxy mesh
cartridges of 5\,cm diameter, separated by a 30\,cm gap. The target consisted
of 1.8\,mm butanol glass beads.  The total amount of target 
material was 1.42\,kg, with a packing fraction of~0.62
and a density of 0.985\,g/cm$^3$ at 77\,K. 
The concentration of paramagnetic electron spins in the target material was
$6.2 \times 10^{19}$\,spins/ml.
In addition to butanol, the target cells contained other material, mostly
the $^3$He\,--$^4$He cooling liquid and the NMR coils for the polarization
measurement (Table~\ref{tabdilu}).

\par
In the 2.5\,T field and at a temperature below 1\,K, the electron spins
are nearly 100\% polarized.
When  their resonance line is saturated at a frequency just above or
below the absorption spectrum centered around the 
frequency of $\nu_e \approx 69.3$\,GHz at 2.5\,T, 
negative and positive proton  polarizations are obtained.
This technique was applied to polarize the material in 
the two target cells in opposite directions.
Modulation of the microwave frequencies with a 30\,MHz amplitude and a 1\,kHz
rate increased the polarization build-up rate by 20\% and  resulted in a
gain in maximum polarization of 6\%.
This method was originally developed to improve the polarization of a 
deuterated butanol target~\cite{SMC95-1}.

\par
The DR~\cite{JK95-2} cools the target material to a temperature below
0.5\,K while absorbing the microwave power applied for DNP.
Once a high polarization is reached, the microwaves are turned off and the
target material is cooled to 50\,mK.
At this temperature the proton spin-lattice relaxation time exceeds
1000 hours at 0.5\,T.
Under these `frozen spin' conditions, the polarization is preserved during
field rotation and during measurements with transverse spin.
To avoid possible systematic errors, the proton polarizations were reversed by
DNP once a week. 

\par
The superconducting magnet system~\cite{DA91,JK95-1} consists
of a solenoid with a longitudinal field of 2.5\,T aligned 
with the beam axis, and a dipole providing a perpendicular `holding'
field of 0.5\,T.
The solenoid has a bore of 26.5\,cm into which the DR with the target 
cells is inserted;
this diameter corresponds to an opening angle of $\pm 65$\,mrad with
respect to the upstream end of the target. 
Sixteen correction coils allow  the field to be adjusted to a
relative homogeneity of $\pm 3.5 \times 10^{-5}$ over  the target volume.
In addition, the trim coils were used to suppress the
super-radiance effect~\cite{YK88}, which can cause losses of the negative
proton polarization while the field is being changed. 
The spin directions were reversed every five hours with relative
polarization losses of less than $0.2\%$.  
This was accomplished by rotating the magnetic field vector of the
superimposed solenoid and dipole fields, with a loss of data-taking 
time of only 10~minutes per rotation~\cite{JL95}.
The dipole field was also used to hold the spin direction transverse to the
beam for the  measurement of  $A_{\perp}$.

\par
The proton  polarization was measured with ten series-tuned Q-meter circuits
with five NMR coils in each target cell~\cite{SMC94-1,NH95}.
The polarization is proportional to the integrated NMR absorption signal
which was determined from consecutively measured response functions of the
circuit with and without the NMR signal.
The latter was obtained by increasing the magnetic field, and thus 
shifting  the proton NMR spectrum outside the integration window.
The calibration  constant was obtained from
a measurement of the thermal equilibrium~(TE) signals at 1\,K, where the
polarization  is known from the Curie law
$P_{\rm TE}={\rm tanh} \left( {h\,\nu_{\rm p}}/{2\,k\,T} \right) 
\simeq 0.002553$;
$T$ is the lattice temperature, $k$ the Boltzmann constant, and
$\nu_{\rm p}$ is the proton Larmor frequency.
The accuracy of the TE calibration signal contributed to the
polarization error by $\Delta  P/P =  1.1\%$~\cite{DK95-2}.
The NMR signals were  measured every minute during data-taking.
The polarizations measured with the individual coils were averaged
for each target cell and over the duration of one data taking run of
typically 30 minutes. 
All measurements inside the same cell agreed to better than 3\%.
To detect a possible radial inhomogeneity, two of the five coils in the
upstream target cell were at the same longitudinal position, but  one was
in the center and the other   at a radius of 1\,cm.
No significant difference was found between the polarizations measured by 
these two coils.

\begin{table}[]
\caption{Quantities (in moles) of the various chemical elements in 
the target volume.
\label{tabdilu}}
\begin{center} 
\begin{tabular}{cccccc}
Element & Quantity & Element & Quantity & Element & Quantity \\
\hline 
$^1$H   & 185.70   & F       & 0.24     & Cu      & 00.36     \\
$^3$He  & ~~6.00     & Na      & 0.17     & O       & 22.70    \\
$^4$He  & ~~23.00     & Cr      & 0.17     & C       & 71.80    \\
Ni      & ~~0.14   &         &          &           &  \\
\end{tabular}
\end{center}
\end{table}

\par 
The characteristic polarization build-up time was
two to three\,hours. However, the highest polarizations of 
$+0.93$ and $-0.94$ were achieved only after several days of DNP. 
The average polarization during the data-taking was 0.86,
and the relative error in the average polarization of the target was 
estimated to be 3\%.

\begin{table*}[t] 
\caption{Detectors of the muon spectrometer. \label{t-detectors}}
\begin{center}
\small{
\begin{tabular}{lccclcccc}
Hodo- & Modules& Pitch & Size & Wire- & Modules& Pitch & Size & Dead \\
scope & $\times$Planes & (cm) & (cm) & chamber &$\times$Planes& (cm) &
(cm) & zone(cm) \\
\hline 
BHA-B     &  2$\times$8 & ~0.4 & 8$\times$8  & 
P0A-E & 5$\times$8 & ~0.1 & $\oslash$ 14 & --- \\
V123      & 5$\times$1 & --- & various     &
PV1 & 1$\times$4 & 0.2 & 150$\times$94~~ & --- \\
H1   &  2   & ~7.0 & 250$\times$130 &
PV2 & 1$\times$6 & 0.2 & 154$\times$100 & $\oslash$~8 \\
H2 cal &  4   & 28.0 & 560$\times$280 &
P123  & 3$\times$3   & 0.2 & 180$\times$80~~ & $\oslash$~13 \\
H3   &  2   & 15.0 & 750$\times$340 &
W12 & 2$\times$8 & 2.0 & 220$\times$120& $\oslash$~12 \\
H4   &  1   & 15.0 & 996$\times$435 &
W45 & 6$\times$4 & 4.0 & 530$\times$260& $\oslash$~13--25 \\
H1',3',4'  & 1   & ~1.4 & 50$\times$50 &
P45 & 5$\times$2 & 0.2 & $\oslash$ 90 & $\oslash$~12 \\
S1,2,4&      1  &---&various& 
ST67 & 4$\times$8 & 1.0 & 410$\times$410& $\oslash$~16 \\
H5    &1$\times$2  & various & 19$\times$20&
P67 & 4$\times$2 & 0.2 & $\oslash$ 90 & $\oslash$~12 \\
H6    &1$\times$2 & various & $\oslash$~14 &
DT67 & 3$\times$4 & 5.2 & 500$\times$420& 83$\times$83 \\
\end{tabular}\vspace{1.cm}
}
\end{center}
\end{table*}

\subsection{Muon spectrometer and event reconstruction}
\par
The spectrometer is similar to the set-ups used by  the EMC~\cite{EMC81}
and the NMC~(Fig.~\ref{f-spectrom}).  
Aging chambers were replaced and  new ones  added to improve the
redundancy of the muon tracking and to extend the kinematic coverage 
to smaller~$x$.
A major new streamer tube detector ST67 was constructed to identify and
measure scattered muon positions downstream of the absorber.
Triggers were optimized for improved kinematic coverage, in particular in 
the region of small~$x$.

\subsubsection{Spectrometer layout}
\par
Three stages of the spectrometer can be distinguished:
tracking of the incident muon, tracking and momentum measurement of the
scattered muon, and muon identification.
The beam tracking section upstream of the target is composed 
of two scintillator hodoscopes (BHA/BHB) and the P0B  MWPC.
A set of veto counters (V1.5, V3, V2.1 and V2) defines the beam spot size. 
Beam tracks are reconstructed with an angular resolution  of 0.1\,mrad and an
efficiency better than 90\% for intensities up to $5 \times
10^7\,\mu/$spill.

\par
The momentum of the scattered muon is measured with a  conventional
large-aperture dipole magnet~(FSM) and  a system of more than 100 planes 
of MWPC (Table~\ref{t-detectors}).
The FSM is operated with bending powers of 2.3  and 4.4\,Tm at 
100\,GeV and 190\,GeV beam energies, respectively, corresponding to a
horizontal beam  deflection of 7 mrad.
The angular resolution for scattered muons is 0.4\,mrad.
The large MWPC are complemented by smaller MWPC with a smaller
wire pitch, to increase the redundancy and the  resolution of
the spectrometer in the high-rate
environment at small scattering angles. 

\par
Scattered muons are identified by the observation of a track 
behind a 2 m thick iron absorber.  The muon identification system consists 
of streamer tubes, MWPC and drift tubes.
To cope with the high beam intensity, the streamer tubes were operated
with voltages at which their pulse heights were close to  the electronic
threshold.  Their efficiencies were thus very sensitive to the ambient 
pressure and temperature, and   
a high-voltage feedback system was developed to stabilize the average
streamer pulse height within 1\%.

\vspace{0.5cm}
\subsubsection{Triggers}
\label{sec:trig}
\par
\begin{figure}[b]
\begin{center}
\vspace{-0.5cm}
\psfig{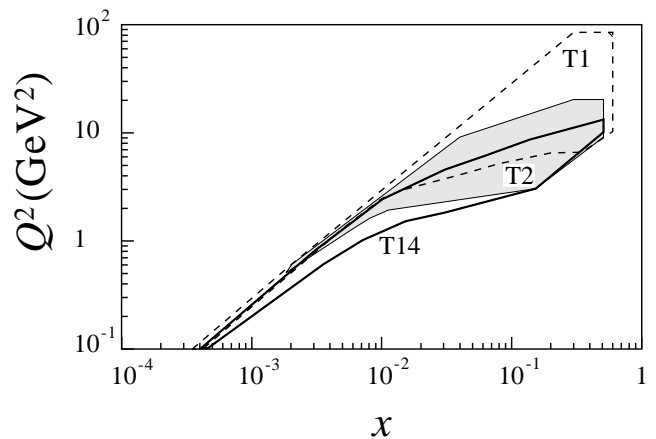}
\caption{Kinematic ranges for triggers T1, T2 and T14 at 190\,GeV.
\label{f-acceptance}}
\end{center}
\end{figure}

The read-out of the detectors was triggered by predefined coincidence
patterns of hits in different planes of scintillation-counter hodoscopes.
Three physics triggers provide a coverage of different $x$ and $Q^2$
ranges (Fig.~\ref{f-acceptance}).
All triggers require that there is no hit in any of the   
beam-defining veto counters.    

\par
The large-angle trigger T1 requires a coincidence pattern of
the hodoscopes H1, H3 and H4.  This trigger has a good acceptance for 
scattering angles $\theta$ larger than 20~mrad.
Target pointing of the scattered muon is also required.
The acceptance decreases for smaller angles, but extends to
$\theta \approx 3$\,mrad.  The small-angle trigger T2 uses the 
smaller hodoscopes H1', H3' and H4'.
This trigger covers the range  $5\,{\rm mrad} \le \theta \le 15$\,mrad. 
It has a more limited $x$-range than T1. However, at a given $x$, T2
selects events with lower $Q^2$ than T1. 
A small-$x$ trigger T14 is provided by the S1, S2 and S4 counters which
are placed close to the beam to cover scattering angles down to 3\,mrad
with good efficiency. The counters for T2 and T14 were located on the 
bending side of the spectrometer magnet. The acceptance of the triggers 
T1 and T14 extends down to $x \simeq 0.5 \times 10^{-3}$ and
thus is sensitive to elastic scattering of muons from atomic electrons,
$x= m_{\rme} / m_{\rm p}$ (Fig.~\ref{f-acceptance}).
The trigger rate per SPS spill was about $200$ for T1, $50$ for
T2 and $100$ for T14.

\par
Other triggers include normalization and  beam-halo triggers which 
were used for calibration, alignment, and efficiency calculations.

\begin{table*}[t]
\caption{Kinematic cuts applied for the $A_\parallel$ and $A_\perp$
analysis. \label{kin_cut}}
\begin{center}

\begin{tabular}{ccccc}    
Kinematic & \multicolumn{2}{c}{$A_\parallel$ analysis} 
          & \multicolumn{2}{c}{$A_\perp    $ analysis    }\\
variable  & \multicolumn{2}{c}{$E_{\mu}$ = 190 {\rm GeV}}
          & \multicolumn{2}{c}{$E_{\mu}$ = 100 {\rm GeV}} \\
\hline
$\nu$     & \multicolumn{2}{c}{$\geq 15~ \rm GeV$} & 
                   \multicolumn{2}{c}{ $\geq 10~ \rm GeV$}\\
$y$       & \multicolumn{2}{c}{$\leq$ 0.9 } & 
                   \multicolumn{2}{c}{ $\leq$ 0.9        }\\
$p^\prime_\mu$   & \multicolumn{2}{c}{$\geq 19~ \rm GeV$}  &
                   \multicolumn{2}{c}{ $\geq 15~ \rm GeV$} \\
$\theta$  & \multicolumn{2}{c}{$\geq$ 9~mrad }  &
                    \multicolumn{2}{c}{ $\geq$ 13~mrad    }  \\
\hline
          & \multicolumn{2}{c}{Final Data Sample for $A_\parallel$ analysis} 
          & \multicolumn{2}{c}{Final Data Sample for $A_\perp    $ analysis}\\
\hline
$x$ range  & $0.003\leq x \leq 0.7$ & $0.0008\leq x \leq 0.7$ 
           & $0.006\leq x \leq 0.6$ & $0.0035\leq x \leq 0.6$ \\
$Q^2$ range& $1 \leq Q^2 \leq 90 $  & $0.2 \leq Q^2 \leq 90 $ 
           & $1 \leq Q^2 \leq 30$   & $0.5 \leq Q^2 \leq 30$  \\
    Events & $4.5 \times 10^6$      & $6.0 \times 10^6$
           & $8.8 \times 10^5$      & $9.6 \times 10^5$    \\
\end{tabular}
\end{center}
\end{table*}

\subsubsection{Event reconstruction}
\par
The track finding starts with the beam-track reconstruction.
The momentum of the incident muons is computed from the hit pattern in
the BMS hodoscopes.
The beam track upstream of the target is found from the hits in 
the BHA and BHB hodoscopes and the P0B wire chamber.
A coincidence is required between the hits in the BMS and those
in the beam hodoscopes.

\par
The reconstruction of the scattered muon tracks starts in
the muon identification system behind the hadron absorber
(ST67, DT67, P67). 
Tracks found in this system are extrapolated upstream and reconstructed in
the MWPC and drift chambers between the absorber and the FSM (W45,
P45, W12, P0E).
The next step in the reconstruction is the track finding in the FSM 
chambers (P123, P0D), starting with the vertical coordinates which are
fitted by  straight lines. 
Horizontal coordinates matching the downstream tracks are searched for
on circular trajectories inside the FSM.
Because of the high track multiplicity in the FSM aperture, each
extrapolation of a downstream track through the magnetic field is tested
with a spline fit and the best track is retained. 
In the vertex chambers (PV12, P0C), hits are selected using the 
extrapolated track  reconstructed in the magnet, and are fitted by a
straight line. 
It is verified that the reconstructed muon track satisfies 
the trigger conditions.

\par
The vertex position in the target is computed as the point of closest
distance of approach between the beam and the scattered-muon tracks. 
Tracks are propagated through the magnetic field in the target 
using a Runge--Kutta method, taking into account energy loss and 
multiple scattering. 
In case of multiple beam tracks, the vertex  with the best space-time
correlation between the beam and the scattered-muon track is chosen.  
The vertex is reconstructed  with resolutions of better than 30\,mm and
0.3\,mm along and perpendicular to the beam direction, respectively.

\vspace{0.5cm}
\subsection{Data-taking}
\vspace{0.25cm}
\par
The data presented in this paper were taken during 134 days of 
the 1993 CERN SPS fixed-target run.
Most  data were taken with longitudinal target polarization, at a beam
energy of 190\,GeV.
For 22 days, data were taken  with the target polarized
transversely to the beam, at a beam energy of 100\,GeV.

\par
A total of 1.6$\times 10^7$ deep-inelastic-scattering events were
reconstructed from the data with a longitudinally polarized target, using
the three physics triggers T1, T2 and T14. 
The integrated muon flux was $ 1.7\times 10^{13}$.

\par
With transverse target polarization, only T1 was used and 
1.6~million events were reconstructed.
The transverse target field was always in the same vertical direction and  
the spin direction was inverted by microwave reversal a total of 10~times.
The integrated muon flux at 100 GeV was $0.2 \times 10^{13}$.

\subsection{Event selection}
\par
Since the $A_{\parallel}$ and $A_{\perp}$ data were recorded at
different  beam energies, they cover different kinematic ranges and are
subject to  different kinematic cuts~(Table~\ref{kin_cut}).
A cut at small $\nu$ rejects events with poor kinematic resolution,
whereas a cut at high $y$ removes events with large radiative corrections.
A cut on the momentum of the outgoing muon reduces the contamination by
muons from $\pi$ and K production in the target and subsequent decay  
to a few~$10^{-3}$. 
The cut on~$\theta$ was only applied for the analysis with
$Q^2 \ge $1~GeV$^2$. It rejects events with poor vertex resolution.  

\par
Cuts were also applied to the beam phase space to ensure
that the beam flux was the same for both target cells.
Fiducial cuts on the target volume reject events from material outside
the target cells (Fig.~\ref{xyz_vertex}).
Less than 10\% of the raw data were discarded because  of instabilities in 
the beam intensity, detector efficiencies, and target polarization.
The size of the final data samples after all cuts
is shown in Table~\ref{kin_cut}.

\section{DATA ANALYSIS}
\label{Ana_asy}

\subsection{Evaluation of cross section asymmetries }
\label{eval_asy}
The two cross section asymmetries $A_{\parallel}$ and 
$A_{\perp}$ (Eq.~(\ref{Asy_cross})) are evaluated from 
counting rate asymmetries.  To determine $A_{\parallel}$ the four measured 
counting rates from the upstream and downstream target cells with the two 
possible antiparallel target spin configurations are used. 
The quantity $A_{\rm T}=A_{\perp}\cos\phi$ is determined separately for the 
upstream and downstream target cells from the four counting rates 
into the upper and lower vertical halves of the spectrometer for the
two transverse  spin directions.

\subsubsection{The $A_{\parallel}$ analysis}
\label{parl_asy}
\par
The number of muons $N_{\rm u}$ and $N_{\rm d}$ scattered in the
upstream and downstream target cells, respectively, is given by
\bmath              
        N_{\rm u}=n_{\rm u}\Phi\, a_{\rm u}\overline{\sigma}(1-f P_{\mu}
 P_{\rm u} A_{\parallel}),\\
        N_{\rm d}=n_{\rm d}\Phi\, a_{\rm d}\overline{\sigma}(1-f P_{\mu}
 P_{\rm d} A_{\parallel}),
\label{yields}
\emath
where $\Phi$ is the integrated beam flux, $P_{\rm u}$ and $P_{\rm d}$ 
are the polarizations in the two target cells, 
$n_{\rm u}$ and $n_{\rm d}$ the area densities of the target nucleons,
and $a_{\rm u}$ and $a_{\rm d}$ are the corresponding spectrometer 
acceptances. 
The dilution factor $f$ accounts for the fact that only a fraction of
the target nucleons is  polarized (Section~\ref{dil_sec}).
The flux $\Phi$ and the spin-independent cross section $\overline{\sigma}$
cancel in the evaluation of the raw counting-rate asymmetries,
$A_{\rm RAW}$ and $A_{\rm RAW}'$,
obtained before and after target polarization reversal:
\begin{eqnarray}
        A_{\rm RAW}=\frac{N_{\rm u}-N_{\rm d}}{N_{\rm u}+N_{\rm d}}, 
        \hspace{1.5cm}
A_{\rm RAW}'=\frac{N_{\rm d}'-N_{\rm u}'}{N_{\rm d}'+N_{\rm u}'}.
\label{A_raw} 
\end{eqnarray}

\begin{figure}[t]
\begin{center}
\psfig{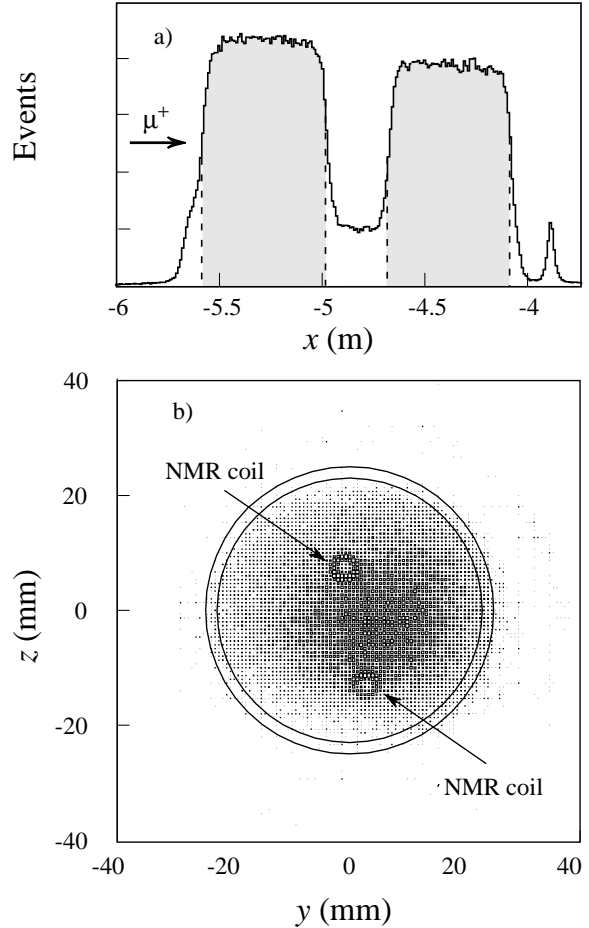}
\caption[]{Vertex distributions of scattered muons after kinematic cuts:
(a) along the beam direction  and (b) in the  plane perpendicular to
the target axis, at the location of one of the NMR coils.
In (a), the dashed lines indicate the fiducial cuts on the target
volume which coincide with the entry and exit windows of
the target cells; most events outside the shaded region 
originate from interactions
with the $^3$He\,--$^4$He cooling liquid.
The small peak at $x\approx -3.9$\,m arises from scattering
in the exit window of the target cryostat.
In (b), the outer circle indicates the wall of the target 
cells, and the inner circle shows the radial cut applied.
Scattering from the tubular NMR coils is clearly visible.
\label{xyz_vertex}}
 \end{center}
\end{figure}

\par
Provided that the ratio
of acceptances is the same before and after polarization reversal,
i.e. $a_{\rm u}/a_{\rm d} = a^{\prime}_{\rm u}/a^{\prime}_{\rm d}$,
and since $n_{\rm u}/n_{\rm d}$ is constant,
the acceptances $a$ and the densities $n$ 
cancel in the average of the raw asymmetries, so that 
\beq
A_{\parallel} =
          - \frac{1}{f P_{\mu}P_{\rm t}}
              \left[ \frac{A_{\rm RAW}+A_{\rm RAW}'}{2} \right].
\label{Araw}
\eeq
If $a_{\rm u}/a_{\rm d} \neq a_{\rm u}'/a_{\rm d}'$
a `false' asymmetry ensues, 
\beq
        A_{\rm false} = -\frac{1}{2f D P_{\mu}P_{\rm t}} 
        \left[\, \frac{r-1}{r+1} - \frac{r'-1}{r'+1} \right].
\label{Af}
\eeq
The virtual photon-proton asymmetry 
$A_1\simeq A_{\parallel}/D$~(Eq.~(\ref{aparall3})) is thus given by:
\beq
A_1 =    - \frac{1}{fD P_{\mu}P_{\rm t}}
    \left[ \frac{A_{\rm RAW}+A_{\rm RAW}'}{2} \right]
   - A_{\rm false}.
\label{A1}
\eeq
In these expressions, $D$ is the depolarization factor~(Eq.~(\ref{D_eq})),
\mbox{$r = n_{\rm u}a_{\rm u}/n_{\rm d}a_{\rm d}$},
{$r' = n_{\rm u}a_{\rm u}'/n_{\rm d}a_{\rm d}'$} and
$P_{\rm t}$  is the weighted average of the target cell polarizations,
\beq
2P_{\rm t}=\frac{\sum |P_{\rm u}|N_{\rm u}+\sum |P_{\rm d}|N_{\rm d}}
           {\sum N_{\rm u}+ \sum N_{\rm d}}
+
           \frac{\sum |P_{\rm u}'|N_{\rm u}'+\sum |P_{\rm d}'|N_{\rm d}'}
               {\sum N_{\rm u}'+ \sum N_{\rm d}'}.
\eeq
Equation~(\ref{A1}) provides an unbiased estimate of the cross section 
asymmetry for large numbers of events. To avoid possible biases for the 
number of events involved,  a maximum likelihood technique     
was developed which allows a common analysis of all events in each 
$x$-bin.
In this method, $A_{\parallel}/D$ is computed from the event
weights $w=fDP_{\mu}$ using the expression
\bmath
\nonumber
A_1 =  &-& \frac{1}{2P_{\rm t}}  
         \left[{
         \left(\frac{\sum w_{\rm u}-\sum w_{\rm d}}
         {\sum w_{\rm u}^2+\sum w_{\rm d}^2}\right)
         +\left(\frac{\sum w_{\rm d}-\sum w_{\rm u}}
         {\sum w_{\rm d}^2+\sum w_{\rm u}^2}\right)^{\prime}}
         ~\right]\\
         &-& A_{\rm false}.
\label{Afd}
\emath
As explained in Section~\ref{dil_sec},
in the actual analysis we use a weight $w=f^\prime D P_\mu$. 
A Monte Carlo simulation confirmed that this method does not introduce
any biases.

\subsubsection{The $A_{\perp}$ analysis}
\label{perp_asy}
\par
A similar formalism applies to the measurement of the transverse 
asymmetry $A_{\perp}$, where the event yields are given by 
$N(\phi) = n \Phi a \overline{\sigma}(1-fP_{\mu}P_{\rm T} 
\cos\phi A_{\perp})$.  
Here, $A_{\perp}$ is obtained for each target cell separately 
from $[N(\phi)-N(\phi-\pi)]/[N(\phi)+N(\phi-\pi)]$ and 
$A_{\perp}/d$ becomes
\bmath
\nonumber
\frac{A_{\perp}}{d} &=&  \frac{-1}{2P_{\mu}\langle P_{\rm t}\rangle}  
   \left[\left(\frac{\sum fd~{\cos\phi}}{\sum (fd~\cos\phi)^2}\right)
 + \left(\frac{\sum fd {\cos\phi}}
  {\sum (fd \cos\phi)^2}\right)^{\prime}\right]\\
 &-& A_{\rm false},
\label{Aperp}
\emath
where $\langle P_{\rm t}\rangle$ is the average target polarization
before and after reversal in absolute value.
To obtain the same statistical accuracy for 
$A_{\perp}/d$ and for $A_{\parallel}/D$ more data are required
for $A_{\perp}/d$ due to its dependence on $\cos\phi$, and also
to a lesser extent to the fact that  $d < D$.

\subsection{Radiative corrections}
\label{rad_corr}
\par
QED radiative corrections are applied  
to convert the measured asymmetries (\ref{Afd}) and (\ref{Aperp})
to one-photon exchange asymmetries. These  corrections 
are calculated using:
\bmath
\label{rad_mul_add}
\overline{\sigma}^{\rm T}&=&v\overline{\sigma}^{1\gamma}
+\overline{\sigma}_{\rm tail},\\ \nonumber
\Delta\sigma^{\rm T}&=&v\Delta\sigma^{1\gamma}
+\Delta\sigma_{\rm tail},
\emath 
where $\overline{\sigma}^{\rm T}$ 
is  the total, i.e.~measured, spin-independent cross-section,
$\overline{\sigma}^{1\gamma}$ is the corresponding one-photon
exchange cross section, and $\overline{\sigma}_{\rm tail}$ is the 
contribution to $\overline{\sigma}^{\rm T}$
from the elastic tail and the inelastic continuum.
The corresponding differences of the cross sections for antiparallel and 
parallel orientations of lepton and target spins are denoted by 
$\Delta\sigma$.
The factor $v$ accounts for vacuum polarization and also includes 
contributions from the inelastic tail close in $x$. 
The decomposition in Eq.~\ref{rad_mul_add} depends on the fraction
of the inelastic tail included in $v$ and is therefore to some 
extend ambiguous. 
Due to a cancelation of the different contributions, $v$ is close to 
unity.
Using the program TERAD~\cite{terad} we find $0.98<v<1.03$ in the
kinematic range of our data. For simplicity we set $v$ to unity in
our analysis and attribute all corrections to 
$\sigma_{\rm tail}$~\cite{smc_dil_note}.

\par
Neglecting $A_2$ and thus implying $A_1=\Delta\sigma/(2D\sigma)$,
the radiative corrections to the one-photon asymmetry,
$A_1^{1\gamma}$, can be written as
\beq
A_1^{\rm T}=\rho ( A_1^{1\gamma}+  A_1^{\rm rc} ),
\label{corr_asy}
\eeq
with $\rho=v\,\overline{\sigma}^{1\gamma}/\overline{\sigma}^{\rm T}$ and
$A_1^{\rm rc}=\Delta{\sigma}_{\rm tail}/2vD\overline{\sigma}^{1\gamma}$.

\par
The ratio $\overline{\sigma}^{1\gamma}/\overline{\sigma}^{\rm T}$ and
the correction $A_1^{\rm rc}$ are evaluated using the program 
POLRAD~\cite{shumeiko,shumeiko2}.  
The asymmetry $A_1^{\rm p}(x)$ required as input is taken from 
Refs.~\cite{As88,SMC94p,E143p} and 
the contribution from $A_2^{\rm p}$ is neglected.
The uncertainty in $A_1^{\rm rc}$ is estimated by varying the
input values of $A_1^{\rm p}$ within the errors.
The factor $\rho$ and the additive  correction $A_1^{\rm rc}$ 
are shown in Table~\ref{a1_t} at the average $Q^2$ of each $x$-bin.

\par
We have incorporated $\rho$ into the evaluation of the
dilution factor, $f'=\rho f$, on an event-by-event basis. Using the
weight $w=f'DP_\mu$ we directly obtain $A_1^{\rm T}/\rho$ on the left-hand 
side of Eq.~\ref{Afd} and thus $A_1^{1\gamma}$ (Eq.~(\ref{corr_asy})).

\par
The radiative corrections to the transverse asymmetry $A_{\perp}^{\rm T}$ are 
evaluated as above, however assuming that $g_2 = g_2^{\rm WW}$~\cite{gww}. 
The additive correction  is much smaller than the statistical error and
has been neglected.

\begin{table*}[t] 
\caption{The virtual photon-proton asymmetry $A_1^{\rm p}$ for 
$Q^2 >1~{\rm GeV}^2$ (top) and $Q^2 > 0.2~{\rm GeV}^2$ (bottom).
In the last column, the first error is statistical and the second
is systematic.
\protect{$\langle A_{\rm RAW} \rangle $} is the straight average of
\protect{$A_{\rm RAW}$} and \protect{$A_{\rm RAW}^\prime$} in 
Eq.~(\ref{Araw}).
The values for $A_1^{\rm p}$ have been corrected for radiative 
effects as described in Section~\ref{rad_corr}.
\label{a1_t}}
\begin{center}
{\footnotesize
\begin{tabular}{ccccccccccc}
$x$ range
    & $\langle x \rangle$ & $\langle Q^2 \rangle$
    & $\langle \protect{P_\mu} \rangle$ 
    & $\langle y \rangle$ & $\langle D \rangle $ 
    & $\langle f \rangle$  
    & $\langle \protect\rho \rangle$  
    &~~ $\langle A_{\rm RAW} \rangle $
    &~~ $A_1^{\rm rc}$
    & $A_1^{\rm  p}$ \\
    & & $({\rm GeV^2})$   & & & & & & & &             \\
\hline
 .003--.006  & .005 &  1.320  &-.79& .791 &  .80 & .070 &1.50 &.004& $.007$ 
 & $  .083$$\pm$$ .041$$\pm$$ .006 $\\
 .006--.010  & .008 &  2.068  &-.78& .748 &  .76 & .081 &1.39 &.003& $.008$ 
 & $  .044$$\pm$$ .037$$\pm$$ .004 $\\
 .010--.020  & .014 &  3.562  &-.78& .704 &  .72 & .090 &1.30 &.003& $.010$ 
 & $  .061$$\pm$$ .032$$\pm$$ .004 $\\
 .020--.030  & .025 &  5.733  &-.78& .660 &  .68 & .096 &1.24 &.003& $.012$ 
 & $  .068$$\pm$$ .044$$\pm$$ .005 $\\
 .030--.040  & .035 &  7.797  &-.78& .634 &  .66 & .099 &1.21 &.002& $.015$ 
 &  $  .041$$\pm$$ .052$$\pm$$ .003 $\\
 .040--.060  & .049 &  10.445 &-.78 & .603 &  .64  & .102 &1.18 &.006& $.017$ 
 & $  .104$$\pm$$ .045$$\pm$$ .007 $\\
 .060--.100  & .077 &  15.011 &-.78 & .551 &  .60  & .106 &1.14 &.009& $.020$ 
 & $  .180$$\pm$$ .045$$\pm$$ .013 $\\
 .100--.150  & .122 &  21.411 &-.78 & .498 &  .55  & .112 &1.10 &.013& $.022$ 
 & $  .289$$\pm$$ .058$$\pm$$ .019 $\\
 .150--.200  & .173 &  27.799 &-.79 & .456 &  .51  & .118 &1.08 &.012& $.022$ 
 & $  .276$$\pm$$ .080$$\pm$$ .019 $\\
 .200--.300  & .242 &  35.542 &-.79 & .417 &  .47  & .127 &1.05 &.010& $.019$ 
 &  $ .246$$\pm$$ .082$$\pm$$ .017 $\\
 .300--.400  & .342 &  45.453 &-.78 & .377 &  .43  & .139 &1.02 &.021& $.010$ 
 & $  .499$$\pm$$ .132$$\pm$$ .036 $\\
 .400--.700  & .482 &  57.089 &-.78 & .337 &  .39  & .156 &0.99 
 &.022&$\!\!\!\!-.006$ & $ .527$$\pm$$ .174$$\pm$$ .041 $\\
\hline
.0008--.0012 & .001 & 0.285 &-.78& .808 & .85 & .044 
 &1.74&$\!\!\!\!-.001$&$.002$&$\!\!\!\!-.032$$\pm$.077$\pm$.004 \\
.0012--.002  & .002 & 0.445 &-.78& .794 & .83 & .054 &1.65& .002 
 &$.003$& .085 $\pm$.055$\pm$.007 \\
 .002--.003  & .003 & 0.686 &-.78& .781 & .80 & .062 &1.56& .001 
 &$.004$& .031 $\pm$.054$\pm$.004 \\
 .003--.006  & .004 & 1.193 &-.78& .763 & .77 & .073 &1.46& .003 
 &$.006$& .059 $\pm$.034$\pm$.005 \\
 .006--.010  & .008 & 2.038 &-.78& .738 & .75 & .082 &1.38& .003 
 &$.008$& .050 $\pm$.036$\pm$.004 \\
\end{tabular}
}
\end{center}
\end{table*}
%
%
\begin{figure}[t]
\begin{center}
\vspace{-0.75cm}
\psfig{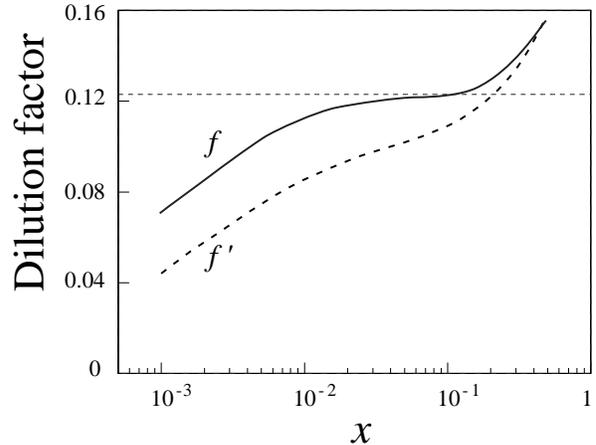}
\caption{The dilution factor  $f$ (solid line) and the
effective dilution factor $f\protect{^\prime}=\rho f$ (dashed line)
as a function of $x$.
\label{dil_f}}
\end{center}
\end{figure}

\subsection{Dilution factor }
\label{dil_sec}
\par
In addition to butanol, the target cells contain  the
NMR coils and the $^3$He\,--$^4$He coolant mixture.
The composition in terms of chemical elements is summarized
in Table~\ref{tabdilu}.
The dilution factor $f$ can be expressed in terms of the number  $n_A$  of
nuclei with mass number $A$ and 
the corresponding total spin-independent cross sections  
$\overline{\sigma}^{\rm T}_A$ per nucleon  for all the elements involved:
\begin{equation}
\label{dil}
f = \frac{n_{\rm H}\cdot\overline{\sigma}^{\rm T}_{\rm H}}
              {\sum_A n_A\cdot\overline{\sigma}^{\rm T}_A}.
\end{equation}
The total cross section ratios 
$\overline{\sigma}_A^{\rm T} / \overline{\sigma}_{\rm H}^{\rm T}$
for D, He, C and Ca are obtained from the  
structure function ratios $F_2^{\rm n}/F_2^{\rm p}$~\cite{nmc_ratio}
and $F_2^A/F_2^{\rm d}$~\cite{nmc_adep}.
The original procedure leading from the measured cross section 
rations $\overline{\sigma}^{\rm T}_A/\overline{\sigma}^{\rm T}_H$
to the published structure function ratios was inverted step by step
involving the isoscalarity corrections and radiative corrections (TERAD).
For unmeasured nuclei the cross section ratios are obtained
in the same way from a
parameterization of $F_2^A(x)/F_2^{\rm d}(x)$ as a function of 
$A$~\cite{emc_adep,arnold,gomez}.

\par
The dilution factor also accounts 
for the contamination from material outside the finite  target cells due to 
vertex resolution.  This  correction is  applied as a function of
the scattering angle, and 
the largest contamination  occurs for the angles between 2 and 9~mrad,
which results in a reduction of the dilution factor by  about 6\%.
The correction needed  because of the NMR  coils (Fig.~\ref{xyz_vertex}) 
is convoluted with the distribution of the  beam intensity profile.
%
%


In the  actual evaluation of Eqs.~(\ref{Afd}) and  (\ref{Aperp}) 
we use an effective dilution factor $f^\prime$~(Fig.~\ref{dil_f}):
\begin{equation}
\label{dil_rad}
f^\prime = \rho f,
\end{equation}
as discussed in Section~\ref{rad_corr}.
The present  procedure guarantees a proper calculation of the
statistical error in the asymmetry, in  contrast to our previous
analysis~\cite{SMC94p,SMC_g2,SMCd93,SMCd95} where all radiative 
effects were included as  an additive radiative correction.
We find an increase in the statistical error
by a factor $1/\rho$ which reaches 5 at small-$x$ (Table~\ref{a1_t}).
However, the central values of the 
asymmetries remain unaffected by the change in the radiative 
correction procedure~\cite{smc_dil_note}.

\par
The  dilution factor is shown in Fig.~\ref{dil_f} 
where it is compared to the `naive' expectation for a mixture of
62\% butanol, ($\rm CH_3(CH_2)_3OH$),
and 38\% helium by volume, $f \simeq 0.123$.
The rise of $f$ at $x>0.3$ is due to the decrease of the ratio
$F_2^{\rm n}/F_2^{\rm p}$, whereas the drop in the low $x$-range 
is due to the  larger contribution of radiative processes from    
elements with mass number much larger
than hydrogen.

\subsection{The longitudinal cross section asymmetry}

\subsubsection{Results for $A_1^{\rm p}$}
 \label{res_A1}
\par
The virtual photon asymmetry $A_1^{\rm p}$ is
calculated  from Eqs.~(\ref{Afd}), (\ref{corr_asy}) and (\ref{dil_rad})
under the assumption that $A_{\rm false}=0$.
The uncertainty introduced by this assumption is estimated using 
Eq.~(\ref{Af}).

\par
\begin{figure}[b]
\begin{center}
\vspace{-0.25cm}
\psfig{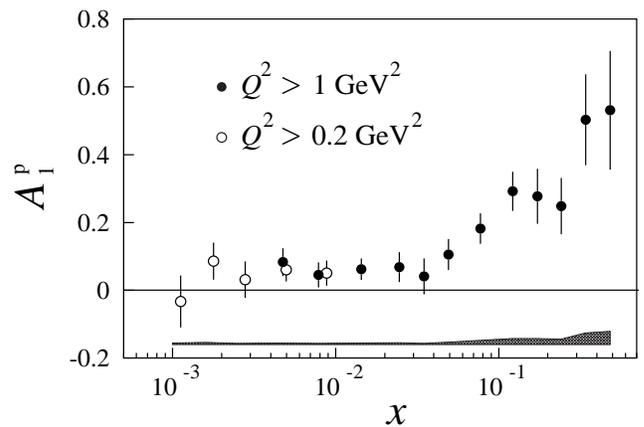}
\caption{The virtual photon asymmetry $A_1^{\rm p}$ as a function 
   of~$x$. The error bars show statistical
   errors only; the systematic errors are indicated by the shaded area.
\label{a1_p}}
\end{center}
\end{figure}

The results for $A_1^{\rm p}$ for $Q^2 \geq 1\,{\rm GeV}^2$
are shown in Table~\ref{a1_t} and in Fig.~\ref{a1_p}.
The kinematic quantities in Table~\ref{a1_t} are mean values within the bins
calculated with the weighting factor $(f^\prime DP_\mu)^2$.
In addition to the results given in Ref.~\cite{SMC94p}, we include
here data obtained with the T14 trigger (Section~\ref{sec:trig}).
In Table~\ref{a1_t} and in Fig.~\ref{a1_p}, we also show data in the
kinematic range  
$0.2~\,\rm GeV^2 \leq Q^2 \leq 1~\,\rm GeV^2$, 
$0.0008 \leq x \leq 0.003$. 
These data are not used to evaluate $g_1^{\rm p}$ 
or its first moment.

\begin{figure}[t]
\begin{center}
\psfig{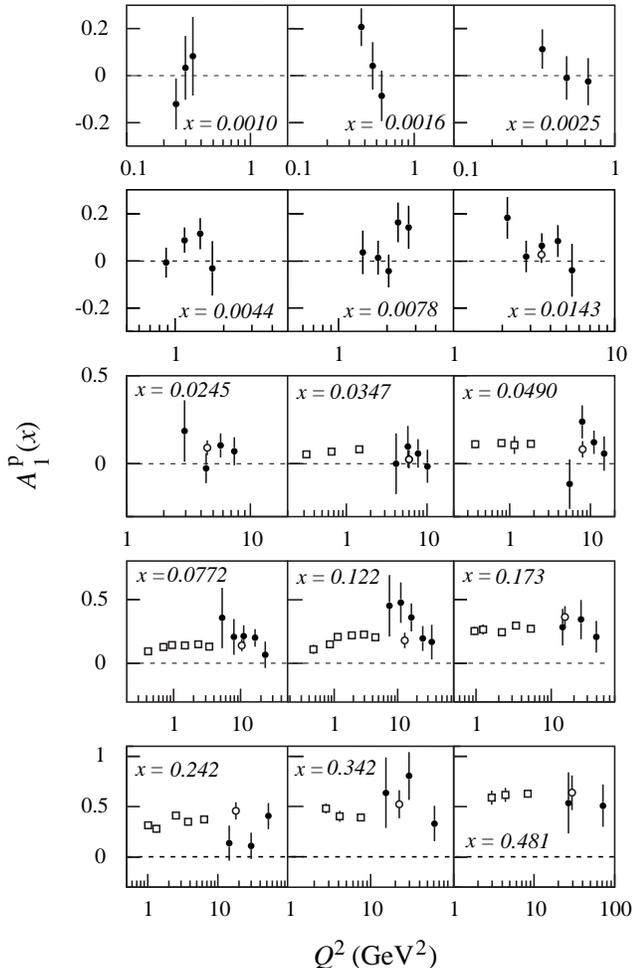}
\caption{The virtual photon-proton  asymmetry $A_1^{\rm p}$ as a function 
of $Q^2$, for constant values of~$x$.
The closed circles are data from this experiment.
The data of the EMC and E143 experiments are also shown as open
circles and squares, respectively.
\label{a1_q2_f}}
\end{center}
\end{figure}

\par
The sources of systematic errors in  $A_1^{\rm p}$  
are time-dependence instabilities of the acceptance ratios $r$ and $r'$, 
uncertainties in the beam and target polarizations, 
in the effective dilution factor $f^\prime$, the radiative corrections,  
and in $R = \sigma_L/\sigma_T$, and the neglect of $A_2$.
The individual errors (Table~\ref{a1_systematics}) 
are combined in quadrature to obtain the total
systematic error (Table~\ref{a1_t}). 

\begin{table}[th]
\caption{Contributions to the systematic errors at the average 
 $Q^2$ of the $x$-bin.  \label{a1_systematics}}
\begin{center}
\begin{tabular}{cccccccc}
$\langle x \rangle$& $\Delta A_{\rm false}$ & $ \Delta P_{\rm t}$ 
& $\Delta P_\mu$ & $\protect{\Delta f^\prime}$ &
$\Delta \rm rc$ & $\Delta A_2$ & $\Delta R$  \\
\hline
0.005&0.0021&0.0025&0.0033&0.0016&0.0012&0.0006&0.0027\\
0.008&0.0019&0.0013&0.0017&0.0008&0.0012&0.0007&0.0012\\
0.014&0.0019&0.0018&0.0024&0.0011&0.0011&0.0009&0.0021\\
0.025&0.0018&0.0020&0.0027&0.0013&0.0010&0.0002&0.0031\\
0.035&0.0018&0.0012&0.0016&0.0008&0.0010&0.0003&0.0016\\
0.049&0.0018&0.0031&0.0041&0.0020&0.0009&0.0003&0.0040\\
0.077&0.0019&0.0054&0.0071&0.0035&0.0009&0.0004&0.0080\\
0.122&0.0019&0.0087&0.0114&0.0058&0.0010&0.0005&0.0112\\
0.173&0.0020&0.0083&0.0109&0.0056&0.0010&0.0005&0.0110\\
0.242&0.0020&0.0074&0.0097&0.0051&0.0009&0.0022&0.0105\\
0.342&0.0020&0.0150&0.0197&0.0107&0.0007&0.0025&0.0236\\
0.482&0.0020&0.0158&0.0208&0.0117&0.0008&0.0030&0.0293\\
\hline
0.0011&0.0032&0.0010&0.0013&0.0011&0.0009&0.0005&0.0017\\
0.0016&0.0027&0.0025&0.0034&0.0026&0.0010&0.0008&0.0035\\
0.0025&0.0024&0.0009&0.0012&0.0009&0.0011&0.0011&0.0012\\
0.0044&0.0021&0.0018&0.0024&0.0014&0.0012&0.0007&0.0023\\
0.0078&0.0020&0.0015&0.0020&0.0010&0.0012&0.0008&0.0015\\
\end{tabular}
\end{center}
\end{table}

\begin{table}[th]
\caption{The virtual photon-proton asymmetry $A_1^{\rm p}$ as a function 
of $x$ and $Q^2$. Only statistical errors are shown. 
\label{a1_q2_t}}
\begin{center}
{\footnotesize
\begin{tabular}{cccccc}
\multicolumn{1}{c}{$\langle x   \rangle$}     &
\multicolumn{1}{c}{$\langle Q^2 \rangle$}     &
\multicolumn{1}{c}{$A_1^{\rm p}$} &
\multicolumn{1}{c}{$\langle x   \rangle$}     &
\multicolumn{1}{c}{$\langle Q^2 \rangle$}     &
\multicolumn{1}{c}{$A_1^{\rm p}$}
\\
\multicolumn{1}{c}{}                &
\multicolumn{1}{c}{$({\rm GeV^2})$} &
\multicolumn{1}{c}{}                &
\multicolumn{1}{c}{}                &
\multicolumn{1}{c}{$({\rm GeV^2})$} &
\multicolumn{1}{c}{}                \\
\hline
0.0009 &  0.25 & $\!\!-0.122 \pm$  0.110  &0.0345 &  7.77 &
   0.058 $\pm$ 0.082  \\
0.0010 &  0.30 &  0.033 $\pm$  0.137  &0.0359 & 10.15 &
 $\!\! -0.012 \pm$  0.095  \\ \cline{4-6}
0.0011 &  0.34 &  0.082 $\pm$  0.169  &0.0474 &  2.94 &
 $\!\! -1.114 \pm$  0.589  \\               \cline{1-3}
0.0014 &  0.38 &  0.209 $\pm$  0.081  &0.0473 &  5.49 &
 $\!\! -0.117 \pm$  0.142  \\
0.0017 &  0.46 &  0.042 $\pm$  0.102  &0.0478 &  7.83 &
   0.241 $\pm$  0.094  \\
0.0018 &  0.55 &$\!\! -0.086 \pm$  0.109  &0.0484 & 10.96 & 
  0.123 $\pm$  0.068  \\               \cline{1-3}
0.0023 &  0.58 &  0.114 $\pm$  0.085  &0.0527 & 14.73 &  
 0.058 $\pm$  0.098  \\ \cline{4-6}
0.0025 &  0.70 &$\!\! -0.009 \pm$  0.094  &0.0738 &  5.33 & 
  0.359 $\pm$  0.239  \\
0.0028 &  0.82 &$\!\! -0.025 \pm$  0.102  &0.0744 &  7.88 &
  0.212 $\pm$  0.142  \\               \cline{1-3}
0.0036 &  0.88 &$\!\! -0.006 \pm$  0.065  &0.0751 & 11.09 &
  0.214 $\pm$  0.088  \\
0.0043 &  1.14 &  0.089 $\pm$  0.054  &0.0762 & 16.32 &  
 0.203 $\pm$  0.068  \\
0.0051 &  1.43 &  0.119 $\pm$  0.067  &0.0855 & 23.04 &  
 0.066 $\pm$  0.105  \\ \cline{4-6}
0.0057 &  1.70 &$\!\! -0.033 \pm$  0.118  &0.1193 &  7.36 & 
 0.456 $\pm$  0.242  \\               \cline{1-3}
0.0070 &  1.42 &  0.037 $\pm$  0.094  &0.1199 & 11.16 & 
 0.480 $\pm$  0.159  \\
0.0072 &  1.76 &  0.014 $\pm$  0.073  &0.1204 & 16.47 &  
 0.364 $\pm$  0.110  \\
0.0077 &  2.04 &$\!\! -0.045 \pm$  0.071  &0.1208 & 24.84 & 
 0.199 $\pm$  0.098  \\
0.0085 &  2.34 &  0.166 $\pm$  0.085  &0.1293 & 34.28 & 
  0.172 $\pm$  0.137  \\ \cline{4-6}
0.0092 &  2.72 &  0.145 $\pm$  0.093  &0.1713 & 14.15 & 
  0.288 $\pm$  0.143  \\               \cline{1-3}
0.0122 &  2.15 &  0.184 $\pm$  0.090  &0.1717 & 24.92 & 
 0.349 $\pm$  0.156  \\
0.0125 &  2.82 &  0.020 $\pm$  0.067  &0.1742 & 39.54 & 
  0.212 $\pm$  0.123  \\ \cline{4-6}
0.0141 &  3.52 &  0.066 $\pm$  0.053  &0.2384 & 14.53 &
   0.139 $\pm$  0.176  \\
0.0165 &  4.43 &  0.085 $\pm$  0.069  &0.2396 & 29.71 &  
 0.110 $\pm$  0.132  \\
0.0184 &  5.43 &$\!\! -0.042 \pm$  0.113  &0.2462 & 52.76 & 
  0.413 $\pm$  0.131  \\ \cline{4-6}   \cline{1-3}
0.0235 &  2.95 &  0.189 $\pm$  0.176  &0.3392 & 15.29 &
  0.644 $\pm$  0.354  \\
0.0236 &  4.38 &$\!\! -0.026 \pm$  0.086  &0.3408 & 29.82 & 
 0.814 $\pm$  0.241  \\
0.0242 &  5.75 &  0.107 $\pm$  0.070  &0.3432 & 61.49 & 
 0.333 $\pm$  0.179  \\ \cline{4-6}
0.0263 &  7.42 &  0.072 $\pm$  0.080  &0.4747 & 26.74 & 
 0.541 $\pm$  0.306  \\               \cline{1-3}
0.0339 &  4.14 &  0.003 $\pm$  0.174  &0.4858 & 71.58 & 
 0.518 $\pm$  0.213  \\
0.0341 &  5.81 &  0.097 $\pm$  0.119  &       &       &   \\ 
\end{tabular}}
\end{center}
\end{table}

\par
Table~\ref{a1_q2_t} and Fig.~\ref{a1_q2_f} show 
$A_1^{\rm p}$ as a function of $Q^2$ and $x$,
including the data with $Q^2 \leq 1~\,\rm GeV^2$.
In Figure~\ref{a1_q2_f}, a small correction is applied to the data
to display them at the same average $x$ in each bin.
A study of the $Q^2$ dependence which includes the SMC
data~\cite{SMC94p,SMCd95} was first made by the E143
collaboration for $0.03 \leq x \leq 0.6$ and 
$Q^2 > 0.3$ $\rm GeV^2$, and showed  no significant $Q^2$ dependence
for $Q^2 > 1~\,\rm GeV^2$~\cite{e143_q2}.
We study here the $Q^2$ dependence for $0.003 \leq x \leq 0.03$.
A parametrization $A_1 = a+b\log Q^2$ is fitted to the data
and $b$ is found to be consistent with zero for all $x$ in this range. 
When fitting a parametrization $a'+c/Q^2$ to account for possible
higher twist effects, we again find no significant $Q^2$ dependence.

\subsubsection{Comparison with earlier  experiments}

\par
\begin{figure}[t]
\begin{center}
\hspace{.75cm}
\vspace{-.5cm}
\psfig{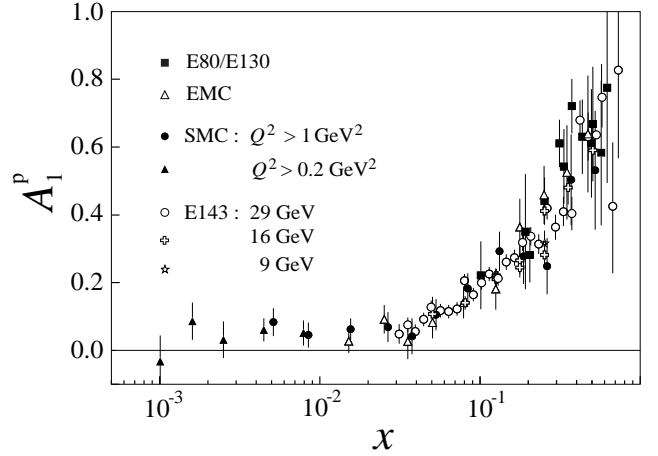}
\vspace{.75cm}
\caption{The virtual photon-proton asymmetry $A_1^{\rm p}$ 
   as a function of~$x$ from this experiment, compared with data 
   from the EMC and the SLAC E80, E130, and E143 experiments.
   For E143, the structure function ratio $g_1^{\rm p}/F_1^{\rm p}$
   is shown instead of $A_1^{\rm p}$.
   The errors are statistical only.
\label{a1_com}}
\end{center}
\end{figure}

In Figure~\ref{a1_com}, we compare our results for
$A_1^{\rm p}$ with data from earlier 
experiments~\cite{e80,As88,E143p,e143_q2}. 
Good agreement is observed in the kinematic region of overlap. 
A consistency test between the SLAC E80/E130,
EMC, SLAC E143 and SMC data yields a $\chi^2 = 11.4$ for 16   
degrees of freedom.
Since the average $Q^2$ of SMC and E143 differ by a factor of 
seven, the good agreement confirms the earlier conclusion that no $Q^2$
dependence is observed within the present accuracy of the data.

\begin{table}[tbh]
\caption{Results on the asymmetry $A_2^{\rm p}$.
Only statistical errors are given. The $A_2^{\rm p}$ values are the
average values from the two target cells.
\label{t_a2}}
\begin{tabular}{cccc}    
$x$ range & $\langle x \rangle$ 
             & $\langle Q^2 \rangle$\,(GeV$^2$) & $A_2^{\rm p}$ \\          
\hline
0.006 -- 0.015 & 0.010 & ~1.4 & $~~0.002\pm 0.109$  \\
0.015 -- 0.050 & 0.026 & ~2.7 & $~~0.041\pm 0.076$  \\
0.050 -- 0.150 & 0.080 & ~5.8 & $~~0.017\pm 0.099$  \\
0.150 -- 0.600 & 0.226 & 11.8 & $~~0.149\pm 0.161$  \\
\hline
0.0035 -- 0.006  & 0.005 & ~0.7 & $-0.066\pm 0.167$ \\
0.006 -- 0.015 & 0.01 & ~1.3 & $~~0.086\pm 0.097$   \\
\end{tabular}
\vspace{.75cm}
\end{table}

\begin{figure}[bth]
\begin{center}
\vspace{-.25cm}
\psfig{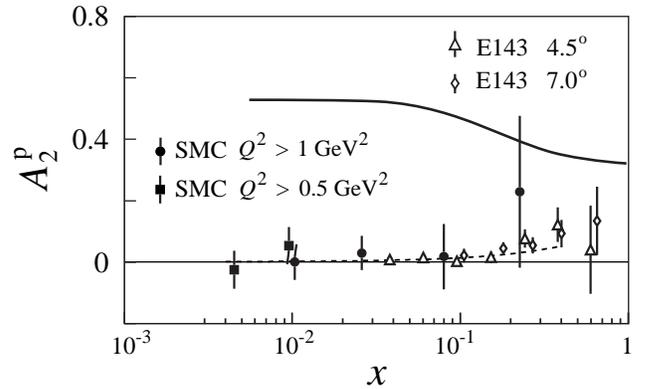}
\caption[]{Results for the asymmetry $A_2^{\rm p}(x)$ extrapolated
to $Q^2_0=5\,\rm GeV^2$ assuming $\sqrt{Q^2} A_2^{\rm p}$ 
scales~\cite{SMC_g2}. 
The solid and dashed curves show the limit $|A_2| < \sqrt{R}$   
and the prediction corresponding to $\bar g_2 = 0$, respectively.
Also shown are data from the E143 experiment~\protect\cite{e143_g2}
extrapolated to the same $Q^2_0$ assuming that $\protect{\sqrt{Q^2} A_2}$ 
scales.  The errors are statistical only.
\label{p_a2}}
\end{center}
\end{figure}

\subsection{The transverse cross section asymmetry}
\label{A2_res}

\subsubsection{Results for $A_2^{\rm p}$}
    
\par
The asymmetry $A_2^{\rm p}$ is obtained from our measurements of 
$A_{\perp}^{\rm p}$~\cite{SMC_g2} and of 
$A_{\parallel}^{\rm p}$~\cite{e80,As88,SMC94p}, using
Eq.~(\ref{q2a2}). It is seen 
from Eq.~(\ref{A12}),  that $A_2$ has an explicit $1/\sqrt{Q^2}$ dependence
and hence it is convenient to evaluate 
$\sqrt{Q^2} A_2^{\rm p}$ assuming that it is independent of $Q^2$
in Eq.~(\ref{Aperp}). 
Our results do not depend on this assumption~\cite{MV_thesis}.

\par
The results for the asymmetry $A_2^{\rm p}$ are shown in
Table~\ref{t_a2} and in Fig.~\ref{p_a2}.
They are significantly smaller than the positivity limit $|A_2| \leq 
\sqrt{R}$ and are consistent with  $A_2^{\rm p}=0$ and with the 
assumption that $g_2 = g_2^{\rm WW}$, i.e. $\bar g_2 = 0$. 
Also shown in Fig.~\ref{p_a2} are the E143 data~\cite{e143_g2}.
They confirm our results, with better statistical accuracy, 
for $x > 0.03$.

\par
The main systematic uncertainties are due to the parametrizations
of $A_{\parallel}^{\rm p}/D$ and $R$. 
The effects due to time variations of the acceptance are negligible  
as  expected,  since the results depend  on the ratio of acceptances
for muons scattered into the top and the bottom halves of the
spectrometer, which should be affected in the same way by typical variations
of chamber efficiencies.
The errors from the dilution factor and the
beam and target polarizations are also very small.
The total systematic error on $A_2^{\rm p}$ is at least  one order of
magnitude smaller than the statistical error at all values of $x$.

\begin{table*}[t]
\caption{Results for the spin-dependent structure 
function \protect$g_1^{\rm p}$. 
The first error is statistical and  the second is systematic.
The third error in the last column is  the uncertainty  associated with the
QCD evolution.
\label{g1_t}}
\begin{center}
\begin{tabular}{ccccc}
 $x$-range & $\langle x \rangle$ & $\langle Q^2({\rm GeV^2}) \rangle$ 
& $g_1^{\rm p}(x,Q^2)$& 
 $g_1^{\rm p}(x,Q^2_0=10\,{\rm GeV^2})$ \\
\hline
 $  0.003$--$ 0.006 $ & $ 0.005$ & $ 1.3 $ & $ 1.97$$\pm$$0.97$$\pm$$0.15$ &
 $ 2.37$$\pm$$0.97$$\pm$$0.15$$\pm$$0.66$\\
 $  0.006$--$ 0.010 $ & $ 0.008$ & $ 2.1 $ & $ 0.73$$\pm$$0.61$$\pm$$0.06 $ 
 & $ 1.03$$\pm$$0.61$$\pm$$0.06$$\pm$$0.17$\\
 $  0.010$--$ 0.020 $ & $ 0.014$ & $ 3.6 $ & $ 0.63$$\pm$$0.33$$\pm$$0.05 $ 
 & $ 0.79$$\pm$$0.33$$\pm$$0.05$$\pm$$0.04$\\
 $  0.020$--$ 0.030 $ & $ 0.025$ & $ 5.7 $ & $ 0.45$$\pm$$0.29$$\pm$$0.03 $ 
 & $ 0.51$$\pm$$0.29$$\pm$$0.03$$\pm$$0.02$\\
 $  0.030$--$ 0.040 $ & $ 0.035$ & $ 7.8 $ & $ 0.20$$\pm$$0.26$$\pm$$0.02 $ 
 & $ 0.22$$\pm$$0.26$$\pm$$0.02$$\pm$$0.01$\\
 $  0.040$--$ 0.060 $ & $ 0.049$ & $10.4 $ & $ 0.38$$\pm$$0.17$$\pm$$0.02 $ 
 & $ 0.37$$\pm$$0.17$$\pm$$0.02$$\pm$$0.00$\\
 $  0.060$--$ 0.100 $ & $ 0.077$ & $15.0 $ & $ 0.42$$\pm$$0.10$$\pm$$0.02 $ 
 & $ 0.40$$\pm$$0.10$$\pm$$0.02$$\pm$$0.01$\\
 $  0.100$--$ 0.150 $ & $ 0.122$ & $21.4 $ & $ 0.41$$\pm$$0.08$$\pm$$0.03 $ 
 & $ 0.39$$\pm$$0.08$$\pm$$0.02$$\pm$$0.01$\\
 $  0.150$--$ 0.200 $ & $ 0.173$ & $27.8 $ & $ 0.26$$\pm$$0.08$$\pm$$0.02 $ 
 & $ 0.25$$\pm$$0.08$$\pm$$0.02$$\pm$$0.01$\\
 $  0.200$--$ 0.300 $ & $ 0.242$ & $35.5 $ & $ 0.15$$\pm$$0.05$$\pm$$0.01 $ 
 & $ 0.15$$\pm$$0.05$$\pm$$0.01$$\pm$$0.01$\\
 $  0.300$--$ 0.400 $ & $ 0.342$ & $45.5 $ & $ 0.15$$\pm$$0.04$$\pm$$0.01 $ 
 & $ 0.17$$\pm$$0.04$$\pm$$0.01$$\pm$$0.00$\\
 $  0.400$--$ 0.700 $ & $ 0.482$ & $57.1 $ & $ 0.06$$\pm$$0.02$$\pm$$0.00 $ 
 & $ 0.08$$\pm$$0.02$$\pm$$0.00$$\pm$$0.00$\\
\end{tabular}
\end{center}
\end{table*}

\section{RESULTS FOR \protect{$\gpone$} AND ITS FIRST MOMENT}
\label{g1_mom}

\subsection{Evaluation of $g_1^{\rm p}(x,Q^2)$}
\par
\begin{figure}[t]
\begin{center}
\psfig{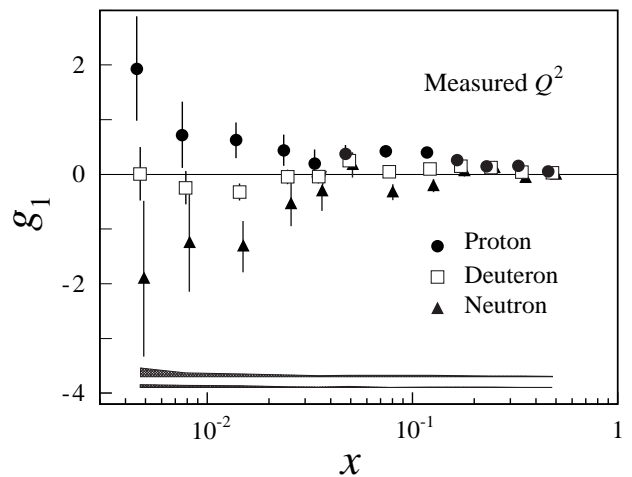}
\caption[]{The structure functions $g_1^{\rm p}$ and $g_1^{\rm d}$ at the 
measured $Q^2$ and the corresponding $g_1^{\rm n}$.
The upper and lower shaded areas represent  the systematic error for
$g_1^{\rm p}$ and $g_1^{\rm d}$, respectively.
\label{g1_f}}
\end{center}
\end{figure}

The spin-dependent structure function $g_1^{\rm p}$ is evaluated from
the virtual photon-proton asymmetry $A_1^{\rm p}$ using Eqs.(\ref{aparall3}) 
and~(\ref{f1f2}).  
This analysis is restricted to $Q^2 > 1\,\rm GeV^2$.
For $F_2$, we use the parametrization of Ref.~\cite{NMC_F295}
and for $R$ the parametrization of Ref.~\cite{r1990}.
The parametrization  of $R$ is based on data for
$x>0.01$  only and therefore must be extrapolated to cover smaller
values of $x$. 
However, the structure function $g_1$ at the average $Q^2$ of the
measurement is nearly independent of $R$ due
to a  partial cancelation between the $R$ dependence of $D$, of $F_2$, and
of the  explicit term $(1+R(x,Q^2))$.
The results for $g_1^{\rm p}$ are shown in Table~\ref{g1_t}
and, together with our deuteron data~\cite{SMCd96},  in Fig.~\ref{g1_f}.

\begin{table*}[t]
\caption[]{Parameters of the polarized parton distributions at 
$Q^2_{\mathrm{ref}} = 1$\,GeV$^2$,  obtained from the QCD fit 
discussed in the text.
\label{evolved}}
\begin{center}
\begin{tabular}{lcccc}
                       &   $a$         &$\alpha$ & $\beta$ & $\eta$ \\
    \hline
$\Delta q^{\rm NS}$ & $~~25.4\pm 39.1$ & $-0.67 \pm 0.25 $ & 
 $2.12\pm 0.28 $ &~~~proton: $1.087 \pm 0.006$ (fixed)\\
 &  &  &  & deuteron: $0.145 \pm 0.002$ (fixed)\\
$\Delta\Sigma$ & $-1.30\pm0.16 $ & ~~$ 0.71\pm0.33$ & $1.56\pm1.00$ 
& $0.40   \pm 0.04   $ \\
$\Delta g$ & $ a_{\Delta\Sigma} $ & $-0.70\pm0.27$ & $4$ (fixed) 
 & $0.98   \pm 0.61   $ \\
\end{tabular}
\end{center}
\end{table*}

\subsection{Evolution of $g_1^{\rm p}$ to a fixed $Q^2_0$}

\label{g1_evo}
\par
To evaluate the first moment $\Gamma_1^{\rm p}=\int_0^1 g_1^{\rm p} {\rm d}x$,
the measured $g_1(x,Q^2)$ must be evolved 
to a common $Q^2_0$ for all $x$.
In previous analyses, $g_1(x,Q_0^2)$ was 
obtained assuming $A_1\simeq g_1/F_1$ to be independent of $Q^2$.
This assumption is consistent with the data. However,  
perturbative QCD predicts the $Q^2$ dependences of $g_1$ and $F_1$ 
to differ by a  considerable  amount at small-$x$. 
The evolution of $g_1/F_1$ is poorly 
constrained by the data in this region,
where the data cover a very narrow  $Q^2$ range. 
Recent experimental and theoretical progress 
allows us  to perform a QCD analysis of 
polarized structure functions in next--to--leading order NLO,  
and therefore a realistic evolution of $g_1$ can be obtained.
Three groups have published 
such analyses~\cite{bfr95b,grv,stirling}.        
They all use the splitting and coefficient functions calculated to NLO in the 
$\overline{\rm MS}$ scheme~\cite{zijl94,mert95,vogelsang},
but the choices made for the reference scales $Q_{\rm ref}^2$ 
at which the polarized parton distributions are parametrized 
and the forms of the parametrization are different. 
Also the selections of data sets used for the fits differ.
In Ref.~\cite{bfr95b} the splitting and coefficient functions are transformed
from the $\overline{\rm MS}$ scheme to different factorization schemes
before the fits are performed. We shall refer to the results obtained in
the Adler--Bardeen scheme.

\par
\begin{figure}[th]
\begin{center}
\psfig{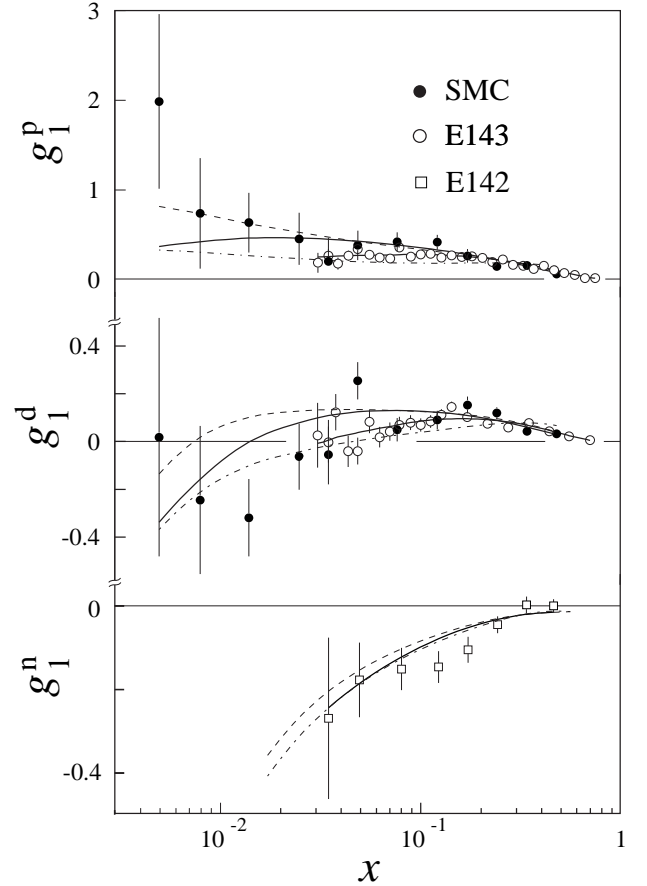}
\caption[]{The  structure functions  $g_1^{\rm p}$,  $g_1^{\rm d}$,
and $g_{1}^{\rm n}$
at the measured $Q^2$ for the SMC~$\cite{SMCd96}$,
E143~$\cite{E143p,E143d}$ and E142~\cite{e142_new} data. 
The solid curves correspond to
our NLO fits at the $Q^2$ of the data points,
the dashed curve  at $Q^2_0$=10~{GeV$^2$}, and 
the dot-dashed  curve at $Q^2_0$=1~{GeV$^2$}.
\label{g1_fit}}
\end{center}
\end{figure}

We used the method \footnote{The code was kindly provided by the authors.}
of Ref.~\cite{bfr95b} to fit the present data and those of 
Refs.~\cite{As88,SMCd93,SMCd95,SMCd96,E143p,e143_q2,E143d}.
The quark singlet, non-singlet and gluon polarized 
distributions are parametrized as
\begin{equation}
\Delta f(x,Q^2_{\rm ref}) = 
 N_f \eta_f x^{\alpha_f}(1 - x)^{\beta_f} (1 + a_f x)~,
\end{equation}
where the normalization factors $N_f$ are chosen such that 
$\int \Delta f {\rm d}x=\eta_f$.
We  have assumed that $a_g=a_{\Delta \Sigma}$.
The normalizations of the non-singlet quark densities are fixed using 
neutron and hyperon $\beta $ decay constants and assuming $\SU3$ flavor 
symmetry. We use $g_A/g_V = F + D = -1.2601 \pm
0.0025$~\cite{par_dat} and $F/D = 0.575 \pm 0.016 $~\cite{cl93}. 
The parameters of the polarized parton distributions obtained from this
fit are given in 
Table~\ref{evolved} and the fit is shown in Fig.~\ref{g1_fit}.
We  have fixed the exponent $\beta$ of the gluon distribution 
to $\beta=4$ as expected from QCD counting rules~\cite{bro95,R90}, 
while the fitted values of $\beta$ for the quark singlet and non-singlet 
components are found to be close to the expectation $\beta=3$.
The $\chi^2$ for the fit is 284 for 295
degrees of freedom.  It is important to note, however, that the fit 
does not converge without  our data points for $x < 0.03$,
where the $Q^2$~range is narrow.
Results of E142 on $g_1^{\rm n}$ were not included in the fit, but used
as a cross check. In Figure~\ref{g1_fit} their data and 
$g_1^{\rm n(\rm fit)}$  calculated from the fit to  $g_1^{\rm p}$
and  $g_1^{\rm d}$ are presented, and found to be in very good agreement.

\begin{figure}[t]
\begin{center}
\psfig{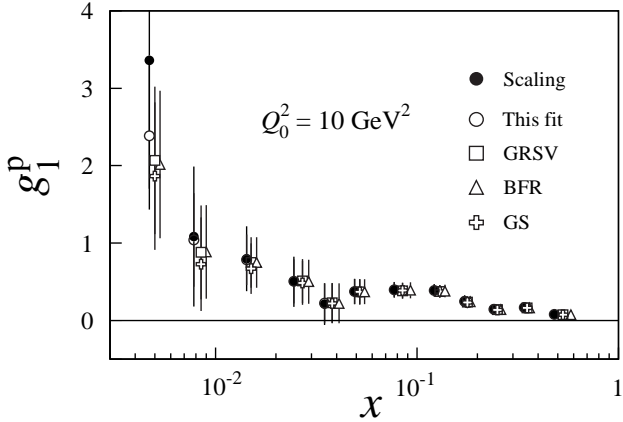}
\vspace{0.5cm}
\caption[]{The  structure function $g_1^{\rm p}$ evolved to $Q^2_0 =  
10~$ GeV$^2$ using the scaling assumption that $g_1/F_1$ is independent 
of $Q^2$, and using   NLO evolution according to our analysis 
and those of  BFR~\cite{bfr95b},
GRSV~\cite{grv} and GS~\cite{stirling}.
\label{g1_f2}}
\end{center}
\end{figure}

\par
The measured $g_1(x,Q^2)$ are then evolved from $Q^2$ to $Q^2_0$ by 
adding the correction
\begin{equation}
  \delta g_1(x,Q^2,Q^2_0) = g_1^{\rm fit}(x,Q^2_0) - g_1^{\rm fit}(x,Q^2)~,
\end{equation}
where $g_1^{\rm fit}$ is calculated by evolving the fitted parton 
distributions. The resulting $g_1^{\rm p}(x,Q^2_0)$ is shown      in 
Table~\ref{g1_t} and Fig.~\ref{g1_f2}.
Also shown is the $g_1^{\rm p}(x,Q^2_0)$ obtained  by using the fits  
of Refs.~\cite{bfr95b,grv,stirling}, and by assuming scaling for $g_1/F_1$.
For the lowest $x$ bin, the latter results in a considerably larger 
value of $g_1$.

\subsection{The first moment of $g_1^{\rm p}$} 
\label{gamma_smc} 

\par
From the evolved structure function $g_1^{\rm p}(x,Q^2_0)$, 
its first moment $\Gamma_1^{\rm p}$
is evaluated at $Q^2_0=10$~GeV$^2$, which is close to the average
$Q^2$ of our data.
The integral over the measured $x$-range is  
\begin{eqnarray}
\label{meas_int}
 \int_{0.003}^{0.7}\!\!\!\!g_1^{\rm p}(x,Q_0^2){\rm d}x 
                     &=& 0.130 \pm 0.013  \pm 0.008 \pm 0.005~,
\end{eqnarray}
where  the first error is statistical, the second systematic 
and the third is the uncertainty due to the $Q^2$ evolution.
The individual contributions to the systematic error are 
summarized in Table~\ref{syserr}.
The error from the evolution is mainly due to the uncertainties in the 
factorization and renormalization scales,
in the parametrizations chosen for the parton distributions, 
the error in $ \alpha_s(M_{\rm Z})$ and mass
threshold effects. In addition we varied the values of $F$ and $D$ 
used as     
inputs to the fit, and of the $A_2$, $A_{\rm false}$, $f$, $P_\mu$, $P_t$,  
$F_2$ and of the radiative corrections used to calculate  $g_1$. 
The uncertainty in the fitted parameters of the parton distributions
is also included, but is found to be relatively  small.
These errors on $\delta g_1(x,Q^2_0)$  are treated as 
correlated from bin to bin, but uncorrelated amongst each other.

\par
The resulting $g_1$ using the  different phenomenological 
analyses of the $Q^2$ evolution~\cite{bfr95b,grv,stirling} are shown in
Fig.~\ref{g1_f2}. Despite their different procedures,
the differences in their   results 
are small and are covered by the error that we 
quote for the evolution uncertainty.

\par
To estimate the integral for $0.7 < x < 1.0$ 
we assume that  $A_1^{\rm p}= 0.7 \pm 0.3$
in this region.
This is consistent with the high-$x$ data and with the
expectation from perturbative 
QCD that
$g_1/F_1 \rightarrow 1$ as $x \rightarrow 1$~\cite{bro95}.
We obtain
\begin{eqnarray}
\label{hi_ext}
 \int_{0.7}^{1.0}g_1^{\rm p}(x,Q_0^2=10\,{\mathrm{GeV}^2}){\rm d}x 
 &=& 0.0015 \pm  0.0007~.
\end{eqnarray}

\par
The results from  our fit shown in   
Fig.~\ref{g1_fit} are used to evaluate  
$\int_{0.003}^{1.0}g_1^{\rm p}(x,Q_0^2){\rm d}x$
and found to be consistent with the sum of Eqs.~(\ref{meas_int})
and (\ref{hi_ext}).

\begin{table}[b] 
\caption {
Contributions to the error of $\Gamma_1^{\rm p}$
\label{syserr}}
\begin{center}
\begin{tabular}{lc}
Source of the  error  & $\Delta \Gamma_1$\\
\hline
Beam polarization  &$0.0048$\\
Extrapolation at low $x$ &$0.0042$\\
Target polarization &$0.0036$\\
Uncertainty on $F_2$ &$0.0030$\\
Dilution factor &$0.0025$\\
Acceptance variation $\Delta r$ &$0.0014$\\
Momentum measurement &$0.0014$\\
Kinematic resolution&$0.0010$\\
Radiative corrections &$0.0008$\\
Extrapolation at high $x$&$0.0007$\\
Neglect of $A_2$ &$0.0004$\\
Uncertainty on $R$ &$0.0000$\\
\hline
Total systematic error  &$0.0087$\\
\hline
Evolution error&$0.0045$\\
\hline
Statistical  error&$0.0125$\\
\end{tabular}
\end{center}
\end{table}

\par
The contribution to the first moment 
from the unmeasured region $0 < x < 0.003$ is evaluated
assuming a  constant $g_1^{\rm p}$ at $Q^2 = 10$~GeV$^2$, 
in agreement with a Regge-type behavior~\cite{He73}.
Using the average of the two lowest $x$ data points in  Table~\ref{g1_t}
we obtain
\begin{eqnarray}
\label{low_ext}
 \int_{0}^{0.003}g_1^{\rm p}(x,Q_0^2=10\,{\mathrm{GeV}^2}){\rm d}x 
 &=& 0.0042 \pm 0.0016~.
\end{eqnarray}
However, to evaluate the systematic error on $\Gamma_1^{\rm p}$ we
have assumed an error of 100\% in this integral (Table~\ref{syserr}).
It should be noted that we have assumed constant Regge-type behavior
at $Q^2=10$\,GeV$^2$. If we apply the same procedure at $Q^2=1$\,GeV$^2$
and then evolve the resulting extrapolation to $Q^2=10$\,GeV$^2$ using the
NLO fits, we obtain a value which is within $1.5\sigma$ of the assumed
error. Other models describing the small-$x$ 
behavior of $g_1$\,(Section~\ref{small_x_sec}) are also considered
to check the sensitivity of our result to the small-$x$ extrapolation. 
A $g_1(x)\approx \ln x$ dependence is compatible with the error
given in Eq.~(\ref{low_ext}), while the $x$ behavior in the diffractive model,
$g_1(x)\approx (x\ln^2x)^{-1}$, gives 
$\int_{0.0}^{0.003}g_1^{\rm p}(x,Q_0^2){\rm d}x = 0.036 \pm  0.016$.
This model results in a larger $\Gamma_1^{\rm p}$, 
but cannot simultaneously accommodate 
the negative values of  $g_1^{\rm n}$ found from our combined
deuteron~\cite{SMCd96} and proton data (Fig.~\ref{g1_f}).
In principle the low-$x$ contribution to the integral can be obtained
from the fit to $g_1$, i.e. $g_1^{\mathrm{fit}}$. However, as known from
unpolarized parton distribution functions, the behavior of the fitted 
distribution below the measured region is unreliable since it 
depends strongly on the choice of the  function, renormalization, 
and factorization scales.

\par
The result for the first moment of $g_1^{\rm p}(x,Q^2_0)$  is
\begin{eqnarray}
\label{g_nlo}
 \Gamma_1^{\rm p}(Q^2_0=10\,{\rm GeV}^2)\!=\!
0.136\!\pm\!0.013\!\pm\!0.009\!\pm\!0.005~.
\end{eqnarray}
Using the results of the NLO evolutions of Refs.\,\cite{bfr95b},
\cite{grv} and \cite{stirling} we find
$\Gamma_1^{\rm p}(Q^2_0)$ between 0.133 to 0.136 (Fig.~\ref{g1_f2}).
If we evaluate $g_1^{\rm p}(x,Q^2_0)$ assuming that 
$g_1/F_1$ is independent of $Q^2$
we obtain $\Gamma_1^{\rm p}(Q^2_0)=0.139 \pm 0.014 \pm 0.010$.
We conclude that within the experimental accuracy of our data 
the different NLO QCD analyses yield consistent results for the
evolution of $g_1$, and that $g_1/F_1$
deviates significantly from scaling at small~$x$.

\begin{table*}[t]
\caption[]{$\Gamma_1^{\rm p}$ and the contributions from different $x$ regions
at $Q_0^2=$5~GeV$^2$. 
The results of our analysis of the SMC and the E143 data, 
as well as the combined analysis of the SLAC-E80/130~\cite{e80}, 
EMC~\cite{As88}, SMC and SLAC-E143~\cite{E143p}  data 
are given with the statistical and systematic errors added in quadrature.
Results of extrapolations are marked with an~($^*$).
\label{ext_com}}
\begin{center}
\begin{tabular}{ccccccc}    
$x$ range&0--0.003&0.003--0.03 & 0.03--0.7 & 0.7--0.8 & 0.8--1 & 0--1 \\
\hline
SMC          & $0.004(2)^*$ & $0.022(7)~$& $0.104(13)$ &
               $0.0018(4)^*$ & 
               $0.0006(2)^*$ & $0.132(17)$ \\
E143         & $0.0012(1)^*$ & $0.010(1)^* $& $0.115(7)$ &
               $0.0020(6)~$ & 
               $0.0006(2)^*$ & $0.129(8)$ \\
ALL          & $0.004(2)^*$ & $0.021(6)~ $& $0.114(6)$ &
               $0.0020(6)~$ & 
               $0.0006(2)^*$ & $0.141(11)$ \\
\end{tabular}
\end{center}
\end{table*}

\begin{table*}[t]
\caption[]{The Ellis--Jaffe sum rule calculated with NLO QCD 
corrections compared to our result for $\Gamma_1^{\rm p}$ 
at $Q^2_0=$10~${\rm GeV^2}$ and 5~${\rm GeV^2}$ and to  
the combined analysis of fit the
E80/E130~\cite{e80}, EMC~\cite{As88}, SMC and 
 E143$~\cite{E143p}$ data
at $Q^2_0=$5~${\rm GeV^2}$.
The Bjorken sum rule calculated with NNLO QCD
corrections and compared to our results on   
$\Gamma_{1}^{\rm p}-\Gamma_{1}^{\rm n}$
from the SMC, the combined analysis of $\Gamma_{1}^{\rm p}$ and 
$\Gamma_{1}^{\rm d}$ (SMC~\cite{SMCd96} and E143~\cite{E143d})
and the combined analysis of $\Gamma_{1}^{\rm p}$,$\Gamma_{1}^{\rm d}$
and $\Gamma_{1}^{\rm n}$ (E142~\cite{e142_new}).
\label{gamma_p}}
\begin{center}
\begin{tabular}{ccccc}    
 Experiment/Theory & $\Gamma_{1}^{\rm p}$ &  $\Gamma_{1}^{\rm n}$ & 
  $\Gamma_{1}^{\rm d}$&
$\Gamma_{1}^{\rm p}-\Gamma_{1}^{\rm n}$ \\
\hline 
\multicolumn{5}{c}{   $Q^2_0=10\,\rm GeV^2$ }\\\hline
 SMC & 
$0.136 \pm 0.016$& 
$-0.046 \pm 0.021$ &
$0.041 \pm 0.007$&  
$0.183 \pm 0.034$ \\
Ellis--Jaffe/Bjorken& 
$0.170 \pm 0.004$& 
$-0.016 \pm 0.004$ &
$0.071 \pm 0.004$&  
$0.187 \pm 0.002$ \\
\hline
\multicolumn{5}{c}{   $Q^2_0=5\,\rm GeV^2$ }\\\hline
 SMC & 
$0.132 \pm 0.017$& 
$-0.048\pm 0.022$& 
$0.039 \pm 0.008$ &
 $0.181 \pm 0.035$ \\
 COMBINED(p,d) & 
$0.141 \pm 0.011$& 
$-0.065\pm 0.017$& 
$0.039 \pm 0.006$& 
$0.199 \pm 0.025$ \\
 COMBINED(p,d,n) & 
$0.142 \pm 0.011$& 
$-0.061\pm 0.016$& 
$0.038 \pm 0.006$& 
$0.202 \pm 0.022$ \\
Ellis--Jaffe/Bjorken &
$ 0.167\pm  0.005 $&
$-0.015\pm  0.004 $&
$ 0.070\pm  0.004 $& 
$ 0.181\pm  0.003 $\\
\end{tabular}
\end{center}
\end{table*}

\subsection{Combined analysis of $\Gamma_1^{\rm p}$}
\label{ej_com}
\par
\begin{figure}[t]
\begin{center}
\psfig{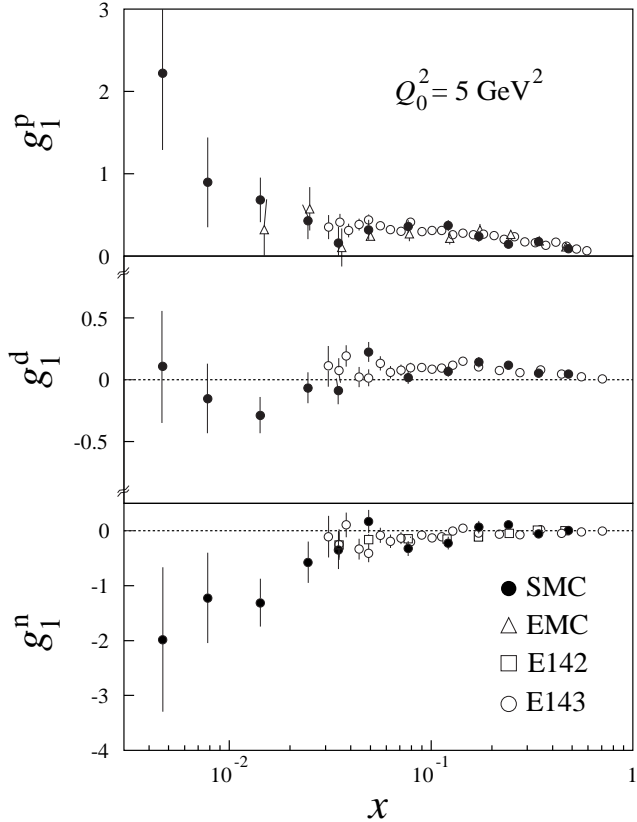}
\caption[]{Measurements of $g_{1}^{\rm p}$,  $g_{1}^{\rm d}$ and  
$g_{1}^{\rm n}$ evolved to $Q^2_0=$5~GeV$^2$. The SMC and E143 $g_{1}^{\rm n}$
data are obtained from  $g_{1}^{\rm p}$ and $g_{1}^{\rm d}$.
\label{g1_all}}
\end{center}
\end{figure}

The combined analysis of $\Gamma_1^{\rm p}$ includes the proton spin 
asymmetries for $Q^2>1~$GeV$^2$ from our data and those of 
Refs.~\cite{e80,As88,E143p}
shown in Fig.~\ref{a1_com}.
The EMC and SMC data were taken at an average $Q^2$ of 10~GeV$^2$, 
while for the SLAC data the average $Q^2$ is 3~GeV$^2$. 
The combined result is evaluated at an intermediate $Q^2$ of 5~GeV$^2$ to 
avoid a large $Q^2$ evolutions. 
Corrections  to $ g_1/F_1$ calculated at NLO are found to be up to 20--25\%.
The evolution of $g_1^{\rm p}$ to $Q^2_0 = 5\,\rm GeV^2$\,(Fig.~\ref{g1_all})
is performed using the procedure of Section~\ref{g1_evo}. 

\par
   The data are combined on a bin-by-bin basis.
The integrals 
$\Delta\Gamma_{1}^i=\int_{\Delta x_i} g_1^{\rm p}(x,Q^2_0)\,{\rm  d}x$
are computed for the $x$-bins of each experiment 
individually, starting from the published asymmetries. 
The $\Delta\Gamma_{1}^i$ which fall into the same SMC $x$-bin are first summed
for each experiment and then the integral for this bin is obtained as 
weighted average of these sums.
The weights are calculated by adding the statistical 
errors and systematic errors uncorrelated between the experiments
in quadrature.
The error and the central value of the integral in the measured region 
is computed using a Monte Carlo method, which  takes into account  
the bin-to-bin correlation of the systematic errors within each 
experiment as well as correlations between the experiments.
These  correlated contributions  are due to
the polarizations of the beam and the target, 
the dilution factor, the neglect of $A_2$, 
the time dependence of the acceptance ratio, the radiative corrections, 
and the parametrizations of $F_2$~\cite{NMC_F295}, of $R$~\cite{r1990}, 
and of the parton distribution functions used to evolve $g_1$. 
Correlations between the experiments arise mainly from the latter three
sources.
The error distributions in the Monte Carlo sampling are assumed to be 
Gaussian.

\par
   The $x$ range of the combined data is $0.003<x<0.8$.
The extrapolations at large and small $x$ are performed using the 
procedures described in Section~\ref{gamma_smc}.
The contributions to the integral from the measured and extrapolated
regions of $x$ are shown in Table~\ref{ext_com}.

\par
The combined result for  the first moment of $g_1^{\rm p}$ is
\beq
\Gamma_1^{\rm p}(Q^2_0 =5\,\rm GeV^2) 
= 0.141 \pm 0.011~~~(\mbox{\it All proton data})~.
\label{G3}
\eeq
If $A_1$ is assumed to be  independent  of $Q^2$, 
we obtain $\Gamma_1^{\rm p}= 0.140 \pm 0.012$. 

\begin{figure}[b]
\begin{center}
\psfig{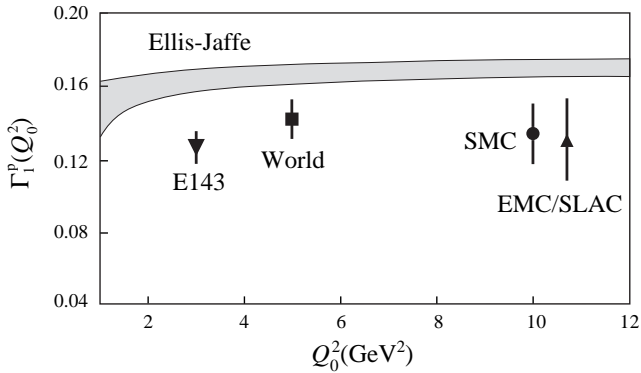}
\vspace{.15cm}
\caption[]{Comparison of the experimental results for $\Gamma_{1}^{\rm p}$
to the prediction of the Ellis--Jaffe sum rule.
\label{th_d_ej} }
\end{center}
\end{figure}

\par
It should be noted that the error quoted by the E143 
collaboration~\cite{E143p} from their data alone and the error 
obtained from our combined analysis are comparable. 
The statistical uncertainties of the SMC data for $0.003<x<0.03$ 
introduce a larger error 
to $\Gamma_1^{\rm p}$ than the uncertainty assumed by the E143 
collaboration for their extrapolation from $x=0.03$ to $x=0$.
We also calculated the extrapolations from the evolved E143 and 
SMC data separately.
The results are compared in Table~\ref{ext_com}.

\par
The results for $\Gamma_{1}^{\rm p}$ from SMC and from  the combined analysis
are compared with the Ellis--Jaffe sum rule in Table~\ref{gamma_p}.
The Ellis--Jaffe prediction is calculated from Eq.~(\ref{gamma_3}).
The higher-order QCD corrections are applied assuming three active quark 
flavors, and using $\alpha_s(5~{\rm GeV}^2)=0.287\pm 0.020$ 
and $\alpha_s(10~{\rm GeV}^2)=0.249\pm 0.015$
corresponding to $\alpha_s(M_{\rm Z}^2)=0.118\pm 0.003$~\cite{par_dat}.
As $Q^2_0=10\,\rm GeV^2$ is close to the charm threshold,
a small uncertainty has been included
to account for the difference between the
perturbative QCD corrections for three and four flavors.
This uncertainty
is  also included   in the error estimate for the Bjorken sum rule prediction
presented in the next section.  

\par
We re-evaluated the first moments for all experiments at their average $Q^2$
using the $g_1$ evolution described in Section~\ref{g1_evo}.
In Figure~\ref{th_d_ej} the results are shown as a function of $Q^2$.
All experimental results are smaller than the Ellis--Jaffe sum 
rule prediction.
From the combined analysis of $\Gamma_1^{\rm p}$ the Ellis--Jaffe sum rule
is violated by more than two standard deviations.  
The implications of this result on 
the spin content of the proton will be discussed in Section~\ref{scon}.

\begin{table*}[t]
\caption[]{Results for the spin-dependent  structure function $g_2^{\rm p}$.
The predicted twist-2 term for  $g_2^{\rm WW}$ [Eq.~(\ref{g2ww_eq})] 
and the upper limit obtained from $|A_2| < \sqrt{R}$ are also given.
Only statistical errors are shown.
\label{g2_ta}}
\small{
\begin{center}
\begin{tabular}{ccccccc} 
$x$ range & $\langle x \rangle$ & $\langle Q^2\rangle$
(GeV$^2$) &$\langle y \rangle$&
$g_2$ &
$g_2^{\rm WW}$& $g_2^{\rm upper}$  \\
\hline
0.006--0.015&  0.010 &  1.36 & 0.72 &
 $~~0.79 \pm 75.84$&$~~0.716 \pm 0.221$ &$429 \pm  61~~$\\
0.015--0.050&  0.026 &  2.66 & 0.57 &
 $~~7.14 \pm 13.92$&$~~0.447 \pm 0.069$ &$101\pm  12~~$\\
0.050--0.100&  0.069 &  5.27 & 0.42 &
 $~~1.06 \pm ~4.77$&$~~0.187 \pm 0.019$ &$17.4 \pm  4.6~~$\\
0.100--0.150&  0.121 &  7.65 & 0.34 &
 $-0.95 \pm ~2.92$&$~~0.037 \pm 0.015$ &$6.1   \pm 2.8 $ \\
0.150--0.300&  0.199 & 10.86 & 0.30 &
 $~~0.20  \pm ~1.66$&$-0.073 \pm 0.007$ &$1.9  \pm  1.2$ \\
0.20 --0.600&  0.378 & 17.07 & 0.25 &
 $~~0.65 \pm ~0.64$&$-0.096 \pm 0.005$ &$0.2  \pm 0.5$  \\
\end{tabular}
\end{center}
}
\end{table*}

\section{RESULTS FOR \protect{$\gptwo$} AND ITS FIRST MOMENT}
\label{g2_mom}

\subsection{Evaluation of $g_2^{\rm p}(x,Q^2)$}
\par
The spin-dependent structure function $g_2^{\rm p}$ is evaluated 
from the $A_2^{\rm p}$ data (Table~\ref{t_a2})
using 
\beq
g_2=\frac{F_1}{2Mx}
\left[  \protect{\sqrt{Q^2}} A_2
   \left(1-\frac{\gamma(\gamma-\eta)}{1+\gamma^2} \right)
 - \frac{A_{\parallel}}{D} \left(\frac{2Mx}{1+\gamma^2} \right)
\right],
\label{find_g2}
\eeq
from Eqs.~(\ref{Asy_cross}) and (\ref{A12}) and a parametrization 
of $A_{\parallel}/D$ from 
Refs.~\cite{As88,SMC94p,E143p}.   
We assume that
$\sqrt{Q^2} A_2^{\rm p}$ and  $A_{\parallel}^{\rm p}/D$ are  
independent of $Q^2$ which is consistent with the data.
The new analyses of $g_1^{\rm p}$
or $F_2$  do not affect the $g_2^{\rm p}$ 
results that we  published in Ref.~\cite{SMC_g2} due to the limited accuracy
of the data.
The  $g_2^{\rm p}$ values are given in 
Table~\ref{g2_ta}. The expected values of ${g_2}^{\rm WW}$ 
and the upper bound of $g_2$,
based on the positivity limit of $A_2$ are also included. 
The statistical accuracy  on $g_2$   is  poor
since the error  is proportional to $1/x^2$ and $\sqrt{Q^2}$, and the data 
are characterized by small-$x$ and large  $Q^2$.
All values are consistent with zero.

\subsection{The first moment of $g_2^{\rm p}$}
\par
The Burkhardt--Cottingham sum rule predicts that the
first moment of $g_2^{\rm p}$ vanishes (Section~\ref{g2_th}).
This integral  is evaluated over the measured $x$ range
at the mean $Q^2$ of the data ($Q^2_0=5$\,GeV$^2)$ 
by assuming a constant
value of $\sqrt{Q^2}A_2(x)$ within each $x$ bin.
We obtain
\beq
-1.0 < \int_{0.006}^{0.6} g_2^{\rm p}(x,Q^2_0)\, {\rm d}x < 2.1~,
\eeq
at 90\% confidence level.
Our measurement     of $g_2$ is  not
accurate enough to perform a meaningful 
extrapolation to $x=0$ using the expected 
$g_2$ Regge behavior,
$g_2(x \rightarrow 0){\sim} x^{-1+\alpha}$~\cite{rev_pol}
and to test the sum rule. 
The first moment  $\Gamma_2(Q^2_0)$ can be  divided into 
$\Gamma_2(Q^2_0)=\Gamma_2(Q^2_0)^{\rm WW}+\overline{\Gamma_2}(Q^2_0)$,
where $\Gamma_2^{\rm WW}$ is obtained 
from  $g_2^{\rm WW}$ (Eq.~(\ref{g2ww_eq})) and
$\overline{\Gamma_2}$ from the $\overline{g_2}$ component.
Using a parametrization of all  $g_1^{\rm p}/F_1^{\rm p}$
data ~\cite{As88,SMC94p,E143p}
we find that the twist-2 part is, as expected, compatible with zero  
($\Gamma_2(Q^2_0)^{\rm WW}\simeq 0.001\pm 0.008.$)
A violation of  the sum rule 
caused by the  $\overline{g_2}$ term cannot be excluded by the present data.

\section{EVALUATION OF \protect{$\gammap-\gamman$}
         AND TEST OF THE BJORKEN SUM RULE}
\label{bj_sec}

\par
We first test the Bjorken sum rule at 
\mbox{$Q^2_0=$ 10~$\rm GeV^2$}  assuming 
\begin{eqnarray}
\label{omega_eq}
  g_{1}^{\rm p}-g_{1}^{\rm n}
=2\,\left(g_{1}^{\rm p}-\frac{g_{1}^{\rm d}}{1-\frac{3}{2}\omega_D}\right).
\end{eqnarray}
For this test we  employ our present proton data and 
our previously published deuteron data~\cite{SMCd93,SMCd95,SMCd96}.  
For the probability of the deuteron to be in a
D-state we have taken  
$\omega_D=0.05\pm0.01$ which covers most of the
published values~\cite{D_state}.
Using the method described in Section~\ref{ej_com}
to account for  the correlations between errors  we obtain
\begin{eqnarray}
 \Gamma_{1}^{\rm p} - \Gamma_{1}^{\rm n} = 0.183 \pm 0.034 \hspace{1.0cm}
       ( \mbox{\it  $Q^2_0$~=~\rm 10\, GeV$^2$\,})~,
\end{eqnarray}
where statistical and systematic errors are combined in quadrature.
The theoretical prediction at the same $Q^2$, including  perturbative
QCD corrections up to ${\cal{O}}(\alpha_s^3)$ and assuming three quark 
flavors (Section~\ref{qcd_cor}),  is
$\Gamma_{1}^{\rm p} - \Gamma_{1}^{\rm n} = 0.186 \pm 0.002 $.

\par
\begin{figure}[t]
\begin{center}
\psfig{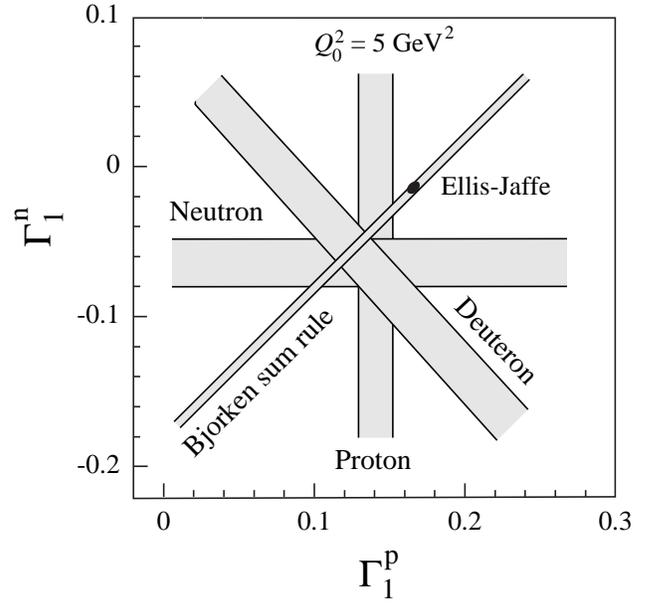}
\vspace{0.25cm}
\caption[]{Comparison of  the combined experimental results 
for $\Gamma_{1}^{\rm p}$, $\Gamma_1^{\rm n}$ and $\Gamma_1^{\rm d}$ with 
the predictions for the Bjorken and the Ellis--Jaffe sum rules.
The Ellis--Jaffe  prediction is shown by the black ellipse inside the 
Bjorken sum rule band.
\label{st_bj}}
\vspace{0.75cm}
\end{center}
\end{figure}

We have also performed a combined analysis of all proton and 
deuteron data at $Q^2_0=5\,\rm GeV^2$  (Fig.~\ref{g1_all}).
The combined $\Gamma_{1}^{\rm d}$ is obtained using the same method as  
described in Section~\ref{ej_com} for $\Gamma_{1}^{\rm p}$. We find
\begin{eqnarray} 
\label{p_d}
\nonumber
\Gamma_{1}^{\rm p}-\Gamma_{1}^{\rm n}=0.199\pm 0.025 ~~~~~
(\mbox{$Q^2_0$=\rm 5\,GeV$^2$,}\\
  \mbox{\it  All proton  and deuteron data})~.
\end{eqnarray}
The corresponding theoretical expectation is    
$\Gamma_{1}^{\rm p} - \Gamma_{1}^{\rm n} = 0.181 \pm 0.003$,
which agrees with the experimental result as shown in Fig.~\ref{st_bj}.

\par
The structure function $g_{1}^{\rm n}$ of the neutron  
has also  been measured 
by scattering polarized electrons on a polarized  $^3$He target~\cite{e142}.
The re-analyzed neutron data on $g_1^{\rm n}$  
from E142~\cite{e142_new} are included in the combined analysis. 
This requires the combination of  $\Gamma_{1}^{\rm p}$,
$\Gamma_{1}^{\rm n}$ and $\Gamma_{1}^{\rm d}$  via a fit constrained by 
the integral of Eq.~(\ref{omega_eq}) and the use of 
a  Monte Carlo method to compute the  $3\times3$ 
correlation matrix between $\Gamma_{1}^{\rm p}$, $\Gamma_{1}^{\rm n}$ 
and $\Gamma_{1}^{\rm d}$.
The  $\Gamma_{1}^{\rm p}$ and $\Gamma_{1}^{\rm d}$  are obtained
as before; $\Gamma_{1}^{\rm n}$  is obtained from the E142 data in their
measured region, but the small-$x$ extrapolation is determined from the 
$g_1^{\rm n}$  values obtained from the SMC proton and deuteron data.
The result is 
\begin{eqnarray} 
\label{p_d_n}
\nonumber
\Gamma_{1}^{\rm p} - \Gamma_{1}^{\rm n}
 = 0.202 \pm 0.022~~~~~~ (\mbox{$Q^2_0$=\rm 5\, GeV$^2$,}\\
       \mbox{\it   All proton, deuteron and neutron data})~.
\end{eqnarray}
As discussed in Ref.~\cite{smc_com},
the central value and the error of $\Gamma_{1}^{\rm n}$ is very
sensitive to the SMC proton and deuteron data.

\par
The relation between $\Gamma_{1}^{\rm p}$, $\Gamma_{1}^{\rm d}$,
$\Gamma_{1}^{\rm n}$ 
and the Bjorken sum rule is illustrated in Fig.~\ref{st_bj} and the
results are given in Table~\ref{gamma_p}.
Proton, deuteron and neutron  results confirm the Bjorken sum but   disagree
with  the Ellis--Jaffe sum rule.

\section{SPIN STRUCTURE OF THE PROTON}  
\label{scon}

\subsection{The $x$ dependence of $g_1^{\rm n}$ and $g_1^{\rm p}$}
\par
In Figure~\ref{g1_f} we show our results for  $g_1^{\rm p}$  and
$g_1^{\rm d}$, together with $g_1^{\rm n}$ obtained from $g_1^{\rm p}$  and
$g_1^{\rm d}$ using Eq.\,(\ref{omega_eq}).
We find that the ratio $g_1^{\rm n}/g_1^{\rm p}$
is close to --1  at small-$x$, in contrast to the ratio 
$F_2^{\rm n}/F_2^{\rm p}$
which is close to +1 for $x<0.01$~\cite{nmc_ratio,E665_RAT,NMC_RAT}. 
In the QPM the difference between $g_1^{\rm p}$ and   $g_1^{\rm  n}$ 
can be written as 
\beq
g_1^{\rm p}-g_1^{\rm n}=\frac{1}{6}[\Delta u_{\rm v}(x)-\Delta d_{\rm v}(x) +
2\Delta \overline{u}(x) - 2\Delta \overline{d}(x) ]~.
\eeq
Under the assumption of flavor symmetry in the polarized sea
$(\Delta \overline{u}=\Delta \overline{d})$\,\cite{kumano1,mulders1}, 
the small-$x$ behavior of
$g_1^{\rm n}/g_1^{\rm p}$ indicates a dominant contribution from 
the valence quarks. This is 
consistent with our results from  semi-inclusive spin 
asymmetries~\cite{SMC_semi} which  show that 
$[\Delta u_{\rm v}(x) - \Delta d_{\rm v}(x)]$  is positive and that 
$\Delta u_{\rm v}(x)$ and  $\Delta d_{\rm v}(x)$ 
have opposite signs.  
Fits of polarized parton distributions in the NLO analysis 
lead to the same conclusion~\cite{grv,stirling}.

\subsection{The axial quark charges}
\label{sec:su3}

\par
When only three flavors contribute to the nucleon spin,
the first moment of $g_1^{\rm p}$ can be expressed in terms 
of the proton matrix elements of the axial vector currents
(Section~\ref{qcd_cor}) 
\beq
\Gamma _1^{\rm p}(Q^2) = \frac{C^{\rm NS}_1(Q^2)}{12} \left [
                    a_3 + \frac{1}{3} a_8 \right ]
                 + \frac{C^{\rm S}_1(Q^2)}{9} \dsigt(Q^2)~.
\eeq

\par
We obtain $\dsigt(Q^2)$ from $\Gamma _1^{\rm p}(Q^2)$ 
and the experimental non-singlet matrix elements $a_3$ and $a_8$, which     
are calculated from $g_{\rm A}/g_{\rm V}$ and $F/D$, as presented 
in Section~\ref{g1_evo}.  The singlet~(non-singlet) 
coefficient function $C^{\rm S}_1~(C^{\rm NS}_1)$ 
is  the same as presented in Section~\ref{qcd_cor},  
and $C^{\rm S}_1$ is computed with the coefficients $c_i^{\rm S}$ 
in the last column of Table~\ref{t_coef}.
If instead the coefficient from the third column were used,
we  would get $\dsigt^{\infty}$.
Numerically, $\dsigt^{\infty}$ is smaller than
$\dsigt(Q^2)$ by 10\% at 
$Q^2=10\,\rm GeV^2$.

\par
From the combined analysis of all proton data we find 
\begin{eqnarray}
\label{delta_con_all}
\nonumber
\dsigt(Q^2_0)= 0.37 \pm 0.11 ~~~~~~~(\mbox{$Q^2_0$~=~\rm 5\, GeV$^2$,}\\
     \mbox{ \it  All proton data })~.
\end{eqnarray}
In Table~\ref{dsigma_t} we compare the results with those 
based on SMC data only.
Calculations in lattice QCD~\cite{dong} agree with the
measured values of both $\dsigt$ and $g_{\rm A}/g_{\rm V}$.
Using $a_3=\au - \ad$, $a_8=\au + \ad-2\as$
and $\dsigt(Q^2)= \au + \ad + \as$,
the individual contributions from quark flavors      are    evaluated from  
\begin{eqnarray}
\label{u_con}
\au
&=& \frac16\,\left[\,2\dsigt(Q^2)+a_8+3a_3\right], \\
\label{d_con}
\ad
&=& \frac16\,\left[2\dsigt(Q^2)+a_8-3a_3\right],\\
\label{s_con}
\as
&=& \frac13\,\left[\dsigt(Q^2)-a_8\right].
\end{eqnarray}
The results are given      in Table~\ref{dsigma_t}.
They indicate that $\as$ is negative, in agreement with 
the measurement  of elastic $\nu$--p scattering~\cite{garvey,ahrens}. 

\par
In the QPM, $a_i= \Delta q_i$. However,
as discussed in Section~\ref{U1}, 
due to the $\U1$ anomaly of the singlet axial vector current
the axial charges receive a gluon contribution.  
In the AB scheme~\cite{bfr95b} used in our QCD fit  for 
three flavors we have
\beq
\label{q_con}
 a_i = \Delta{q}_i -\frac{\alpha_s(Q^2)}{2\pi}\Delta g(Q^2)
  \;\;\;\; (i=u,d,s)~.
\eeq
In this scheme $\Delta q_i$
is independent of $Q^2$. For this reason some authors
consider this to be the correct scheme when assuming 
$\Delta s=0$~\cite{ano_1,ano_2,ano_11}.

\begin{table*}[t]
\begin{center}
\caption[]{Results for  $\dsigt$ and individual
quark contributions from proton data.  
The results based on SMC data only are given at the average $Q^2$ of the data,
$Q^2_0 =$ 10~ GeV$^2$, and at $Q^2_0 =$ 5~GeV$^2$ for a direct comparison with
the combined analysis of all proton data.
\label{dsigma_t}}
\begin{tabular}{ccccc} 
Data used                 & 
  $\dsigt$&
  $\au$&
  $\ad$&
  $\as$\\
\hline
~SMC $\Gamma_1^{\rm p}(10~{\rm GeV}^2)$ &  
$0.28\pm 0.16$ &$ 0.82\pm 0.05$ & $-0.44\pm 0.05$& $-0.10\pm 0.05$ \\
\hline
 SMC $\Gamma_1^{\rm p}(5~{\rm GeV}^2)$ & 
$0.28\pm 0.17$&$0.82\pm0.06$ & $-0.44\pm 0.06$&  $-0.10\pm 0.06$ \\
\hline
 All~~~ $\Gamma_1^{\rm p}(5~{\rm GeV}^2)$ &  
$0.37\pm 0.11$&  $ 0.85 \pm 0.04$& $-0.41\pm 0.04$& $-0.07\pm 0.04$\\
\end{tabular}
\end{center}
\end{table*}

\par
The relation between the matrix element $a_3$ 
and  the neutron $\beta$-decay constant $g_{\rm A}/g_{\rm V}$
relies   only on the  assumption of  isospin invariance. However,
in order to relate $a_8$ to the semileptonic hyperon decay constants  
$F$ and $D$, we assume $\SU3$ flavor symmetry and hence
conclusions on $\dsigt$
depend on its validity.
$\SU3$ symmetry breaking effects do not vanish at first order
for axial vector matrix elements~\cite{ga84},
as they do for vector matrix elements~\cite{ad64}.
It has been suggested that in order to reproduce the
experimental values of $F$ and $D$, the QPM requires large relativistic 
corrections which depend on the quark masses;
since the s quark mass is much larger than that of u and d quarks,
these corrections should break $\SU3$ symmetry.
Similarly, the relations between the baryon magnetic moments
predicted by $\SU3$ are badly broken~\cite{dz91}.

\par
The uncertainty on $a_8$ propagates into $\dsigt$ and $\as$ according to
\bmath
\frac{\partial \dsigt}{\partial a_8}&=&
   -\frac{C_1^{\rm NS}}{4 C_1^{\rm S}} \simeq -0.23~,\\
\frac{\partial\as}{\partial a_8} &=&
   -\frac{C_1^{\rm NS} +4 C_1^{\rm S}}{12 C_1^{\rm S }} \simeq -0.44~.
\emath
The smaller magnitude of $a_s$ and its larger derivative with respect
to $a_8$ make it much more sensitive to uncertainties  
in $a_8$ than $\dsigt$~\cite{li95}.
For instance, the experimental test of $\SU3$ from the compatibility
of different hyperon $\beta$ decays allows for
a 15\% modification of $a_8$;
this would change  $\as$ by as much as 50\%,     
while  $\dsigt$  changes by less than 10\%.

\par
A result for $a_8$ has been obtained from a leading-order
$1/N_c$ expansion~\cite{man_nc} which is much smaller than 
the value based on the  $\SU3$ analysis. 
The use of this  smaller value of  $a_8$ causes
$\dsigt$  to become larger, while $\as$ becomes
positive.

\par 
In principle another source of uncertainty arises from the 
possible contributions
of heavier quarks.  The heavy quark axial current
has a non-zero matrix element because it can mix
with light quark operators~\cite{man_heavy}. 
This mixing is closely related to the $\U1$
anomaly and is directly calculable in QCD~\cite{WHITTEN,ABBOTT},
where the heavy quark contributions can be expressed in terms of light
quark contributions.  Following the analysis of Ref.~\cite{man_heavy}  
and using the result for $\dsigt$
of Eq.~(\ref{delta_con_all}), the expected values
for $\ab$ and $\ac$ for $Q^2 \ll m_b^2$  are
$-0.003\pm 0.001$ and $-0.006\pm 0.002$, respectively.
In view of the current accuracy for  $\dsigt$ and of  the $Q^2$ range covered
by the data, the contribution from heavier quarks can be neglected.

\subsection{The spin content of the proton}
\label{spin_content}

\par
The nucleon spin can be written:
\beq
\label{ang_sum}
S_z=\frac{1}{2} \Delta\Sigma + L_q + \Delta g + L_g =\frac{1}{2}~,
\eeq
in which $\Delta \Sigma = \Delta u +\Delta d +\Delta s$ and $\Delta g$
are the contributions of the quark and gluon spins to the nucleon spin, and 
$L_q$ and $L_g$ are the components of the 
orbital angular momentum of the quarks and the
gluons along the quantization axis~\cite{man_jaffe}.
The $Q^2$ dependence of the angular momentum terms analyzed in
LO was studied in Ref.~\cite{ji_lz2}. It is observed that
the asymptotic limit ($Q^2 \rightarrow \infty$) of the terms
($\frac{1}{2} \Delta\Sigma + L_q$) and ($\Delta g + L_g$) are about the same
and equal to $\sim 1/4$. 

\par 
\begin{figure}[t]
\begin{center}
\psfig{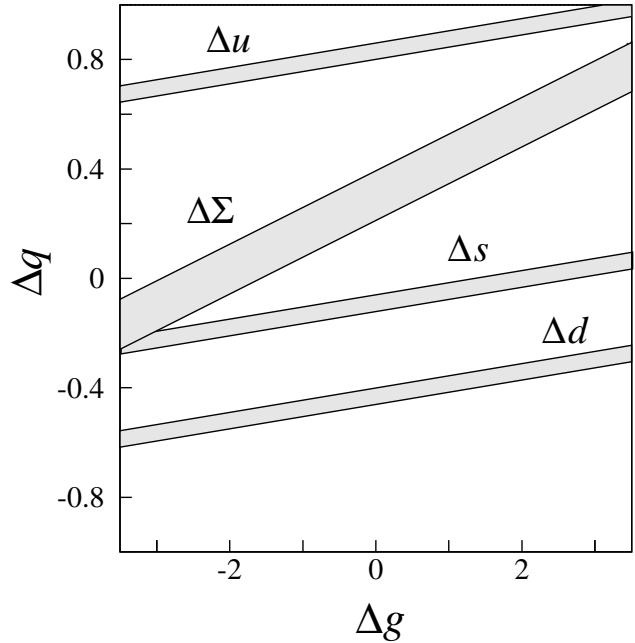}
\caption[]{Quark spin  contributions to the proton  spin as a function of the 
gluon contribution at  $Q^2$ = 5 GeV$^2$ in the Adler--Bardeen scheme.
\label{delta_g}}
\end{center}
\end{figure}

\par
In the naive QPM, $\Delta g=L_g=0$ and $\Delta{\Sigma}=\dsigt$.
In this framework earlier 
experiments  concluded that only a small fraction of 
the nucleon spin is carried
by the quark spins and that the strange quark spin contribution 
is negative. This conclusion     is  in disagreement with
the Ellis--Jaffe assumption of  $\Delta s=\as=0$, 
which corresponds to  $\Delta{\Sigma}=a_8\simeq 0.57$ 
with $L_q$ carrying  about half of the total angular momentum.
The Skyrme Model also assumes  $\Delta{g}=L_g=0$.
In a recent version of this model, where 
$g_{\rm A}/g_{\rm V}$ is calculated to within  4\% of the experimental value,
$\Delta{\Sigma}$ is found to be between 0.18 and 0.32~\cite{Eisen}.

\par
In QCD $a_0$ differs from $\Delta \Sigma$
in a scheme dependent way.
In the AB scheme the determination of $\Delta\Sigma$
and the various $\Delta q_i$
from the measured $a_0$ and $a_i$ requires an input value for $\Delta g$.
The allowed values for  $\Delta{\Sigma}$ and for the $\Delta q_i$
are shown in Fig.~\ref{delta_g}  as a function of $\Delta g$
(Eq.~(\ref{q_con})).
We see that a value of $\Delta s=0$ and
$\Delta{\Sigma}\sim0.57$  corresponds  to 
$\Delta g(Q^2)\approx 2$ at $Q^2_0 = 5\,\rm GeV^2$.
However, the gluon contribution $\Delta g$ could be smaller than indicated
in Eq.~(\ref{q_con}) due to finite quark masses and a possibly
non-negligible contribution from charm, 
according to the authors of Ref.~\cite{Thomas}.
In the absence of direct measurements 
of $\Delta g$ our results can only be compared with the estimate 
of $\Delta g(Q^2)$ 
obtained from NLO GLAP fits to the  $g_1$ data as in Section~\ref{g1_evo}.
Different estimates of $\Delta g(Q^2)$ have been obtained.
The factorization scheme used in the fit of Ref.~\cite{bfr95b}  
and  Section~\ref{g1_evo}
provides $\Delta\Sigma$ and $\Delta g(Q^2)$, while $a_0(Q^2)$ and
$\Delta g(Q^2)$ are obtained in the scheme used for the fits
of Refs.~\cite{grv} and~\cite{stirling}.  While the singlet distribution
depends on the factorization scheme, the gluon distribution is the same
in both~\cite{cheng}.  For $Q_0^2=5\,\mathrm{GeV}^2$ we find
$\Delta g(Q^2_0) = 1.7 \pm 1.1 $ and Refs.~\cite{Fo_paris} and~\cite{grv}
find $\Delta g(Q^2_0)=2.6$ and $0.76$, respectively.  The results of
Ref.~\cite{Fo_paris} are based on the method of Ref.~\cite{bfr95b}.
Similarly, at $Q_0^2=10\,\mathrm{GeV}^2$ it is found 
that  $\Delta g(Q^2_0)$ is equal to 
$2.0\pm 1.3$, $3.1$, and $0.89$, respectively.

\subsection{Combined analysis of  $\dsigt$ from all 
proton, neutron  and deuteron data} 

\par
The~analysis~used~to~test~the~Bjorken sum rule can be extended to evaluate
$\dsigt$, giving \newline
\vspace{0.25cm}
( \mbox{\it Proton, deuteron and neutron  data,
  $Q^2_0$~=~\rm 5\, GeV$^2$})~,
\vspace{-0.25cm}
\begin{eqnarray}
\label{delta_con_dp_all}\nonumber
\dsigt= ~~0.29 \pm 0.06~&,&\hspace{0.5cm}
\au=   ~~0.82\pm 0.02~,\\\nonumber
\ad= -0.43\pm 0.02~&,&\hspace{0.5cm}
\as=  -0.10\pm 0.02~.
\end{eqnarray}

\par
An analysis of  $\dsigt$ 
based on a different selection and treatment of experimental data
has been presented in 
Ref.~\cite{ek95}, with similar results.

\section{CONCLUSION}
\label{conclusion}

\subsection{Summary}
\par
We have presented a complete analysis of our measurement of 
the spin-dependent structure
function $g_1$ of the proton from deep-inelastic scattering of
high-energy polarized muons on a polarized target. 
The data cover the kinematic range $0.003 < x < 0.7$ for
$Q^2 > 1 \:{\rm GeV}^2$, with an average $Q^2 = 10\:{\rm GeV}^2$.
In addition to these data we have also
shown for the first time virtual photon-proton asymmetries
in the kinematic range $0.0008 < x < 0.003$ and $Q^{2} > 0.2$ GeV$^{2}$.
In the kinematic range $x < 0.03$ our data are the 
only available measurements 
of the spin-dependent asymmetries.

\par
The virtual photon asymmetry $A_1^{\rm p}\simeq g_1^{\rm p}/F_1^{\rm p}$ 
shows no $Q^2$ dependence over the $x$ range of our data within
the experimental uncertainty.
This observation holds when we combine our results with those
from electron scattering experiments performed at smaller $Q^2$. However, 
$g_1$ and $F_1$ are predicted to evolve differently  and the difference should
be observable at small-$x$ in future precise measurements.

\par
From our data on $g_{1}^{\rm p}$ together with our deuteron data we find 
that the ratio 
$g_1^{\rm n}/g_1^{\rm p}$ is close to --1 at 
small-$x$ ($\sim 0.005$), in contrast to
$F_2^{\rm n}/F_2^{\rm p}$ which approaches +1.
This suggests that either the valence quarks give a significant contribution
to the net quark polarization in this region, or that
the spin distribution functions of the $u$ and $d$ 
sea quarks are different, i.e. 
$\Delta \overline{u} (x) \neq \Delta \overline{d} (x)$.
The data suggest a rise in $g_1^{\rm p}(x)$ as $x$ decreases from $0.03$
to $0.0008$.  A low-$x$ extrapolation of $g_{1}$ beyond the measured region 
is necessary to compute its first moment $\Gamma_{1}$ and test 
sum rule predictions. 
Precise data at low $x$ are crucial 
for constraining this extrapolation.

\par
The new data have initiated much theoretical activity in recent years,
resulting in an extensive discussion of the NLO QCD analyses of the $x$
and $Q^2$ dependence of $g_1$ and of the interpretation of $\dsigt$ in
terms of the spin content of the nucleon.  As a result, we have used
new methods for the evaluation of 
the structure function $g_1$ at fixed $Q^2$.
From this evolved structure function we determined the 
first moment of $g_{1}$ 
and confirmed the violation of 
the Ellis--Jaffe sum rule for the proton by more than $2 \sigma $.
We obtain for the singlet axial charge
of the proton $\dsigt(Q^2_0) = 0.28 \pm 0.16$ at $Q^2_0 = 10\:{\rm GeV}^2$.
From the fit to all currently available data, we obtain
$\Delta g(Q^2_0) = 2.0 \pm 1.3$ which 
in the Adler-Bardeen renormalization scheme implies 
the value $\Delta\Sigma\simeq 0.5$.
The new data and theoretical developments now afford a first glimpse of
the polarized gluon distribution and its first moment.

\par
The Bjorken sum rule is fundamental and must hold in perturbative QCD.
When corrections up to ${\cal O}(\alpha_{s}^{3})$ are included, it 
predicts $\Gamma_1^{\rm p} - \Gamma_1^{\rm n} = 0.186\pm 0.002$ 
at $Q^2_0 = 10\:{\rm GeV}^2$.
Using the first moments of the structure functions $g_{1}$ evaluated from
our proton and deuteron data, we find 
$\Gamma_1^{\rm p} - \Gamma_1^{\rm n} = 0.183\pm 0.034$ at $Q^2_0 = 
10\:{\rm GeV}^2$ in excellent agreement with the theoretical prediction.  
Combining our data with all available data 
results in a somewhat more precise confirmation of the Bjorken sum rule.

\subsection{Outlook}

\par
New data on the spin-dependent structure functions $g_1$ and $g_2$  
of the nucleon are expected in the next two  years 
from the SMC, the E154 and  E155 collaborations at SLAC, and 
from the HERMES collaboration 
at HERA. However,  further knowledge is needed of the low $x$ behavior of 
$g_1$ and of the polarized gluon 
distribution $\Delta g(x)$ due to the limited coverage 
in $x$ and $Q^2$ of these experiments.

\par
Future experiments are planned at various experimental facilities, including
semi-inclusive polarized proton--proton scattering by 
RHIC SPIN~\cite{RHICSPIN} 
at BNL, semi-inclusive polarized muon--nucleon scattering by 
COMPASS~\cite{COMPASS} at CERN, 
and a similar semi-inclusive polarized electron--nucleon  
experiment at SLAC~\cite{nSLAC}.
Furthermore, a polarized electron--proton collider experiment  at HERA
to study the inclusive and semi-inclusive 
scattering is also under consideration~\cite{HERA}.
The non-Regge behavior of the unpolarized structure function $F_{2}$
has been observed at HERA in agreement with perturbative QCD 
predictions~\cite{f2_h1,f2_zeus}. The corresponding behavior predicted 
for the polarized spin structure function $g_1$  
is particularly interesting due to 
the fact that the higher-order corrections in the polarized case 
are expected to be stronger~\cite{Fo_paris,hoc}. 
Also, unlike  the unpolarized case where only the 
gluon distribution is important at low~$x$, in the polarized 
case the singlet quark, the non-singlet quark, and the gluon 
distributions all play a role.

\par
In conclusion, the study of the spin structure of the nucleon  appears
certain to remain active well into the next century.

\section{ACKNOWLEDGMENT}
\par
We wish to thank our host laboratory CERN for providing
major and efficient support for our experiment and an exciting and
pleasant environment in which to do it. 
In particular, we thank J.V.~Allaby, P.~Darriulat, F.~Dydak, L.~Foa,
 G.~Goggi, H.J.~Hilke and H.~Wenninger for substantial support and
constant advice. We also wish to thank L. Gatignon and the SPS Division for
providing us with an excellent beam,  the LHC-ECR group for efficient
cryogenics support, and J.M.~Demolis for all his technical support.       

We also thank all those people in our home institutions who have
contributed to the construction and maintenance of our equipment,
especially A.~Da\"el, J.~C.~Languillat and C.~Cur\'e from DAPNIA/Saclay
for providing us with the high performance target superconducting magnet,
Y. Lef\'evre and J. Homma from NIKHEF for their contributions to the
construction of the dilution refrigerator, and E. Kok for his contributions
to the electronics and the data taking.

It is a pleasure to thank G.~Altarelli, R.D.~Ball, J.~Ellis,
S.~Forte and  G.~Ridolfi, for valuable discussions.





\end{document}